\documentclass[fleqn,usenatbib]{mnras}

\usepackage{newtxtext,newtxmath}

\usepackage[T1]{fontenc}

\DeclareRobustCommand{\VAN}[3]{#2}
\let\VANthebibliography\thebibliography
\def\thebibliography{\DeclareRobustCommand{\VAN}[3]{##3}\VANthebibliography}

\usepackage{graphicx}	
\usepackage{amsmath}	

\def\colibre{COLIBRE}

\title[Kennicutt-Schmidt relation in \colibre]{Kennicutt-Schmidt relation of galaxies over 13 billion years in the \colibre\ hydrodynamical simulations}

\author[C.D.P. Lagos et al.]{
\parbox[t]{\textwidth}{
\vspace{-0.5cm}
Claudia del P. Lagos$^{1,2}$\thanks{E-mail: claudia.lagos@icrar.org}, 
Joop Schaye,$^{3}$
Matthieu Schaller,$^{3}$
Danail Obreschkow,$^{1}$
Yannick M. Bahé,$^4$
Alejandro Benítez-Llambay,$^5$
Evgenii Chaikin,$^{3}$
Camila Correa,$^3$ 
Timothy A. Davis,$^6$ 
Carlos S. Frenk,$^7$ 
Filip Hu{\v{s}}ko,$^{3}$ 
Melanie Kaasinen,$^8$ 
Robert J. McGibbon,$^{3}$
Kyle Oman,$^7$ 
Sylvia Ploeckinger,$^{9}$ 
Alexander J. Richings,$^{10,11}$  
James W. Trayford,$^{12}$ 
Jing Wang,$^{13}$ 
Ruby J. Wright$^{1}$}
\\
$^{1}$International Centre for Radio Astronomy Research (ICRAR), M468, University of Western Australia, 35 Stirling Hwy, Crawley, WA 6009, Australia.\\
$^{2}$Cosmic Dawn Center (DAWN), Denmark.\\
$^3$Leiden
Observatory, Leiden University, PO Box 9513, 2300 RA Leiden, the Netherlands.\\
$^4$School of Physics and Astronomy, University of Nottingham, University Park, Nottingham NG7 2RD, UK.\\
$^5$Dipartimento di Fisica G. Occhialini, Università degli Studi di Milano Bicocca, Piazza della Scienza, 3 I-20126 Milano MI, Italy.\\
$^6$Cardiff Hub for Astrophysics Research \& Technology, School of Physics \& Astronomy, Cardiff University, Queens Buildings, Cardiff CF24 3AA, UK.\\
$^7$Department of Physics, Institute for Computational Cosmology, Science Laboratories, Durham University, South Road, Durham, DH1 3LE, UK.\\
$^8$Research School of Astronomy and Astrophysics, Australian National University, Canberra, ACT 2611, Australia.\\
$^9$Department of Astrophysics, University of Vienna, Turkenschanzstrasse 17, 1180 Vienna, Austria.\\
$^{10}$Centre for Data Science, Artificial Intelligence and Modelling, University of Hull, Cottingham Road, Hull, HU6 7RX, UK.\\
$^{11}$E. A. Milne Centre for Astrophysics, University of Hull, Cottingham Road, Hull, HU6 7RX, UK.\\
$^{12}$Institute of Cosmology and Gravitation, University of Portsmouth, Dennis Sciama Building, Burnaby Road, Portsmouth PO1 3FX, UK.\\
$^{13}$Kavli Institute for Astronomy and Astrophysics, Peking University, Beijing 100871, China.
}

\date{Accepted XXX. Received YYY; in original form ZZZ}

\pubyear{\the\year{}}

\begin{document}
\label{firstpage}
\pagerange{\pageref{firstpage}--\pageref{lastpage}}
\maketitle

\begin{abstract}
We investigate the correlation between star formation rate (SFR) surface density and gas surface density (known as the Kennicutt–Schmidt, KS, relation) at kiloparsec (kpc) scales across cosmic time ($0\le z \le 8$) {for galaxies with stellar masses $>10^9\,\rm M_{\odot}$}, using the \colibre\ state-of-the-art cosmological hydrodynamical simulations. These simulations feature on-the-fly non-equilibrium chemistry coupled to dust grain evolution and detailed radiative cooling down to $\approx 10$~K, enabling direct predictions for the atomic (\ion{H}{i}) and molecular (H$_2$) KS relations. At $z\approx 0$, \colibre\ reproduces the observed (spatially-resolved) KS relations for \ion{H}{i} and H$_2$, including the associated scatter, which we predict to be significantly correlated with stellar surface density, local specific SFR (sSFR), and gas metallicity. We show that the \ion{H}{i} KS relation steepens for lower-mass galaxies, while the H$_2$ KS relation shifts to higher normalisation in galaxies with higher sSFRs. The H$_2$ depletion time decreases by a factor of $\approx 20$ from $z = 0$ to $z = 8$, primarily due to the decreasing gas-phase metallicity. This results in less H$_2$ and more \ion{H}{i} being associated with a given SFR at higher redshift. We also find that galaxies with higher sSFRs have a larger molecular gas content and higher star formation efficiency per unit gas mass on kpc scales. The predicted evolution of the H$_2$ depletion time and its correlation with a galaxy's sSFR agree remarkably well with observations in a wide redshift range, $0\le z\le 5$. 
\end{abstract}

\begin{keywords}
galaxies: ISM -- galaxies: star formation -- galaxies: evolution -- galaxy: formation
\end{keywords}



\section{Introduction}

The connection between star formation and the interstellar medium (ISM) of galaxies is one of the most fundamental questions in extragalactic astronomy. Many surveys have been dedicated to understanding how the star formation rate (SFR) of galaxies, either in a global or in a spatially resolved manner, is connected to the different phases of the ISM (see \citealt{Tacconi20} and \citealt{Schinnerer24} for recent reviews).  

In the local Universe the SFR correlates strongly with the molecular hydrogen (H$_2$) content, with the intrinsic scatter being significantly smaller than for the correlations obtained with either atomic hydrogen (\ion{H}{i}) or total neutral hydrogen (\ion{H}{i}+H$_2$) (e.g. \citealt{Leroy08,Bigiel08,Schruba11,Pessa21}). These relations have been studied at scales from a few hundred parsecs (pc) to a kilo pc (kpc) through correlating the surface densities of SFR, $\Sigma_{\rm SFR}$, and gas, $\Sigma_{\rm gas}$, in many tens of galaxies in the local Universe. This relation is known as the Kennicutt-Schmidt (KS) relation, named after the seminal works of \citet{Schmidt59} and \citet{Kennicutt83,Kennicutt89}. Establishing how the KS relation evolves to higher redshift, $z$, for either \ion{H}{i} or H$_2$, has been challenging, with the samples in which a resolved KS relation can be measured being limited to $\lesssim 10$ galaxies (e.g. \citealt{Freundlich13,Sharon13,Hodge15,Sharon19,Nagy23,Bethermin23,Zanella24,Gomez25}). However, from global unresolved measurements of SFR and H$_2$ mass in thousands of galaxies, it has been possible to establish that the two quantities continue to correlate well up to at least $z\approx 5$ (e.g. \citealt{Tacconi20}).

Local Universe observations have also helped establish that the KS relation, either in its atomic, molecular or total neutral gas form, displays a scatter that  correlates with other galaxy properties; for instance, with the stellar surface density (e.g. \citealt{Shi11,Shi18,Dey19,Ellison20,Lin22,Wang24,Wong24}), suggesting that the stellar contribution to the local dynamical state plays a role in the gas-to-star conversion (e.g. \citealt{Blitz06,Ostriker10}). Even when very dense gas traced by HCN (hydrogen cyanide) is used to measure the KS relation, observations continue to show scatter that correlates with other local properties \citep{Usero15,Bigiel16}, confirming that the underlying process of star formation is  sensitive to local physical conditions, such as the stellar surface density (e.g. \citealt{Gallagher18,Jimenez-Donaire19}).

The physical drivers of the atomic and molecular KS relations are not understood. For instance, it is not clear to what extent molecular gas is a prerequisite for star formation as opposed to being a by-product of the cold and dense gas phase normally associated with star formation \citep{Schaye04,Glover12}. Several state-of-the-art cosmological galaxy formation models and simulations assume that H$_2$ is required to form stars (e.g. \citealt{Lagos10,Somerville15,Lacey15,Lagos18,Xie17,Dave19,Feldmann23,Lagos24}); while others only consider dense gas, whether it is molecular or not, and typically assume a fixed threshold density (and/or temperature) for star formation (e.g. \citealt{Tremmel17,Pillepich18,Semenov18,Agertz21}). Going beyond a fixed density threshold for star formation, \citet{Schaye14} attempted to encapsulate the transition from warm to cold ISM into a metallicity-dependent threshold density.

On smaller scales than usually probed by the KS relation results, observations of thousands of individual molecular clouds show very inefficient star formation, with only $\simeq 1$\% of the gas being converted into stars per dynamical time (see \citealt{Schinnerer24} for a review). This inefficiency is challenging to explain \citep{Federrath15}, with a combination of high turbulence (e.g. \citealt{Krumholz05,Federrath12}) and efficient early stellar feedback (e.g. \citealt{Agertz13,Segovia25}) likely needed to cause a quick dispersal of clouds shortly after they have started to form stars. 

It is still an open question how to combine the small scale, almost constant efficiency of gas-to-star conversion at the molecular cloud level, with the emergence of the atomic and molecular KS power law on larger scales, at $\approx 1$~kpc. Cosmological hydrodynamical simulations have recently attempted to shed light on this problem. \citet{Kraljic24} used the {\sc NewHorizon} cosmological hydrodynamical simulation box to study the emergence of the KS relation, focusing primarily on low-mass galaxies, which are more readily captured in their small cosmological volume of $(16\,\rm Mpc)^3$. They found the local gas turbulence to be a key driver of the scatter in the KS relation. This is perhaps unsurprising given the explicit dependence of the star formation efficiency on gas turbulence in their underlying star formation model. They also found that the observed slope of the KS relation at $z=0$ is only established in their simulation at $z\approx 2-3$. More studies like this are needed to understand how general these results may be.

In this paper, we use a new suite of cosmological hydrodynamical simulations, \colibre\ \citep{Schaye25,Chaikin25a} to study the emergence of the atomic, molecular and total neutral gas KS relations, including their scatter. \colibre\ is especially well suited for this study as it includes a sophisticated model of the ISM that takes into account non-equilibrium chemistry of H and He coupled with a live dust model and gas cooling that tracks gas temperatures down to $\approx 10$~K and accounts for self-shielding and local sources of radiation \citep{Ploeckinger25,Trayford25}. This means that the \ion{H}{i} and H$_2$ contents of the gas are {\it predictions} of the simulation. Previous generations of cosmological hydrodynamical simulations did not directly model the formation of \ion{H}{i} and H$_2$ which could only be included in post-processing, imposing chemical and thermal equilibrium (e.g. \citealt{Lagos15,Diemer17,Dave20}). \citet{Richings22} show that these assumptions fail catastrophically at the high densities and low temperatures typical of the H$_2$-dominated gas, with differences up to an order of magnitude in the expected abundance of \ion{H}{i} and H$_2$ in that regime. 
\colibre\ also includes a range of volumes and resolutions that allow us to probe a very large dynamic range in stellar mass, from dwarf galaxies to the most massive galaxies in the universe. \colibre\ uses a star formation model that assumes a fixed efficiency of gas conversion per free fall time of $1$\%, and hence the emergence of molecular and atomic KS relations is not at all trivial.

This paper is organised as follows. Section~\ref{colibre} presents the \colibre\ simulations used in this work, briefly describing the physical models included. Section~\ref{colibre} also describes how we measure the KS relation and some of the other resolved properties of interest, such as gas metallicity and stellar surface density. 
Section~\ref{KSrelationLocalUSection} presents a detailed study of what \colibre\ predicts for the atomic, molecular and total neutral KS relations at $z=0$, including correlations that arise between other resolved properties and the scatter of the KS relation.  Section~\ref{KSrelationLocalUSection} also presents a comprehensive comparison with observations of the local Universe that probe the zero point and scatter of the KS relation. Section~\ref{sec:zevo} focuses on the predicted redshift evolution of the KS relation from $z=8$ to $z=0$, comparing with observations where possible, and exploring the correlation between the scatter of the KS relation and third parameters, such as gas metallicity and stellar surface density. Section~\ref{sec:conclusions} presents our conclusions. In Appendix~\ref{convergence}, we present extensive convergence tests of the KS relation.

\section{The \colibre\ simulations}\label{colibre}

\colibre\ is a suite of cosmological hydrodynamical simulations of varying volume and resolution. A key aspect of \colibre\ is the leap in the sophistication of the modelling of several ``subgrid'' modules compared to the previous generation of simulations in large cosmological volumes. Subgrid modules encapsulate the physics that happens below the resolution scale of the simulation and include (i) radiative cooling and heating; (ii) the formation, growth and destruction of dust grains; (iii) star formation from unstable gas; (iv) stellar mass loss and the rates of supernova (SN) Ia; (v) stellar winds, radiation pressure from starlight, and HII regions, i.e. stellar
feedback processes that start earlier than SN feedback; (vi) SN feedback; (vii) black holes (BHs); (viii) Active Galactic Nuclei (AGN)  feedback; (ix) 
turbulent diffusion. Below we summarise the main characteristics of the subgrid physics modules. We refer to \citet{Schaye25} for a full description:

\begin{itemize}
    \item {\it Numerical setting.} The \colibre\ simulation suite uses the SWIFT code \citep{Schaller24} with the {\sc Sphenix} Smoothed Particle Hydrodynamics (SPH) formulation for hydrodynamics \citep{Borrow22} and a fourth-order fast multipole method for gravity \citep{Dehnen14}. It includes both baryonic and dark matter (DM) particles, with four times more DM than baryonic particles, which helps suppress spurious transfer of energy from DM to stellar particles and hence improves the effective resolution for the baryonic component \citep{Ludlow19,Ludlow21,Ludlow23}. Neutrino effects are modelled semi-linearly to account for their influence on structure growth \citep{Ali-Haimoud13}.

\item {\it Radiative cooling} is allowed down to $\approx 10$~K. Cooling is computed using an updated version of the {\sc Chimes} chemical network \citep{Richings14a,Richings14b}, integrated into the {\sc hybrid-chimes} framework \citep{Ploeckinger25}. The model treats non-equilibrium hydrogen and helium chemistry explicitly and uses equilibrium tables for metal cooling, corrected for non-equilibrium free-electron densities. It includes molecular hydrogen formation, photoionisation, photodissociation, photoelectric heating by dust, and shielding by gas and dust, using a local Jeans column approximation \citep{Schaye01,Ploeckinger20}. Radiation sources include the Cosmic Microwave Background (CMB), the UV/X-ray background \citep{Faucher-Giguere20}, and a local interstellar radiation field \citep{Ploeckinger25}.

\item {\it Dust evolution} is modelled on-the-fly using six dust species (three compositions, two sizes), following \citet{Trayford25}. Dust grains are produced by Asymptotic Giant Branch (AGB) stars \citep{DellAgli17} and core-collapse supernovae (CCSNe; \citealt{Zhukovska08}), and grow via gas accretion following \citet{Hirashita14}. Destruction occurs via sputtering \citep{Tsai95}, stellar feedback, and astration. Grain-grain collisions lead to shattering and coagulation \citep{Aoyama17,Granato21}. The dust physics is coupled to the chemistry and cooling modules.

\item {\it Star formation:} follows a gravitational instability criterion \citep{Nobels24}, based on the local thermal and turbulent velocity dispersions. Specifically, gas particles are only eligible for star formation if they satisfy 

\begin{equation}
    \alpha\equiv \frac{\sigma^2_{\rm th} + \sigma^2_{\rm turb}}{G\,\langle N_{\rm ngb}\rangle^{2/3}\, m_{\rm g}^{2/3} \, \rho_{\rm g}^{2/3}}\, < 1,\label{alphacrit}
\end{equation}

\noindent where $\sigma_{\rm th}$ and $\sigma_{\rm turb}$ are the particle’s 3D thermal and turbulent velocity dispersions, respectively. $N_{\rm ngb}$ 
  is the weighted mean number
of neighbours in the SPH kernel, which is about 65 for a quartic spline SPH kernel adopted in \colibre. $m_{\rm g}$ is the mean, initial gas particle mass (which we list in Table~\ref{TableSimus} for the runs used in this work). $\rho_{\rm g}$ is
the gas density. 
$\sigma_{\rm th}$ and $\sigma_{\rm turb}$ are computed from a gas particle's temperature and the local inter-particle velocity (peculiar plus Hubble) dispersion. 
Gas that meets the criterion in Eq.~\ref{alphacrit} is stochastically converted into stars using a \citet{Schmidt59} law ($\rho_{\rm SFR}\propto \rho_{\rm gas}/\tau_{\rm ff}$) with a fixed efficiency of $1$\% per free-fall time, $\tau_{\rm ff}$. The free-fall time scales as $\rho^{-0.5}_{\rm gas}$, so $\rho_{\rm SFR}\propto \rho^{3/2}_{\rm gas}$.

Because the instability criterion is evaluated at the resolution limit, the instability requirement in Eq.~\ref{alphacrit} enables higher-resolution runs to probe denser ISM conditions.
This also means that as the resolution increases, a larger fraction of the SFR happens in denser gas compared to lower resolution runs (which we show in \S~\ref{comparisonobsandcharac}). \citet{Nobels24}, using a suite of idealised galaxy simulations, demonstrated that coarse-grained quantities, such as the observable Kennicutt-Schmidt law on kpc scales, can still converge with the resolution. We demonstrate that this is indeed the case using the fully cosmological simulations \colibre\ (see Appendix~\ref{convres}). 

\item {\it Chemical enrichment} models mass loss from Asymptotic Giant Branch (AGB) stars, core collapse SNe, and SNIa using updated nucleosynthetic yields \citep{Correa26}. Tracked elements include the $11$ cooling-dominant species \citep{Wiersma09}, s-process elements (Ba, Sr), and r-process element (Eu). The SNIa rate follows an exponential delay-time distribution.
The model uses metallicity-dependent stellar lifetimes \citep{Portinari98}, with SNIa yields from the W7 model of \citet{Leung19}, core collapse SNe yields from \citet{Nomoto06, Nomoto13}, and pre-SN winds from \citet{Kobayashi06}. C and Mg yields are scaled by $1.5$ for massive stars to match APOGEE observations \citep{Jonsson18}. 

\item {\it Turbulent diffusion} is included to model small-scale mixing of metals and dust, following a diffusion approach motivated by Kolmogorov turbulence \citep{Martinez-Serrano08,Greif09,Shen10}. The diffusion coefficient is tied to the local velocity shear, with the strength calibrated to reproduce Milky Way abundance patterns \citep{Correa26}.

\item {\it Stellar feedback:} includes early feedback from stellar winds, radiation pressure, and HII regions \citep{Benitez-Llambay25}. Core collapse SN feedback includes a stochastic thermal component (\citealt{DallaVecchia12}, with updates) and a kinetic component to drive turbulence \citep{Chaikin23}. SNIa feedback uses the same thermal model to prevent numerical overcooling.

\item {\it Black hole growth and AGN feedback:} The fiducial simulations use the black holes and thermal AGN feedback following \citet{CBooth09} and \citet{Bahe22}, with modifications to improve the sampling of feedback from low-mass black holes. A subset of runs uses a hybrid AGN feedback model with both thermal and kinetic jets, and black hole spin tracking \citep{Husko25}.
\end{itemize}

All feedback parameters are calibrated to reproduce the galaxy stellar mass function, galaxy sizes, and black hole masses at redshift zero, as described in \citet{Chaikin25a}.


The adopted cosmology is consistent with the DES Y3 ``3×2pt + All Ext.'' $\Lambda$CDM constraints \citep{Abbott22}. This assumes a flat universe with key parameters: $H_{\rm 0}= 68.1\,\rm  km/s/cMpc$, $\Omega_{\rm m}=0.306$, $\Omega_{\rm b} = 0.0486$, $\Omega_{\Lambda}=0.693922$, $\sigma_{8}=0.807$, $n_{\rm s}=0.967$, $\Sigma\,m_{\nu}\,c^2=0.06$~eV (corresponding to one massive plus two massless neutrinos). The initial baryonic composition assumes $X = 0.756$ and $Y = 0.244$. 

Structure identification proceeds in three stages:
\begin{itemize}
    \item Friends-of-Friends (FoF): Identifies DM haloes using a standard linking length of $0.2$ times the mean DM particle separation. Baryonic particles are linked to the nearest DM particle in a halo. Groups with $<32$ particles are discarded.
    \item HBT-HERONS \citep{ForouharMoreno25} Tracks subhaloes across snapshots using a hierarchical bound-tracing approach, improving on HBT+ \citep{Han12,Han18}. It excels at identifying satellite subhaloes near halo centres and avoids issues like artificial mass fluctuations at pericentres \citep{ForouharMoreno25}, which can affect the reconstruction of merger trees \citep{Chandro-Gomez25}. 
    \item SOAP \citep{McGibbon25}: Calculates halo and galaxy properties within various 3D and projected apertures. The default aperture for galaxy properties used in this work is a sphere of radius $50$ proper kpc (pkpc), centred on the most bound particle. Spherical over-density masses are also computed.
\end{itemize}

The combination of FoF, HBT-HERONS, and SOAP provides a detailed and robust catalogue of halo and galaxy properties, that enables a consistent post-processing of \colibre\ simulation data.
From the SOAP catalogues, we use 
 stellar masses, instantaneous SFRs, \ion{H}{i} and H$_2$ masses. We also use these properties to define the specific SFR, $\rm sSFR \equiv \rm SFR/M_{\star}$.


%
In this study, we select all galaxies with $M_{\star}>10^9\,\rm M_{\odot}$ and $\rm SFR>0$ to construct their resolved KS relation. We apply the same selection to snapshots from $z=0$ to $z=8$. This selection yields $106,572$ galaxies at $z=0$ in the L200m6 box, with the number of galaxies increasing to $120,398$ at $z\approx0.5$, followed by a steady decrease to $289$ at $z=8$. Appendix~\ref{convres} presents a convergence study that explains our reasoning for choosing $M_{\star}>10^9\,\rm M_{\odot}$ as our minimum stellar mass for this analysis.

{We make reference throughout this work to gas particles belonging to the hot ionised medium (HIM), warm neutral medium (WNM), and cold neutral medium (CNM). For simplicity, these phases are defined purely by gas temperature, acknowledging that this classification does not uniquely map onto ionisation state (e.g. the WNM may contain ionised gas and the HIM may contain neutral gas; see Fig.~\ref{PhaseSpacez0}). Specifically, we define the phases as follows:}

\begin{align}
{\rm HIM}&: T\ge 10^{4.5}\,\rm K \nonumber\\
{\rm WNM}&: 10^3\,\rm K <{\it T}<10^{4.5}\,\rm K,\nonumber\\
{\rm CNM}&: T\le 10^{3}\,\rm K.\label{CNMdef}
\end{align}

\subsection{\colibre\ runs used in this work}

\begin{table*}
\begin{center}
  \caption{\colibre\ simulations used in this paper. The columns list:
    (1) the name of the simulation, (2) comoving box size ($L$), (3) number
    of initial baryon ($N_{\rm b}$) and (4) DM particles ($N_{\rm DM}$), (5) initial mean particle masses of gas and (6) dark
    matter, (7) comoving gravitational softening length ($\epsilon_{\rm com}$) and (8) the maximum proper gravitational softening length ($\epsilon_{\rm prop}$). Here, ``Thermal'' refers to runs using the thermal AGN feedback model, while ``Hybrid'' refers to runs using the AGN feedback model of \citet{Husko25}. Our fiducial resolution is m6, {and unless otherwise stated, L200m6 is the default simulation used for the analysis in this paper. To make this decision visually clear, we highlight the name of that simulation in bold.}}\label{TableSimus}
\begin{tabular}{l c c c c c c c}
\\[3pt]
\hline
(1) & (2) & (3) & (4) & (5) & (6) & (7) & (8)\\
\hline
Name & $L$ & $N_{\rm b}$ & $N_{\rm DM}$ & gas particle mass & DM particle mass & $\epsilon_{\rm com}$ & $\epsilon_{\rm prop}$ \\
Units & $[\rm cMpc]$   &                &  &   $[\rm M_{\odot}]$ &  $[\rm M_{\odot}]$ & $[\rm ckpc]$ & $[\rm pkpc]$\\
\hline
L025m5 (Thermal) & $25$  &  ~$752^3$    & $4\times 376^3$ & $2.30\times 10^5$ & $3.03\times 10^5$   & $0.9$ & $0.35$  \\
L025m6 (Thermal) & $25$ &   $376^3$  & $4\times 376^3$ & $1.84\times 10^6$   &   $2.42\times 10^6$ & $1.8$ & $0.7$  \\
L025m7 (Thermal) & $25$ &  $188^3$ & $4 \times 188^3$ &    $1.47\times 10^7$   &  $1.94\times 10^7$ & $3.6$ & $1.4$  \\
L050m6 (Thermal) & $50$ &   $752^3$  & $4\times 752^3$ & $1.84\times 10^6$   &   $2.42\times 10^6$ & $1.8$ & $0.7$  \\
L050m6 (Hybrid) & $50$ &   $752^3$  & $4\times 752^3$ & $1.84\times 10^6$   &   $2.42\times 10^6$ & $1.8$ & $0.7$  \\
{\bf L200m6 (Thermal)} & $200$ &   $3008^3$  & $4\times 3008^3$ & $1.84\times 10^6$   &   $2.42\times 10^6$  & $1.8$ &  $0.7$\\
\hline
\end{tabular}
\end{center}
\end{table*}

Table~\ref{TableSimus} presents the \colibre\ runs used in this work. Throughout this paper, we will be primarily using the L200m6 run (using a cubic volume of 200 comoving Mpc on a side, and an initial mean particle mass of $10^6\,\rm M_{\odot}$). When extending the lower stellar mass limit from $10^9\,\rm M_{\odot}$ to $10^8\,\rm M_{\odot}$, we also employ the higher resolution run L025m5 (a cubic volume  of 25 comoving Mpc on a side, and an initial mean particle mass of $10^5\,\rm M_{\odot}$). In Appendix~\ref{convres}, we use the L025 runs of different resolutions to study how well converged the predicted KS relations are with the numerical resolution. Throughout this work, we primarily use the ``Thermal AGN feedback'' runs of \colibre. \citet{Schaye25} present the full suite of \colibre\ runs, which include more boxes and variations with the ``hybrid AGN feedback'' model. AGN feedback in \colibre\ primarily affects massive galaxies, and the varying AGN feedback models have the greatest impact on the population of massive, quenching/quenched galaxies. We confirm in Appendix~\ref{convagn} that the Thermal and Hybrid AGN feedback models predict KS relations at $z=0$ that are almost indistinguishable from each other. Thus, throughout this paper we only analyse runs that adopt the Thermal AGN feedback model.



\subsection{Constructing resolved maps}\label{ResolvedMapsMethods}
\begin{figure*}
\begin{center}
\includegraphics[trim=0mm 0mm 0mm 0mm, clip,width=0.33\textwidth]{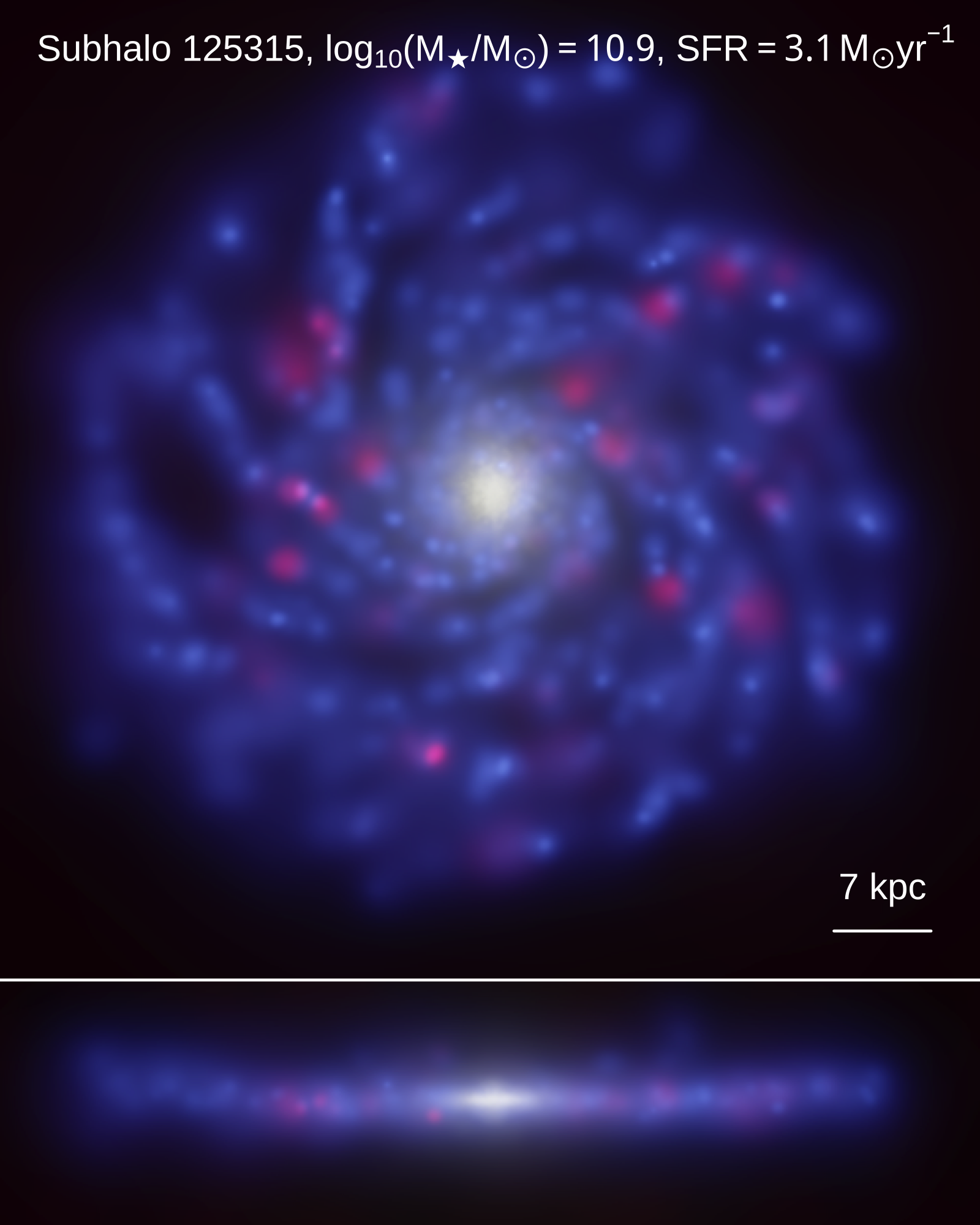}
\includegraphics[trim=0mm 0mm 0mm 0mm, clip,width=0.33\textwidth]{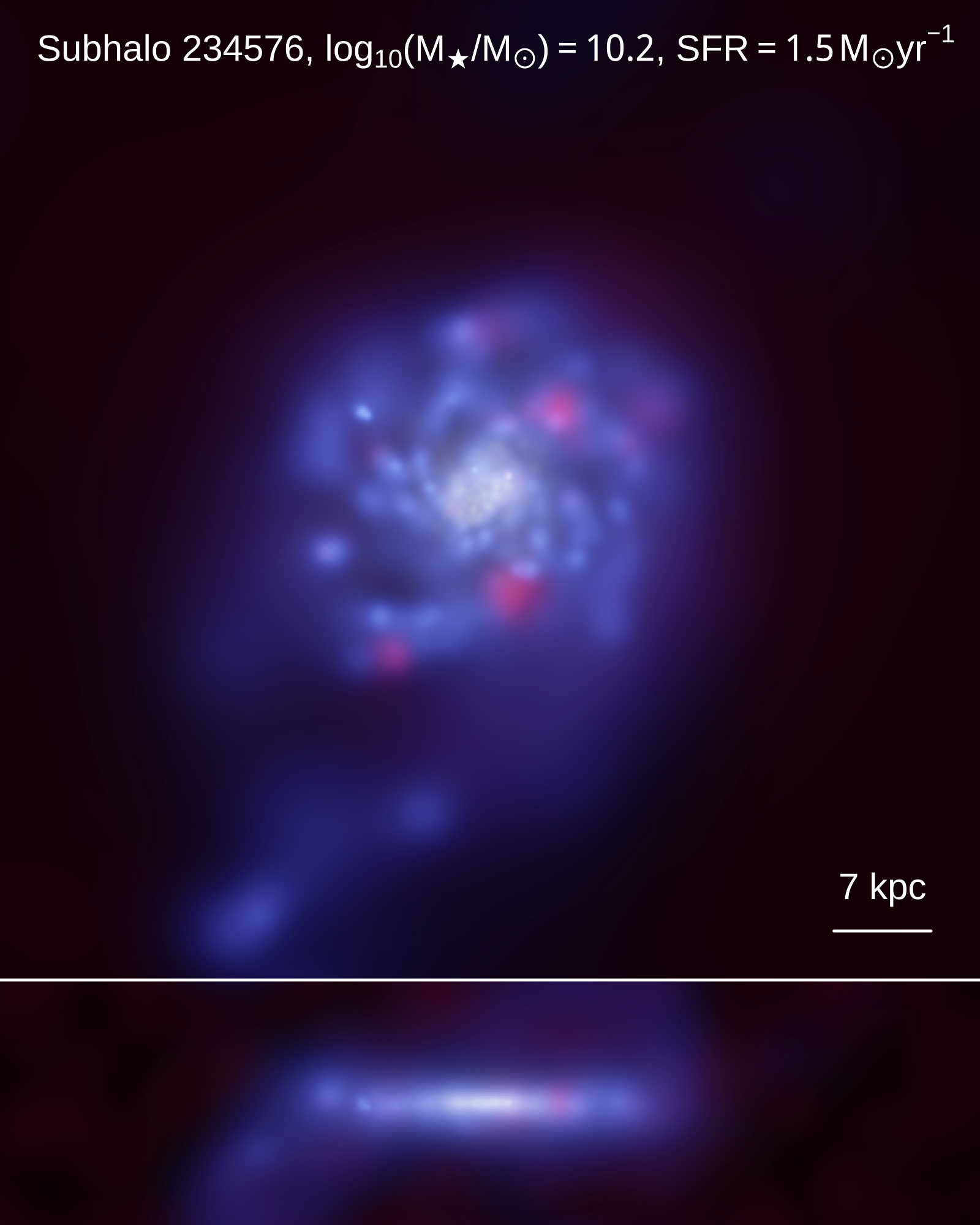}
\includegraphics[trim=0mm 0mm 0mm 0mm, clip,width=0.33\textwidth]{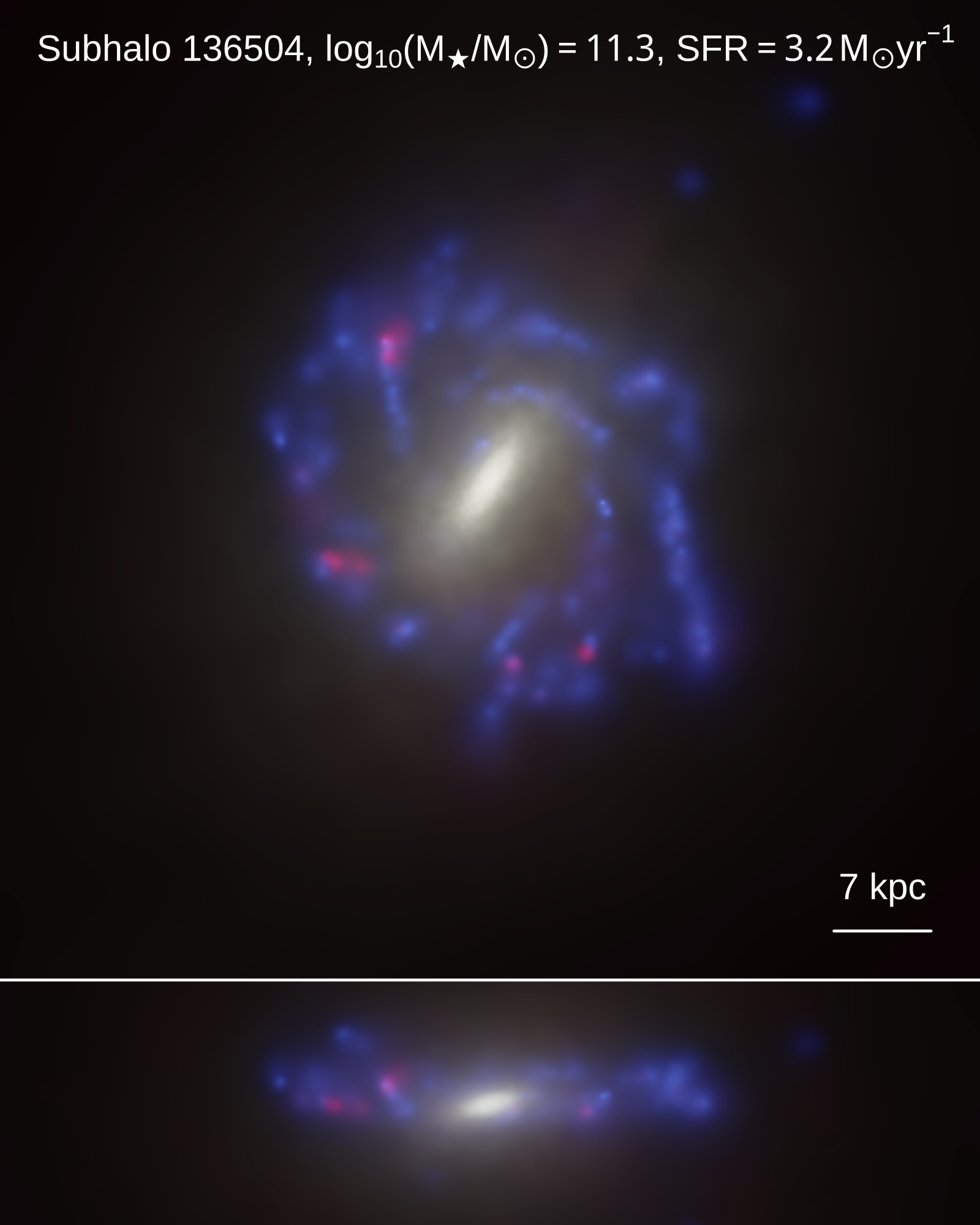}
\hspace*{-0.8cm}\includegraphics[trim=5mm 5mm 3mm 1mm, clip,width=0.37\textwidth]{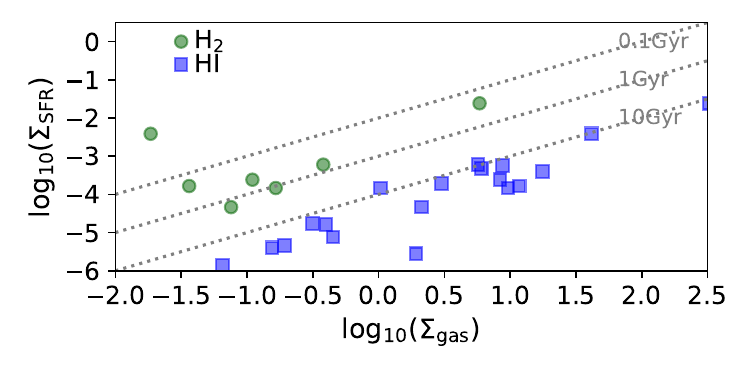}
\hspace*{-0.3cm}\includegraphics[trim=10mm 5mm 3mm 1mm, clip,width=0.35\textwidth]{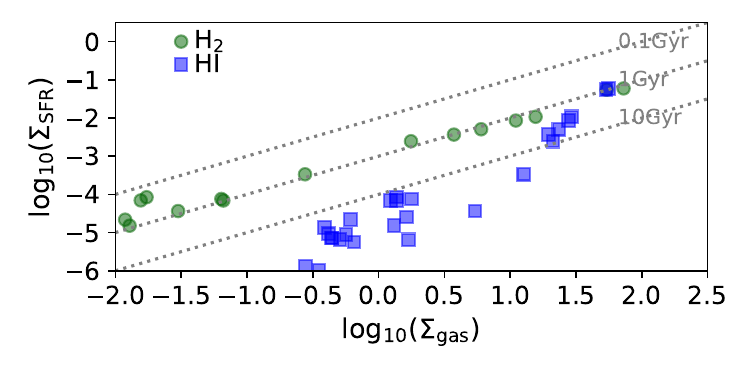}
\hspace*{-0.3cm}\includegraphics[trim=10mm 5mm 3mm 1mm, clip,width=0.35\textwidth]{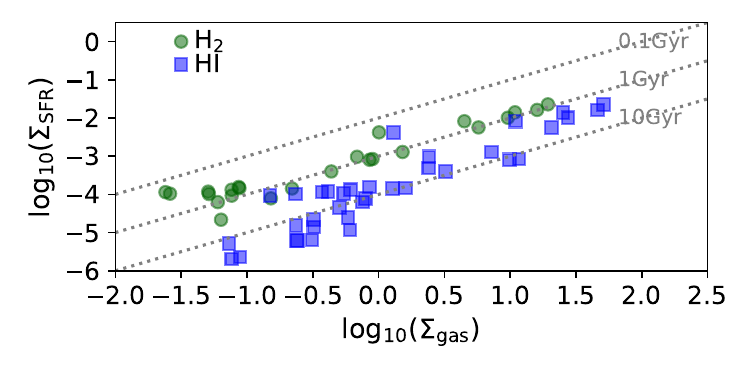}
\caption{Three example galaxies at $z=0$ in the L025m6 \colibre\ simulation used in this work. The top and middle panels show face-on and edge-on views of the gas and stellar disk, which were produced using R-package {\sc swift}. The gas is coloured by the ionisation fraction, with blue indicating neutral gas (\ion{H}{i} + H$_2$), and red ionised gas (primarily HII regions when the red compact regions are in the disk). The stellar component is shown with a gold colour, with the intensity of the colour scaling on the stellar mass density. Each figure shows at the top the stellar mass and SFR of the subhalo and the subhalo index within the SOAP catalogue. We also indicate with a white bar a physical scale for reference. 
The bottom panels show the individual \ion{H}{i} (blue squares) and H$_2$ (green circles) KS tracks of the galaxies at the top. The tracks are computed using our fiducial {\sc annuli-face} method. In the bottom panels, $\Sigma_{\rm SFR}$ and $\Sigma_{\rm gas}$ are in units of $\rm M_{\odot}\,yr^{-1}\,kpc^{-2}$ and $\rm M_{\odot}\,pc^{-2}$, respectively. The dotted lines indicate fixed depletion times of $0.1$, $1$ and $10$~Gyr as labelled. The left and middle galaxies are examples of normal star-forming galaxies, while the one on the right is an example of a strong bar galaxy that has emptied most of its inner region.} 
\label{IndividualGalsexample}
\end{center}
\end{figure*}

We implement several different ways of computing resolved maps of HI, H$_2$, SFR, stellar mass, gas metallicity, and gas velocity dispersion, which we describe and compare in Appendix~\ref{convmeth}. This is done for the purpose of testing the dependence of the KS relation on the way in which radial profiles are computed. We find very good agreement of the KS relation among different methods, and hence we limit ourselves to using the one described below.

 {\it Annuli on face-on galaxy:} We first orient galaxies to be face-on using their stellar angular momentum vector. The latter is computed using all bound stellar particles within the $50$~pkpc spherical aperture, centred on the most bound particle. We then take concentric rings of width $\Delta r$, and those are extracted out to $50$~pkpc. From the particles belonging to each ring, we measure the HI, H$_2$, stellar mass, SFR and dust content of the ring, by summing their masses and SFRs. For the total dust mass, we sum the masses in the $6$ dust species. For the metallicity, we compute the total number of oxygen atoms and divide by the total number of hydrogen atoms associated with gas particles in the ring that have a temperature $<10^{4.5}\,\rm K$ and  $>0.1\,\rm cm^{-3}$, respectively, excluding metals locked in dust. We then multiply by a factor of $12$ to obtain the metallicity in observer units of $12+\rm log_{10}(O/H)$. We also save the number of particles that are used for these calculations in each ring and include only annuli with a number of gas particles larger than $N_{\rm min}$. 
 
 Our fiducial parameters are $\Delta r=1$~proper kpc and $N_{\rm min}=10$ particles. Appendix~\ref{convmeth} shows that our results are insensitive to the precise values if they are within $\Delta r=0.5-2$~kpc and $N_{\rm min}=1-100$. 
 Note that here we do not place constraints on the vertical distance from the disk plane and simply take all the particles that are bound to the subhalo. To turn masses into surface densities, we simply divide them by the area of the ring. We refer to this method as {\sc annuli-face} in Appendix~\ref{convmeth} when comparing with other methods. 

 The resolved maps above are computed for every galaxy in the L200m6 volume with $M_{\star}>10^9\,\rm M_{\odot}$ and $\rm SFR>0$ at different snapshots, but because we place a $N_{\rm min}$ threshold for the minimum number of gas particles per ring, some rings in galaxies do not contribute to the measurement of the KS relation as they are too gas poor. 

With these resolved maps, we define the surface densities of SFR ($\Sigma_{\rm SFR}$), stellar mass ($\Sigma_{\star}$), \ion{H}{i} ($\Sigma_{\rm HI}$), H$_2$ ($\Sigma_{\rm H_2}$), total neutral gas ($\Sigma_{\rm HI+H_2}$), dust ($\Sigma_{\rm dust}$), the local gas metallicity $\rm 12 + log_{10}(O/H)$ and local specific SFR ($\rm s\Sigma_{\rm SFR}=\Sigma_{\rm SFR}/\Sigma_{\star}$). We also compute the cool gas vertical velocity dispersion profile, by selecting only the gas particles with a temperature $T<10^{4.5}\,\rm K$ in each annulus and calculating their mass-weighted velocity dispersion:

\begin{equation}
    \sigma_{\rm cool} = \sqrt{\frac{\Sigma \left(v^2_{\rm z,i} + \sigma^2_{\rm th}\right) \,m_i}{ \Sigma\, m_{i}}}, \label{eq.sigmacool}
\end{equation}

\noindent where $v_{\rm z,i}$ is the velocity component of gas particle $i$ parallel to the stellar angular momentum vector in the centre-of-mass rest frame of the galaxy, $\sigma_{\rm th}=\sqrt{k_{\rm B}\,T/\mu\, m_{\rm H}}$ is the one-dimensional thermal velocity dispersion, $m_i$ is its mass, $T$ its temperature, $k_{\rm B}$ is Boltzmann's constant, $\mu$ is the atomic weight and $m_{\rm H}$ is the hydrogen atom's mass. 
We tested including limits on the distance of the gas particles to the midplane of the galaxy disk, following the method presented in \citet{Jimenez23} for the {\sc Eagle} simulations, but we do not see a noticeable effect on our results. For simplicity, we thus do not include a limit on distance to the midplane. 

Fig.~\ref{IndividualGalsexample} shows examples of $z=0$ galaxies in the L025m6 simulation with their corresponding \ion{H}{i} and H$_2$ KS relations. Face-on and edge-on images (top and middle panels) were built with R-package {\sc swift}\footnote{The code is publicly available at \url{https://github.com/obreschkow/swift}.}.
Individual symbols in the bottom panels correspond to individual rings in the example galaxies using the {\sc annuli-face} method to measure the KS relation. In Fig.~\ref{IndividualGalsexample2} we show additional examples of star-forming and passive galaxies, with the latter still having some low-level SFR. In general, we find that in a single galaxy, the \ion{H}{i} KS relation has a lower amplitude and a similar or steeper slope than the H$_2$ KS relation. 
We will quantify these trends in \S~\ref{KSrelationLocalUSection}.  

 We measure the correlation between $\Sigma_{\rm SFR}$ and other spatially resolved properties with the deviations from the primary KS relations. The idea is to understand how the scatter of the KS relation correlates with other resolved properties and to pinpoint the strongest correlations. This has been done in the past in observations to tease out the dependence of the scatter of one resolved scaling relation on another resolved property (e.g. \citealt{Ellison20}). With this in mind, we define:

 \begin{itemize}
     \item $\Delta\,\rm SFE$: we fit power laws to the relations between $\rm \Sigma_{\rm SFR}$ and $\rm \Sigma_{\rm HI}$, $\rm \Sigma_{\rm H_2}$ or $\rm \Sigma_{\rm HI+H_2}$, using a $\chi^2$-minimisation in log-space. We refer to these fits as $\rm KS_{\rm HI}$, $\rm KS_{\rm H_2}$ and $\rm KS_{\rm HI+H_2}$. Using these fits, we then define $\Delta\rm\, SFE = \rm log_{10}(\Sigma_{\rm SFR}) - KS_x$, with $x$ being HI, H$_2$ or \ion{H}{i}$\rm +H_{\rm 2}$ (i.e. deviations from the primary relations). 
     \item $\Delta\,\rm SFR-prop$: characterises the correlations between the deviations from the main relation between $\rm log_{10}(\Sigma_{\rm SFR})$ and another property. For our purpose, the properties we study are: $\rm log_{10}(\Sigma_{\star})$, $\rm log_{10}(\Sigma_{\rm dust})$ and $\rm log_{10}(O/H)$. We also use a linear fit to characterise all these correlations. 
 \end{itemize}


Appendix~\ref{convres} presents a convergence study of the KS relation for HI, H$_2$ and \ion{H}{i}+H$_2$, using the L025m5, L025m6, L025m7 simulations. For the H$_2$ KS relation, we see convergence at higher gas surface densities only, $\gtrsim 3 \,\rm M_{\odot}\, pc^{-2}$. Comparing L025m5 and L025m6, we see that the H$_2$ KS relation converges at $\Sigma_{\rm H_2} \gtrsim 1 \,\rm M_{\odot}\,pc^{-2}$ for the m6 resolution at $z=0$. The surface density at which the H$_2$ KS relation at m6 resolution appears to be converged decreases to $0.1 \,\rm M_{\odot}\,pc^{-2}$ at $z=2$. The median \ion{H}{i} KS relation is very well converged at $z=0$ at all gas surface densities, whereas the scatter tends to increase with increasing resolution. At $z=2$, however, the \ion{H}{i} KS relation at m6 starts to deviate from the results at m5 at $\lesssim 10\,\rm M_{\odot}\,pc^{-2}$. We will use these surface density thresholds when fitting the KS relation of galaxies in the L200m6 box. We also show that selecting a minimum number of particles per annulus of $10$, $50$ or $100$ has little impact on the resulting KS relation. Unless otherwise stated, throughout this paper, we require a minimum number of gas and stellar particles per annuli of $10$.

\section{Kennicutt-Schmidt relation at $z=0$}\label{KSrelationLocalUSection}

\begin{figure*}
\begin{center}
\includegraphics[trim=20mm 0mm 30mm 10mm, clip,width=0.95\textwidth]{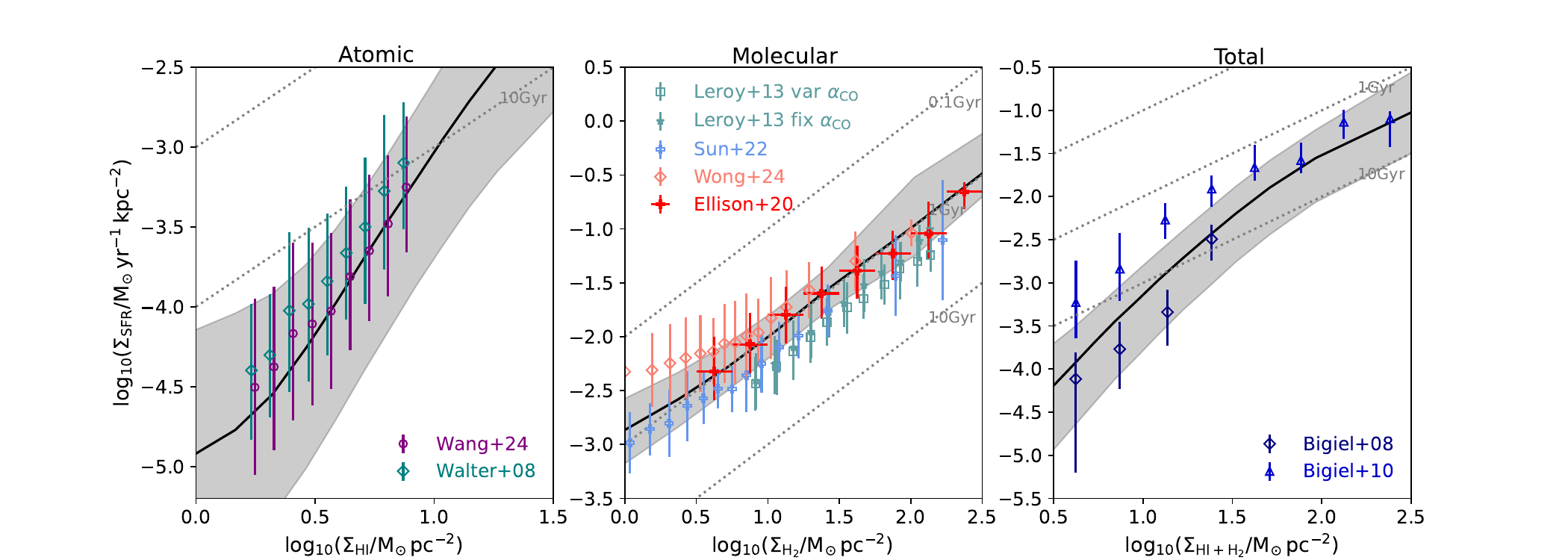}
\caption{KS relation for \ion{H}{i} (left panels), H$_2$ (middle panels) and total neutral hydrogen (\ion{H}{i}$\rm +H_2$) (right panels) for galaxies at $z=0$ {with $M_{\star}\ge 10^9\,\rm M_{\odot}$} in the \colibre\ simulation L200m6 (see Table~\ref{TableSimus}). The KS relations are calculated as azimuthal averages in rings with a width of 1~kpc. Black lines and shaded regions show the median and $16^{\rm th}-84^{\rm th}$ percentile ranges, respectively, for \colibre. The bins are chosen to have a similar number of rings per bin, and always $\ge 10$. For reference we show in each panel dotted lines of constant gas depletion time ($0.1$, $1$ and $10$~Gyr), as labelled. We show observations from \citet{Walter08,Wang24} for HI, \citet{Leroy13,Ellison20,Sun22,Wong24} for H$_2$ and \citet{Bigiel08,Bigiel10} for total neutral gas, as labelled in each panel. For \citet{Leroy13} we show two sets of values, using a constant and a variable CO-to-H$_2$ conversion factor ($\alpha_{\rm CO}$) to demonstrate the typical level of systematic uncertainty introduced by this conversion. Note that the $x$- and $y$-axes ranges are different in each panel and chosen to span the range over which observations are reported.} 
\label{KS_CompObs}
\end{center}
\end{figure*}

\begin{figure}
\begin{center}
\includegraphics[trim=2mm 4mm 3mm 5mm, clip,width=0.48\textwidth]{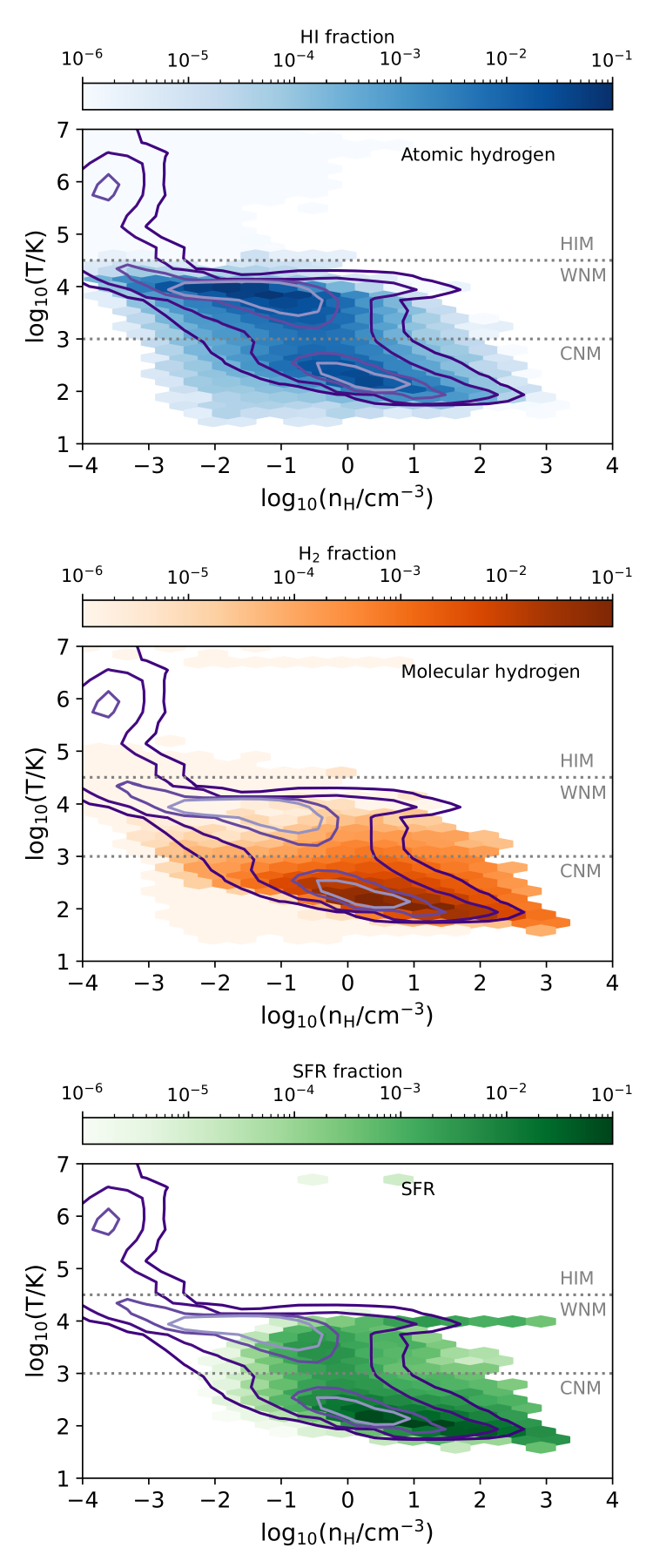}
\caption{The distribution of all gas particles within $50$~pkpc of the centre of mass of galaxies at $z=0$ with $M_{\star}\ge 10^9\,\rm M_{\odot}$ and $\rm SFR>0$ in the temperature-density plane. Contours enclose the regions where $99$\%, $95$\%, $68$\% and $50$\% of the particles are. The coloured hexbins show the \ion{H}{i} (top), H$_2$ (middle) and SFR (bottom) contribution from each bin to the total HI, H$_2$ and SFR, respectively. The totals are obtained by summing these quantities of all the galaxies. The horizontal dotted lines mark the regions associated with the hot ionised (HIM), warm neutral (WNM) and cold neutral (CNM) media, as labelled, {and as defined in Eq.~\ref{CNMdef}}.} 
\label{PhaseSpacez0}
\end{center}
\end{figure}

\begin{figure}
\begin{center}
\includegraphics[trim=2mm 3mm 3mm 1mm, clip,width=0.45\textwidth]{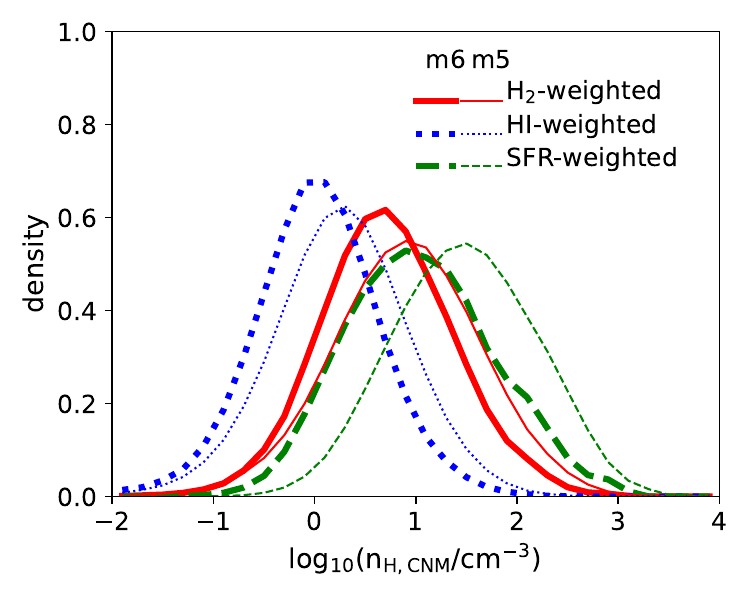}
\caption{Probability density function of the gas density of gas particles in the CNM {(as defined in Eq.~\ref{CNMdef})}, weighted by the H$_2$, \ion{H}{i} mass or the SFR, as labelled. This is shown for $z=0$ for gas particles in the L200 m6 (thick lines) and L025m5 (thin lines) simulations, that are in the CNM of galaxies at $z=0$ with $M_{\star}\ge 10^9\,\rm M_{\odot}$ and SFR$>0$.} 
\label{densities_CNM}
\end{center}
\end{figure}

In this section, we explore what \colibre\ predicts for the KS relation at $z=0$ and present a comprehensive comparison with observations. Where necessary, we subtract the contribution of helium from the observational data (typically applied as a constant multiplicative factor) to compare with our predicted surface densities of HI, H$_2$ and \ion{H}{i}+H$_2$. We also convert from a \citet{Salpeter55} or \citet{Kroupa01} IMF to our adopted \citet{Chabrier03} IMF where necessary. We do this following the conversion factors between different SFR tracers from different IMFs tabulated in Appendix~A of \citet{Gonzalez-Perez13}. For stellar masses, we divide the masses derived using an \citet{Salpeter55} and \citet{Kroupa01} IMF by $2$ and $1.14$, respectively, to convert them to a \citet{Chabrier03} IMF.

\subsection{KS relation: comparison with observations and characterisation}\label{comparisonobsandcharac}

We first aim to establish how well \colibre\ matches the local Universe observations of the KS relation. Fig.~\ref{KS_CompObs} shows the KS relation at $z=0$ predicted by \colibre, separating the dependence on HI, H$_2$ and total neutral hydrogen. Here we use the {\it annuli on face-on galaxy} method. Each galaxy contributes up to $50$ data points (i.e. rings) to each plane. However, because we set a minimum number of gas particles per ring to consider it in our analysis, we find that at $z=0$ the median number of rings contributed to the measurement of the KS relation per galaxy is only $7$. 
Observational data from \citet{Walter08}, \citet{Bigiel08}, \citet{Bigiel10}, \citet{Wang24}, and \citet{Ellison20} correspond to pixel-wise measurements of sizes $1$, $0.75$, $0.6$, $1$ and $0.3$~kpc, respectively. Data from \citet{Leroy13}, \citet{Sun22} and \citet{Wong24} correspond to annuli of width $1$~kpc, which exactly mimics our method. 
We note, however, that \citet{Bigiel08} demonstrated that for individual galaxies measurements of the KS relation in pixels vs radial annuli returned indistinguishable results. This agrees with what we show for \colibre\ in Appendix~\ref{convmeth}.

\colibre\ is in remarkable agreement with observations for all three gas phases. For HI, \colibre\ tends to prefer a lower amplitude, very close to what \citet{Wang24} reported. The difference between \citet{Walter08} and \citet{Wang24} is that the latter includes the more diffuse \ion{H}{i} which tends to be missed by interferometric surveys. 
For H$_2$, we find that the \colibre\ amplitude is closest to  the observational constraints from \citet{Ellison20} and \citet{Wong24}, which correspond to a depletion time of or slightly above $1$~Gyr. We note however, that the differences between different observations are within the typical systematic uncertainties associated with the CO-to-H$_2$ conversion factor ($\alpha_{\rm CO}$). 
For the total neutral hydrogen, we find that \colibre's predictions are in between the observational estimates at surface densities $\lesssim 10^{1.5}\,\rm M_{\odot}\,pc^{-2}$, while at higher densities, there is agreement with \citet{Bigiel10}, which is the only data set available at those surface densities.

The agreement with observations was not guaranteed. In \colibre, the only free parameter in the star formation model is the fraction of gas mass converted into stars per free-fall time, provided the gas satisfies the instability criterion of Eq.~\ref{alphacrit}. The adopted value is taken from observations of local universe molecular clouds (e.g. \citealt{Schinnerer24}) rather than to ensure agreement with the observed KS relation, and it does not know how much \ion{H}{i} or H$_2$ a gas particle has. The latter can then lead to large variations in the KS law with gas metallicity for example, which we return to in \S~\ref{resolveddeps}.

The predictions of \colibre\ presented here agree with those in \citet{Nobels24} at $\rm log_{10}(\Sigma_{\rm H_2}/M_{\odot}\,pc^{-2})\gtrsim 0.5$. At lower values, 
the simulation results in \citet{Nobels24} drop towards longer depletion times. \colibre\ instead tends to continue on the constant depletion time line or even flatten a bit towards shorter depletion times. This difference is not necessarily surprising as there are many components to the physical model of \colibre\ that were not present in \citet{Nobels24}, including the non-equilibrium chemistry component, live dust and chemical evolution model and the overall cosmological context, which is lacking in \citet{Nobels24}. Regarding the non-equilibrium chemistry, by applying {\sc Chimes} to the FIRE simulations, \citet{Richings22} show that explicitly solving for the chemistry of the gas leads to varying \ion{H}{i} and H$_2$ mass fractions at low densities $(n_{\rm H}\lesssim 1 \,\rm cm^{-3})$, which will then change the relation between SFR and \ion{H}{i} or H$_2$ in that regime. In the future, we will quantify the effect the different subgrid physics modules and parameters have on the KS relation and its scatter.

To understand what gives rise to the different relations between $\Sigma_{\rm SFR}$ and the \ion{H}{i} and H$_2$ gas content, it is informative to study where these gas phases are located in the phase-space diagram. We show this in Fig.~\ref{PhaseSpacez0} for galaxies with stellar masses $>10^9\,\rm M_{\odot}$ and $\rm SFR>0$. From the \ion{H}{i} distribution we see the bimodality that arises from \ion{H}{i} being primarily in the {WNM and CNM (see Eq.~\ref{CNMdef})} of the ISM. Most of the \ion{H}{i} mass is however contributed by the WNM ($\approx 64$\%). For H$_2$, we see that very little is in the WNM, with most of it being in the CNM ($\approx 99$\%). Note, however, that most of the gas mass in the CNM is atomic ($\approx 70$\%), while only $\approx 30$\% is molecular. 
For the SFR, we see that $\approx 7$\% is associated with the WNM and $93$\% with the CNM. 

Even though most of the CNM consists of atomic rather than molecular gas, the correlation of the SFR is stronger with H$_2$ than \ion{H}{i} or total neutral gas (as seen by the smaller scatter of the H$_2$). The reason for this is that even within the CNM, the densities traced by H$_2$ and SFR are much higher than those traced by the HI. This is shown in Fig.~\ref{densities_CNM} for the L200m6 and L025m5 simulations. Even in comparison to H$_2$, the SFR is concentrated in higher density gas regardless of the resolution. 
Thus, the larger fraction of SFR associated with the high-density tail of the CNM, where the H$_2$ resides is the cause of the stronger correlation between $\Sigma_{\rm SFR}$ and $\Sigma_{\rm H_2}$ than between  $\Sigma_{\rm SFR}$ and $\Sigma_{\rm HI}$ or $\Sigma_{\rm HI + H_2}$ .
Comparing the m5 and m6 curves, we see that the SFR-weighted density distribution moves to higher values as the resolution increases. This is because the criterion for gravitational instability is evaluated at the resolution limit. The \ion{H}{i}- and H$_2$-weighted distributions also move to higher densities, but to a lesser extent than the SFR-weighted gas density distribution. 

Fig.~\ref{PhaseSpacez0} shows a population of particles with $\rm SFR>0$ and $T\approx 10^4\,\rm K$, which is neither associated with significant H$_2$ nor HI. These particles correspond to those that were evaluated to have non-zero star formation rate at the beginning of the timestep, but received energy from stellar feedback at the end of the timestep. 
This is a rare occurrence and these particles contribute only $\approx 0.7$\% to the total instantaneous SFRs of galaxies in our sample. In any case, the star formation model in \colibre\ is such that these particles would not turn into stellar particles because they would be classified as non star-forming at the beginning of the next timestep.  

\subsection{KS relation: characterising the scatter}

\subsubsection{Dependence on resolved galaxy properties}\label{resolveddeps}

\begin{figure*}
\begin{center}
\includegraphics[trim=20mm 3mm 30mm 10mm, clip,width=0.95\textwidth]{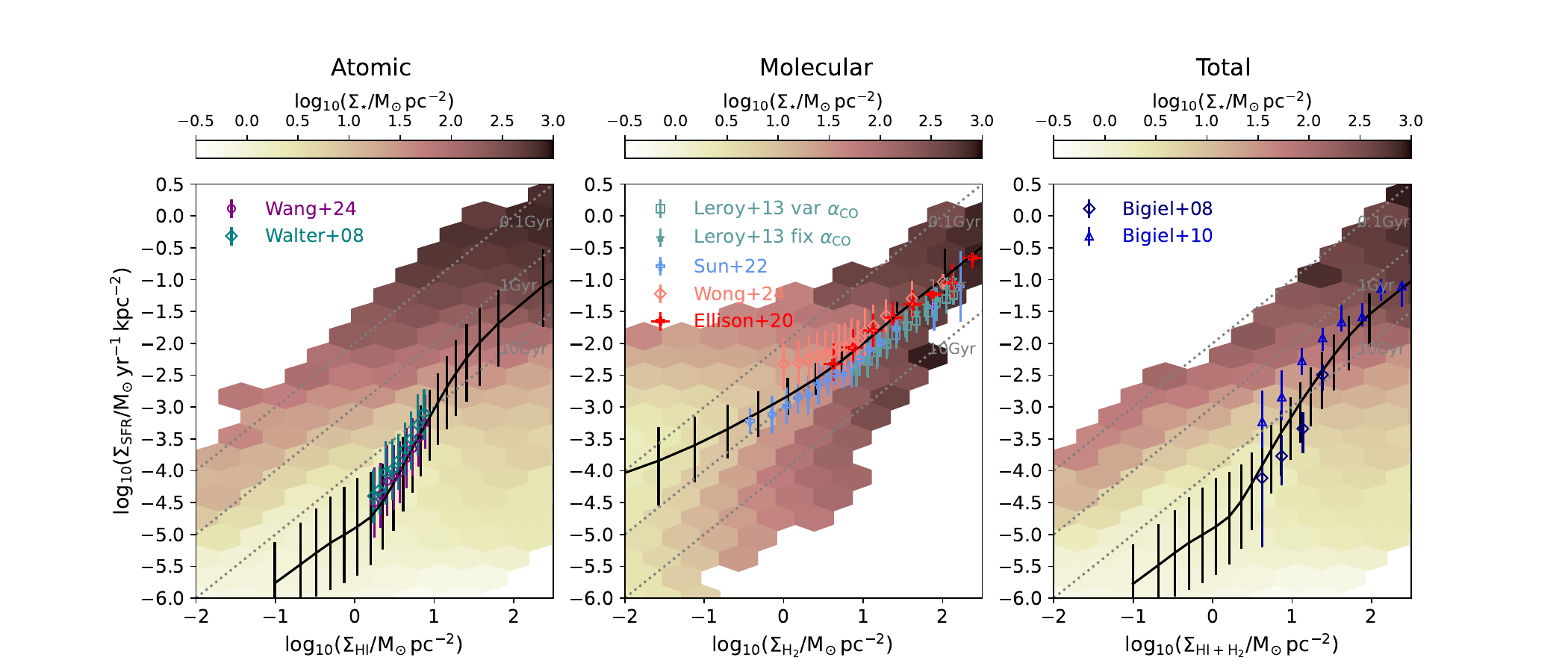}
\includegraphics[trim=20mm 3mm 30mm 20mm, clip,width=0.95\textwidth]{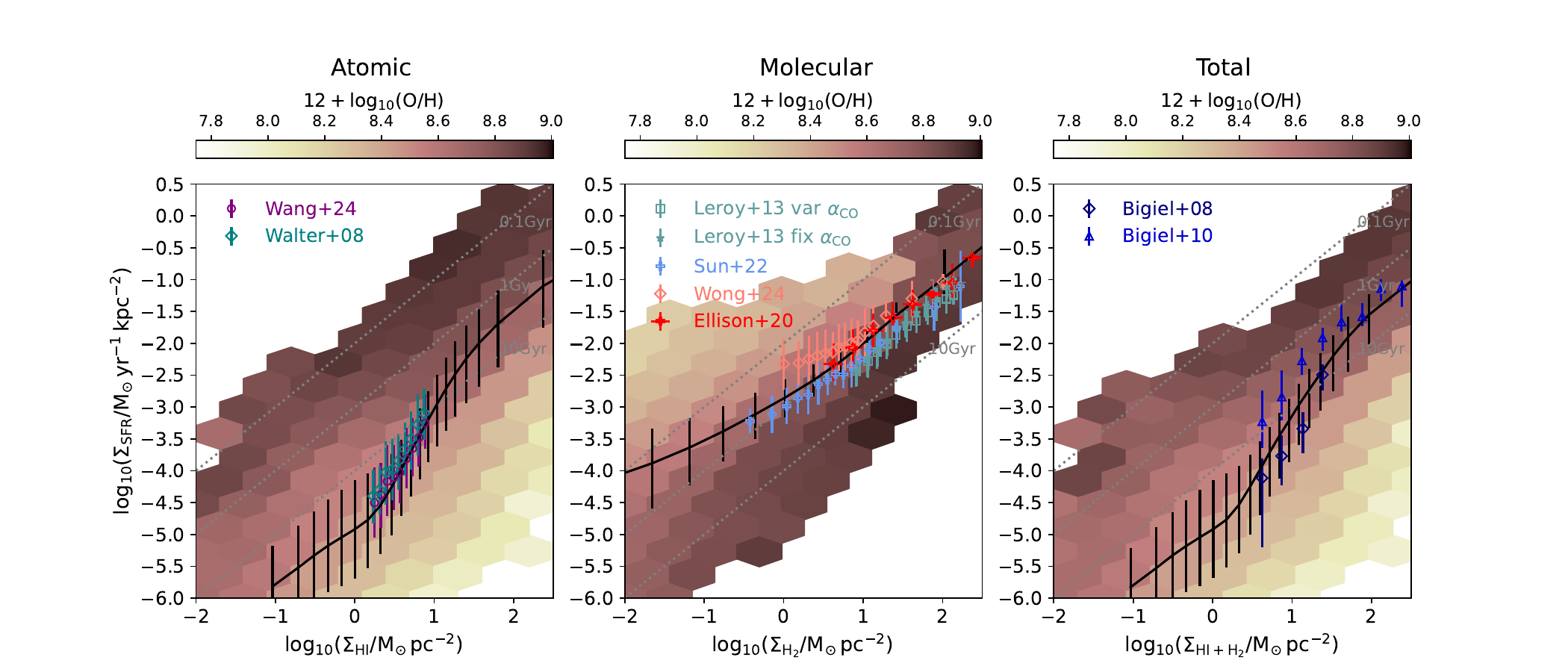}
\includegraphics[trim=20mm 3mm 30mm 20mm, clip,width=0.95\textwidth]{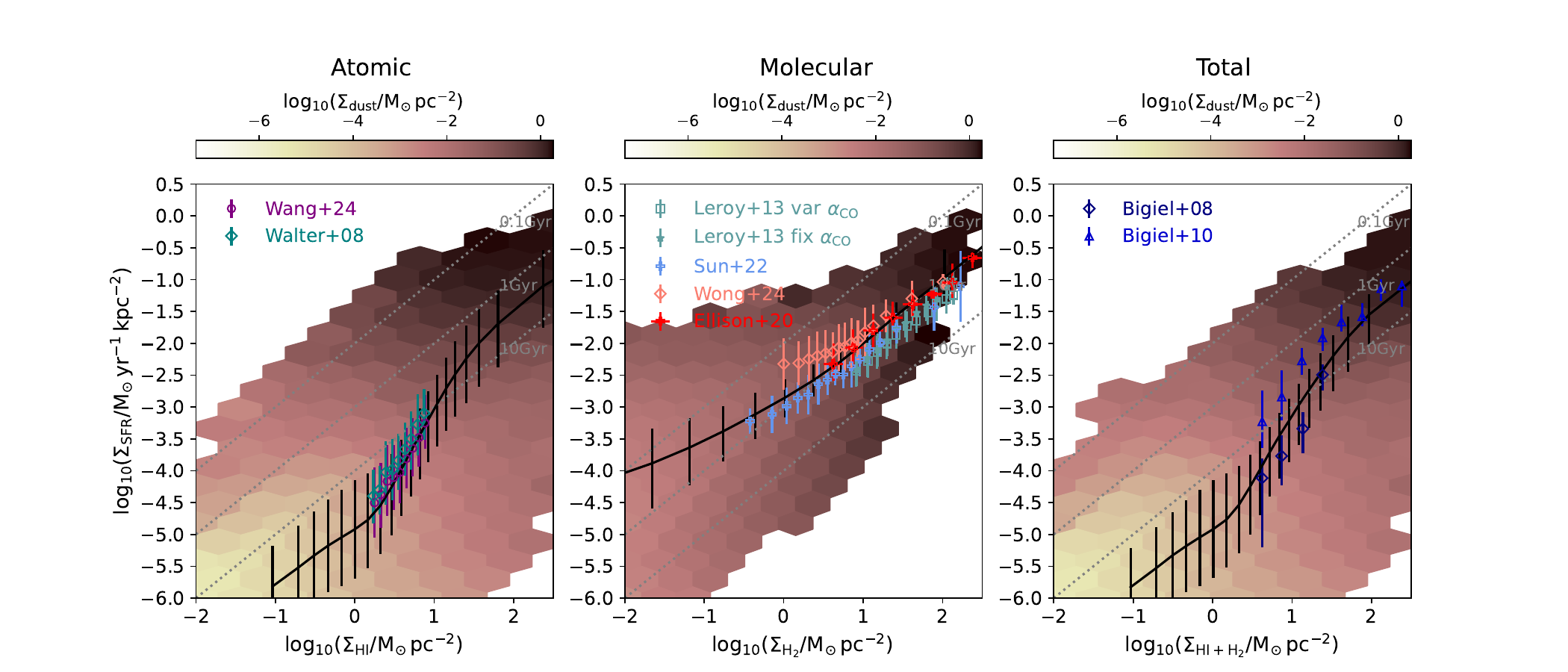}
\caption{As Fig.~\ref{KS_CompObs} but with the coloured bins showing the median stellar surface density ($\Sigma_{\star}$; top panels), gas metallicity ($\rm log_{10}(O/H)$; middle panels) and surface density of dust ($\Sigma_{\rm dust}$, lower panels). Black lines with error bars in each panel show the median and $16^{\rm th}-84^{\rm th}$ percentile ranges, respectively. Only hexbins with $\ge 5$ objects are included.} 
\label{KS_ResolvedDeps}
\end{center}
\end{figure*}

\begin{figure}
\begin{center}
\includegraphics[trim=4mm 3mm 5mm 3mm, clip,width=0.45\textwidth]{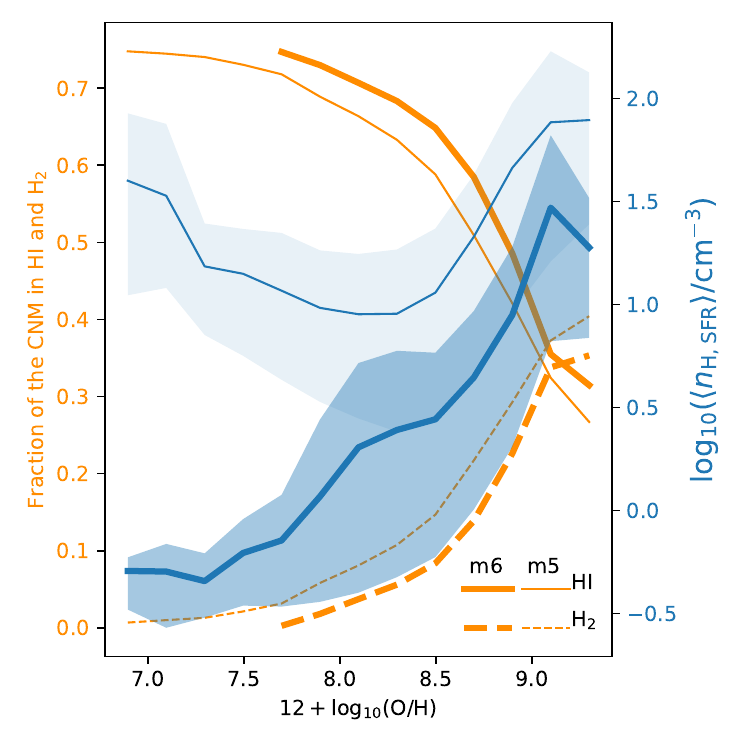}
\caption{The fraction of the CNM in the form of \ion{H}{i} (orange solid line) and H$_2$ (orange dotted line), and the SFR-weighted gas density (blue line with shaded region), as a function of gas metallicity. This is constructed with all the gas particles that are within $50$~pkpc from the centres of galaxies with $M_{\star}\ge 10^9\,\rm M_{\odot}$ and $\rm SFR>0$ at $z=0$ in the L200m6  (thick lines) and L025m5 (thin lines) simulations. The \ion{H}{i} fraction saturates at a CNM fraction of $\approx 0.75$, as the rest of the gas is in the form of helium and metals. For the gas density, we show the median and the $16^{\rm th}-84^{\rm th}$ percentile ranges.} 
\label{FracCNMz0}
\end{center}
\end{figure}

\begin{figure}
\begin{center}
\includegraphics[trim=3mm 2mm 2mm 3mm, clip,width=0.48\textwidth]{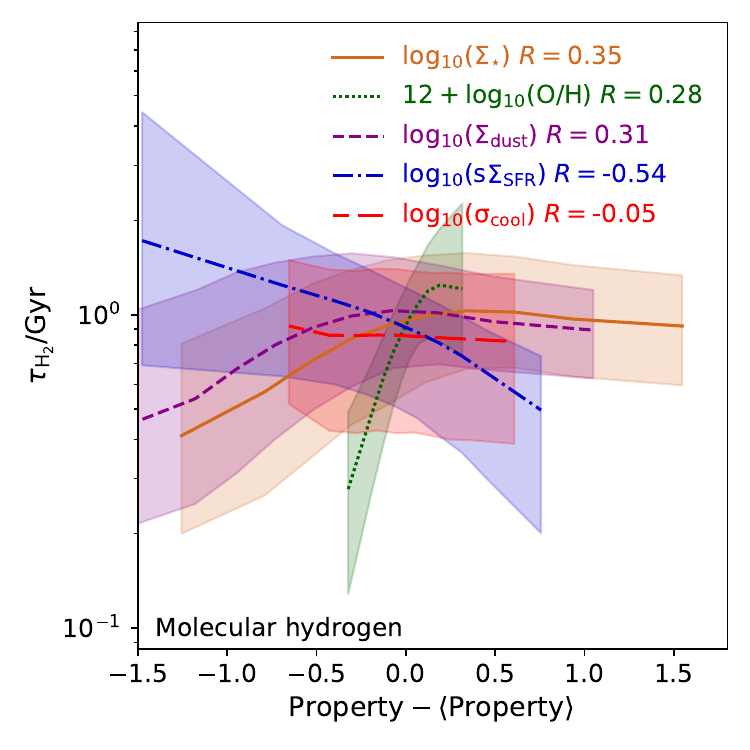}
\includegraphics[trim=0mm 5mm 4mm 3mm, clip,width=0.48\textwidth]{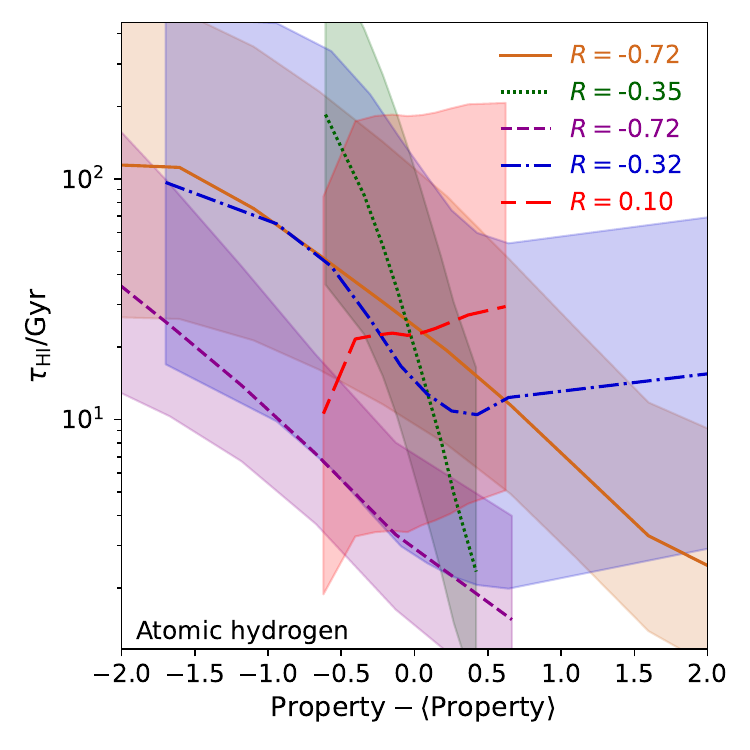}
\caption{The correlation between the H$_2$ (top) and \ion{H}{i} (bottom) depletion times, defined as $\Sigma_{\rm gas}/\Sigma_{\rm SFR}$, with gas being either H$_2$ or HI, and the stellar surface density, cold gas metallicity, surface density of dust, local sSFR and cool gas velocity dispersion (as defined in Eq.~\ref{eq.sigmacool}), as labelled. 
Lines and shaded regions show medians and $16^{\rm th}-84^{\rm th}$ percentile ranges, respectively, {for galaxies at $z=0$ with $M_{\star}\ge 10^9\,\rm M_{\odot}$.}
To plot all the properties on the same x-axis scale, we subtract from every property its median. We also show the Spearman correlation coefficient, $R$, of each correlation, as labelled.} 
\label{Delta_correlations}
\end{center}
\end{figure}

\begin{figure}
\begin{center}
\includegraphics[trim=0mm 0mm 7mm 13mm, clip,width=0.49\textwidth]{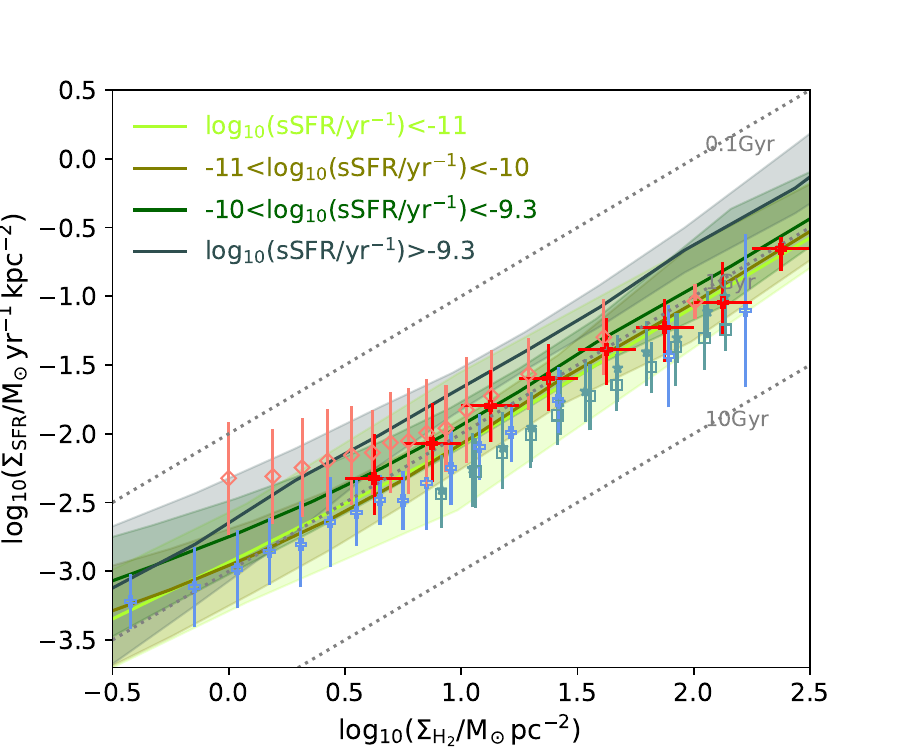}
\includegraphics[trim=0mm 0mm 7mm 13mm, clip,width=0.49\textwidth]{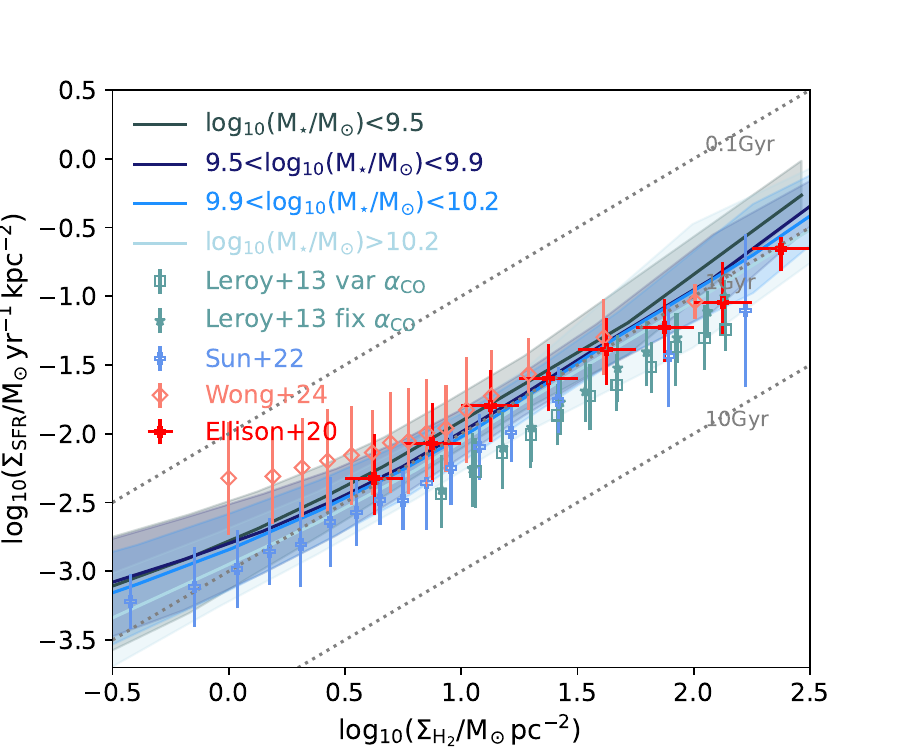}
\caption{The H$_2$ KS relation at $z=0$ for galaxies selected based on their sSFR (top panels) and stellar mass (bottom panel). The median and $16^{\rm th}-84^{\rm th}$ percentiles are shown with solid lines and shaded regions, respectively.
All galaxies included here have $M_{\star}\ge 10^9\,\rm M_{\odot}$ and $\rm SFR>0$.}
\label{KS_GlobalDeps}
\end{center}
\end{figure}
The KS relation predicted by \colibre\ and in the observations has a significant scatter. In this section we explore how the scatter is correlated with other resolved properties of galaxies, which were calculated using the particles tagged to belong to the same annuli in the {\sc annuli-face} method. 

Fig.~\ref{KS_ResolvedDeps} shows how the scatter in the KS relation at $z=0$ for HI, H$_2$ and total neutral hydrogen depends on the stellar surface density (top), gas metallicity (middle) and dust surface density (bottom). Overall, we find that in all gas phases, the scatter around the KS relation tends to decrease with increasing gas density, even for HI. However, the decrease is stronger for H$_2$ and total neutral gas. 

For $\Sigma_{\star}$ (top panels in Fig.~\ref{KS_ResolvedDeps}), we find that as $\Sigma_{\star}$ decreases, the local H$_2$ depletion time, computed as $\Sigma_{\rm H_2}/\Sigma_{\rm SFR}$ decreases, which is clear at $\Sigma_{\rm H_2}\lesssim10\,\rm M_{\odot}\,pc^{-2}$. At higher H$_2$ surface densities, there is little dependence of the H$_2$ KS relation's scatter on $\Sigma_{\star}$
For HI, lower $\Sigma_{\star}$ are associated with lower HI-to-SFR conversion efficiencies at fixed $\Sigma_{\rm HI}$, opposite to the trend displayed in the H$_2$ case.  
The trend between the KS relation scatter and $\Sigma_{\star}$ at fixed gas surface density appears stronger for \ion{H}{i} than H$_2$. We will come back to this point later. The total neutral gas appears similar to what is observed for the \ion{H}{i} KS relation. This is not surprising as at $z=0$ the overall gas content of galaxies tends to be dominated by \ion{H}{i} rather than H$_2$.

For the gas metallicity (i.e. metals not locked in dust; middle panels in Fig.~\ref{KS_ResolvedDeps}), we see again opposite trends in \ion{H}{i} and H$_2$, the lower metallicity gas tends to be associated with a higher (lower) conversion efficiency between H$_2$ (\ion{H}{i}) and stars. 
The dust surface density trends (lower panels in Fig.~\ref{KS_ResolvedDeps}) resemble those of the gas metallicity: lower $\Sigma_{\rm dust}$ is associated with more (less) efficient conversion from H$_2$ (\ion{H}{i}) into stars. 

The physical drivers of the relations between the scatter of the KS relation and the gas metallicity and dust content are the same. As the gas becomes more metal and dust poor, it needs to reach higher densities for it to be cold and hence likely to form stars. In the absence of dust (or at least if it is scarce), the cold, dense gas will struggle to form H$_2$. 
Hence, as the gas becomes more metal and dust-poor, less H$_2$ and more \ion{H}{i} are associated with the same amount of SFR, leading to the appearance of a higher SFR conversion efficiency for H$_2$, and a lower one for HI. 
%
%
\citet{Orr18} found a similar tendency for low-metallicity gas to be associated with longer depletion timescales for the total \ion{H}{i}+H$_2$ KS relation in the FIRE simulations. They did not however explore the effect of gas metallicity on the \ion{H}{i} and H$_2$ KS relation separately, partially because their model does not explicitly follow the formation of hydrogen species. 

We show how the gas metallicity influences the \ion{H}{i} and H$_2$ composition of the CNM in \colibre\ in Fig.~\ref{FracCNMz0}. 
The \ion{H}{i} mass fraction in the CNM saturates at $\approx 0.75$ at low metallicity, as the rest of the gas is in helium and metals. The figure shows that the dependence of the \ion{H}{i} fraction in the CNM on gas metallicity is qualitatively the same in the L200m6 and the L025m5 simulations, showing that the result is robust against resolution. We also show the typical gas density at which stars form in \colibre\ for the m6 and m5 resolution. At m6 resolution, the density consistently decreases with decreasing metallicity, even dropping below $n_{\rm H}\approx 1 \,\rm cm^{-3}$ for the very metal-poor gas. This is, however, driven by resolution, as for the m5 run we see that star formation consistently happens at densities $>10 \,\rm cm^{-3}$, even when \ion{H}{i} overwhelmingly dominates the mass of CNM. We remind the reader that this difference between resolutions is expected and intentional, as the instability criterion for star formation is evaluated at the resolution limit. 
At the Milky-Way metallicity of $12+\rm log_{10}(O/H)\approx 8.75$, \colibre\ predicts the fraction of \ion{H}{i} in the CNM to be $\approx 0.5$, which is similar to the value observed in the Milky-Way \citep{McClure-Griffiths23}. This is indicative only, as we are not selecting galaxies to be of Milky-Way mass in Fig.~\ref{FracCNMz0}.
At lower metallicities, \citet{Park25} find that there is more \ion{H}{i} in the CNM, which they interpret as the gas remaining \ion{H}{i} as it gets colder due to the difficulty in forming H$_2$ in the absence of dust. This is in qualitative agreement with what Fig.~\ref{FracCNMz0} shows for \colibre. 

Fig.~\ref{Delta_correlations} explores the strength of the correlations between the gas depletion time, measured in a spatially resolved manner, $\tau_{\rm gas}\equiv \Sigma_{\rm gas}/\Sigma_{\rm SFR}$, and the resolved properties in galaxies (including those shown in Fig.~\ref{KS_ResolvedDeps}).
To show all the correlations together, we simply subtract from every property its corresponding median for the entire sample of Fig.~\ref{KS_ResolvedDeps}. For example, for the stellar surface density, the x-axis shows $\rm log_{10}(\Sigma_{\star})-\langle log_{10}(\Sigma_{\star}\rangle$). 

For $\tau_{\rm H_2}$, we find that the strongest correlation is the negative correlation with specific $\Sigma_{\rm SFR}$ (s$\Sigma_{\rm SFR}$; which is related to the correlation with $\Sigma_{\star}$), and slightly weaker positive correlations appears with $\Sigma_{\rm dust}$ and local gas metallicity. 
%
The strong anti-correlation between $\tau_{\rm H_2}$ and resolved sSFR means that the high sSFR is not just due to a higher concentration of cold gas but also that the cold gas is being converted into stars more efficiently. 
In the case of the gas metallicity and dust surface density, the presence of a moderately strong correlation with $\tau_{\rm H_2}$ is not unexpected, as the accumulation of metals and dust directly affects the efficiency of gas cooling and H$_2$ formation, but the fact that the correlations are positive is more surprising. In \colibre, dust acts both as a catalyst for the formation of H$_2$ and helps shield the gas. In this case, more of the hydrogen gas is in the form of H$_2$ and thus associated with a given SFR, giving the appearance of a longer H$_2$ depletion time.
We test these predictions against observations in \S~\ref{CompObsSec}.
%
%

In the case of $\tau_{\rm HI}$ (bottom panel in Fig.~\ref{Delta_correlations}), we find that the strongest correlations are with the dust and stellar surface densities (both with a Spearman correlation coefficient $\approx -0.75$). The latter dependence has also been reported in observations; e.g. \citet{Wang24} find $\tau_{\rm HI}\propto \Sigma^{-0.85 \pm 0.09}_{\star} \Sigma_{\rm SFR}/\Sigma_{\rm HI}$. In \colibre, we find $\tau_{\rm HI}\propto \Sigma^{-0.72}_{\star}$, in relatively agreement with these observations (see \S~\ref{CompObsSec} for a more detailed comparison with \citealt{Wang24}). 
The strong anti-correlation between $\tau_{\rm HI}$ and $\Sigma_{\rm dust}$ is expected. The more dust there is, the easier it is for the hydrogen to transition from atomic to molecular, resulting in a relative lower fraction of star formation associated with atomic hydrogen. This leads to an apparent higher efficiency of \ion{H}{i} to SFR conversion. 

Interestingly, the cool gas ($T<10^{4.5}\,\rm K$) one-dimensional velocity dispersion (Eq.~\ref{eq.sigmacool}) is only weakly correlated with $\tau_{\rm HI}$, while the correlation with $\tau_{\rm H_2}$ is almost absent. 
\citet{Schaye25} show that in \colibre, 
stars tend to form from cold gas with supersonic turbulent velocities. Our results show that, even though at the resolution limit the velocity dispersion (especially the turbulent component) play an important role in the star formation model, at kpc scales, star formation appears to be poorly correlated with gas velocity dispersion. 
We also tested different definitions of the gas velocity dispersion (weighting by H$_2$ or \ion{H}{i} mass, and using different selections in distance from the midplane ranging from 3 to 10~kpc), and found that the weak correlation holds for different definitions. 

The correlation with $\sigma_{\rm cool}$ manifests in a way that a higher velocity dispersion is associated with a longer \ion{H}{i} depletion time. The star formation model in \colibre\ indirectly depends on the local gas velocity dispersion, since a higher value increases the Toomre instability parameter, making the gas less likely to collapse and be available for star formation (see \citealt{Nobels24} for a detailed discussion). The weaker correlation with $\sigma_{\rm cool}$ compared to some of the other local properties is in part due to $\sigma_{\rm cool}$ displaying less variations within and in-between galaxies, compared with other local properties. This is clear from the dynamic range covered by $\sigma_{\rm cool}$ which only extends over $1$~dex, while $\Sigma_{\star}$, $\Sigma_{\rm dust}$ and $s\Sigma_{\rm SFR}$ tend to display variations of $\gtrsim 1.5$~dex within our sample. Note, however, that the gas metallicity also varies over a small range compared to some of the other properties, while still displaying stronger correlations with both $\tau_{\rm H_2}$ and $\tau_{\rm HI}$ than what is obtained for $\sigma_{\rm cool}$. 

The total neutral hydrogen (\ion{H}{i}+H$_2$) behaves very much like $\tau_{\rm HI}$, with the correlations only becoming a bit weaker than what is reported in the bottom panel of Fig.~\ref{Delta_correlations}. This simply means that \ion{H}{i} dominates the neutral gas reservoir in most galaxies at $z=0$ in \colibre. 

In Appendix~\ref{convergence}, we test the convergence of the correlations in Fig.~\ref{Delta_correlations} with numerical resolution. We find that all correlations are very well converged between the m6 and m5 resolution, with the exception being the correlation between $\tau_{\rm H_2}$ and local gas metallicity. Although the correlation is similarly strong in the L025m5 compared to the L025m6 simulation (see Table~\ref{tab:corrfactors}), there is an offset in normalisation. This is not surprising given that the stellar mass-gas metallicity relation is not well converged in \colibre\ at $z=0$ (see discussion in \citealt{Schaye25}).

\subsubsection{Dependence on global galaxy properties}\label{globalprops}

Fig.~\ref{KS_GlobalDeps} presents the molecular gas KS relation at $z=0$, separating the galaxy population by their global sSFR (top) and stellar mass (bottom). We show these properties because they are the ones that exhibited the strongest correlations with the scatter of the KS relation. In addition to the trends in Fig.~\ref{KS_GlobalDeps}, we also explored a possible dependence of the KS relation scatter on global gas metallicity, stellar age, galaxy type (whether the galaxies are defined as central or satellites) and galaxy morphology, and found only weak or no correlations.

The top panel of Fig.~\ref{KS_GlobalDeps} shows a strong dependence of the molecular KS relation's normalisation on the global sSFR of the galaxy, in a way that galaxies with higher sSFR have shorter molecular gas depletion times. 
\colibre\ galaxies can have high sSFRs from a combination of increased gas reservoirs and a more efficient conversion of cold gas into stars. We quantify this effect by measuring the Spearman correlation coefficient $R$, between the resolved molecular depletion time and the global sSFR, at fixed global molecular and stellar masses. We find  $R\approx -0.6$ with negligible p-values, indicating strong anti-correlations. The dependence across bins of stellar and molecular gas masses is similar and takes the shape $\tau_{\rm H_2}\propto \rm sSFR^{\beta}$, with $\beta\approx [-0.8,-0.9]$. Thus, \colibre\ predicts that changes in the global sSFR are accompanied by changes in the local conversion efficiency of molecular gas to stars.

The bottom panel of Fig.~\ref{KS_GlobalDeps} shows a weak dependence on stellar mass, so that lower stellar mass galaxies have slightly elevated $\Sigma_{\rm SFR}$ at fixed $\Sigma_{\rm H_2}$, with the differences being more clearly seen at $\Sigma_{\rm H_2}\gtrsim 10 \,\rm M_{\odot}\,pc^{-2}$. 
We find that the dependence on stellar mass in the bottom panel of Fig.~\ref{KS_GlobalDeps} is driven by the sSFR dependence, as the sSFR has a negative dependence on stellar mass for main sequence galaxies (see Fig.~16 in \citealt{Schaye25}).

\subsubsection{Comparison with observations}\label{CompObsSec}

\begin{figure}
\begin{center}
\includegraphics[trim=3mm 2mm 1mm 1mm, clip,width=0.49\textwidth]{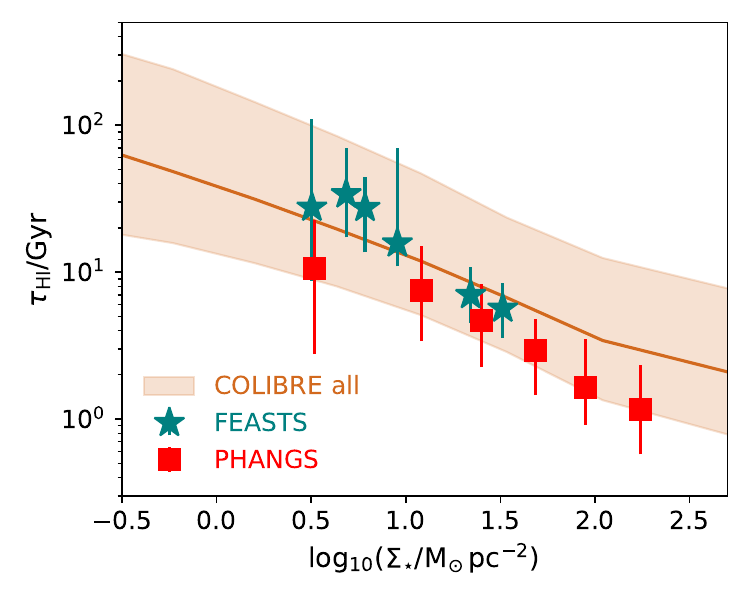}
\caption{The dependence of the (resolved) \ion{H}{i} depletion time, $\tau_{\rm HI}$ on the local stellar surface density, $\Sigma_{\star}$, for \colibre\ galaxies at $z=0$ with $M_{\star} \ge 10^9\,\rm M_{\odot}$ compared with the FEASTS \citep{Wang24} and PHANGS \citep{Sun22} surveys. The lines and symbols show the median of \colibre\ and the observations, respectively, while the shaded region and error bars show the $16^{\rm th}-84^{\rm th}$ percentile ranges. 
This threshold matches well the minimum $\Sigma_{\rm HI}$ value in FEASTS and PHANGS.} 
\label{KS_TauHIWang}
\end{center}
\end{figure}

\begin{figure}
\begin{center}
\includegraphics[trim=0mm 12mm 11mm 12mm, clip,width=0.49\textwidth]{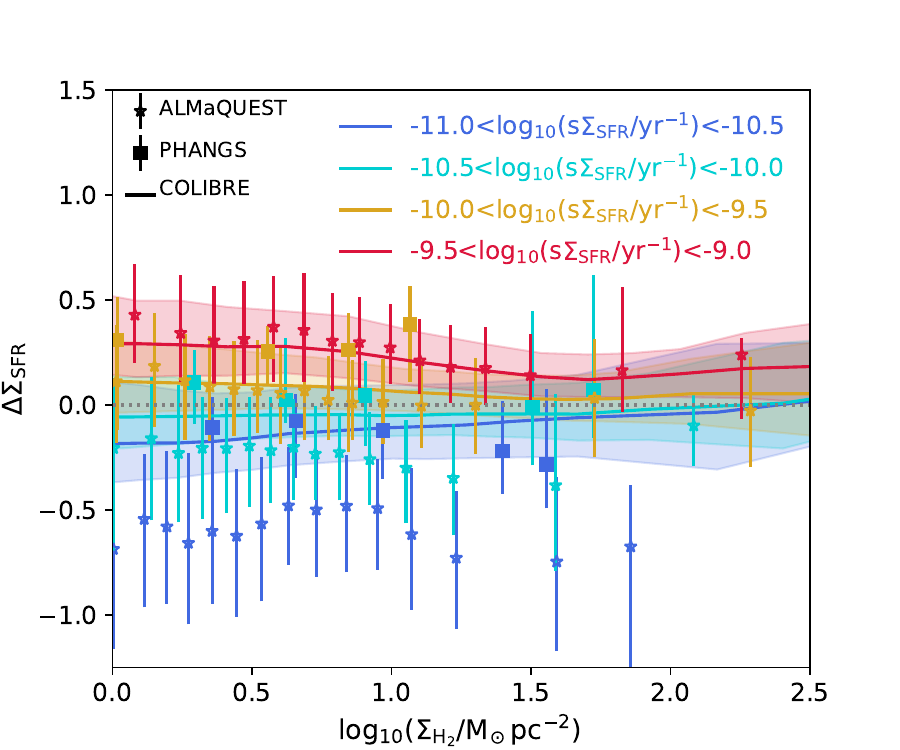}
\includegraphics[trim=0mm 2mm 11mm 6mm, clip,width=0.49\textwidth]{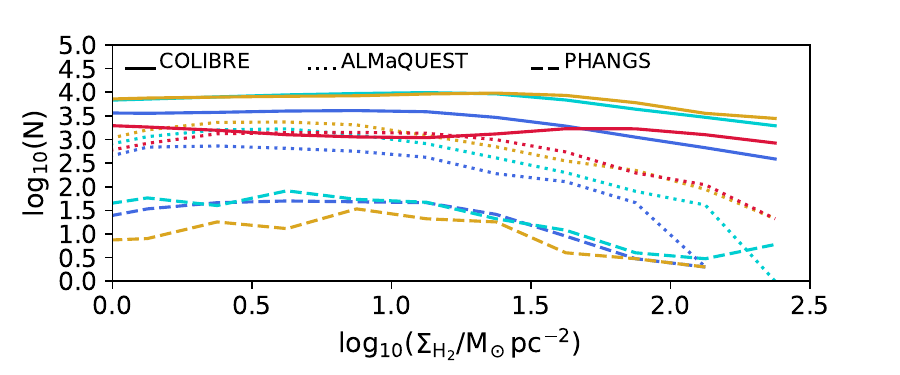}
\caption{{\it Top panel:} Deviations from the median molecular gas KS relation ($\rm log_{10}(\Sigma_{\rm SFR}) - \langle log_{10}(\Sigma_{\rm SFR})\rangle$) at $z=0$ as a function of H$_2$ surface density, separating regions by their local sSFR ($\rm s\Sigma_{\rm SFR}\equiv \Sigma_{\rm SFR}/\Sigma_{\star}$, for observations (symbols) and \colibre\ (lines). Symbols and lines show medians, while error bars and shaded regions show the $16^{\rm th}-84^{\rm th}$ percentile ranges. For ALMaQUEST \citep{Lin19}, we include all galaxies and spaxels; for PHANGS \citep{Sun22}, we include all galaxies and annuli; while for \colibre\ we include $z=0$ galaxies with $M_{\star}>10^{9.7}\,\rm M_{\odot}$ and $\rm SFR>0$. Note that no PHANGS data are presented for the highest $\rm s\Sigma_{\rm SFR}$ bin, as there were not enough statistics to produce reliable measurements for the medians and scatter.
The agreement between \colibre\ and the observations is remarkable. At low values of sSFR, \colibre\ agrees better with PHANGS. 
%
%
{\it Bottom panel:} histogram of $\Sigma_{\rm H_2}$ for each of the samples of the top panel. The observational and simulated samples follow similar distributions, with most regions having $\rm log_{10}(\Sigma_{\rm H_2}/M_{\odot}\,pc^{-1}) \lesssim 1.4$. This is also the regime in which the differences between sSFR samples are clearest in both observations and simulations.} 
\label{KS_ResolvedDepsALMAQUEST}
\end{center}
\end{figure}

\begin{figure}
\begin{center}
\includegraphics[trim=0mm 12mm 11mm 12mm, clip,width=0.49\textwidth]{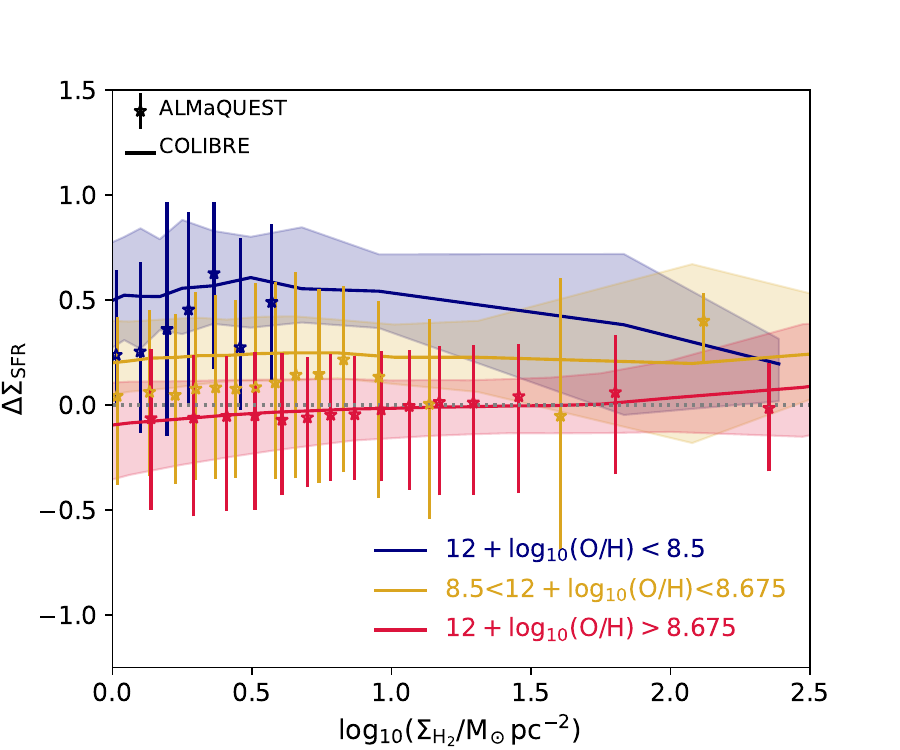}
\includegraphics[trim=0mm 2mm 11mm 6mm, clip,width=0.49\textwidth]{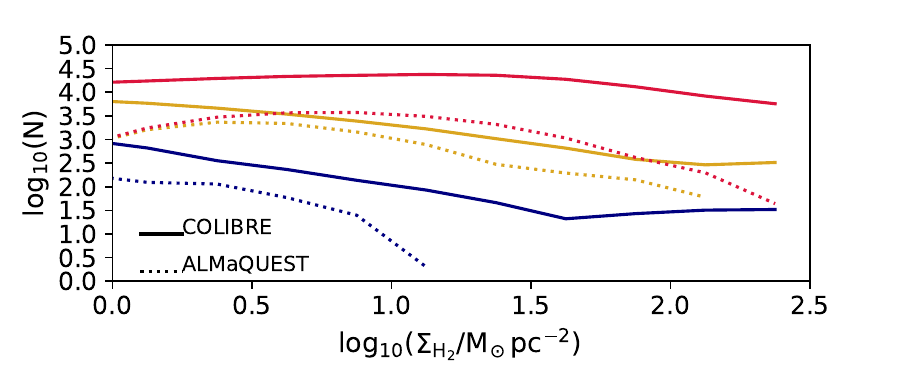}
\caption{As Fig.~\ref{KS_ResolvedDepsALMAQUEST} but for regions binned by the local gas metallicity, $12+\rm log_{10}(O/H)$, and comparing only with ALMaQUEST. 
The sampling of the metallicity bins is similar in both observations and simulations, with a similar distribution of $\rm log_{10}(\Sigma_{\rm H_2}/M_{\odot}\,pc^{-1})$ in each metallicity bin.} 
\label{KS_ResolvedDepsALMAQUEST2}
\end{center}
\end{figure}

\begin{figure}
\begin{center}
\includegraphics[trim=0mm 0mm 11mm 12mm, clip,width=0.49\textwidth]{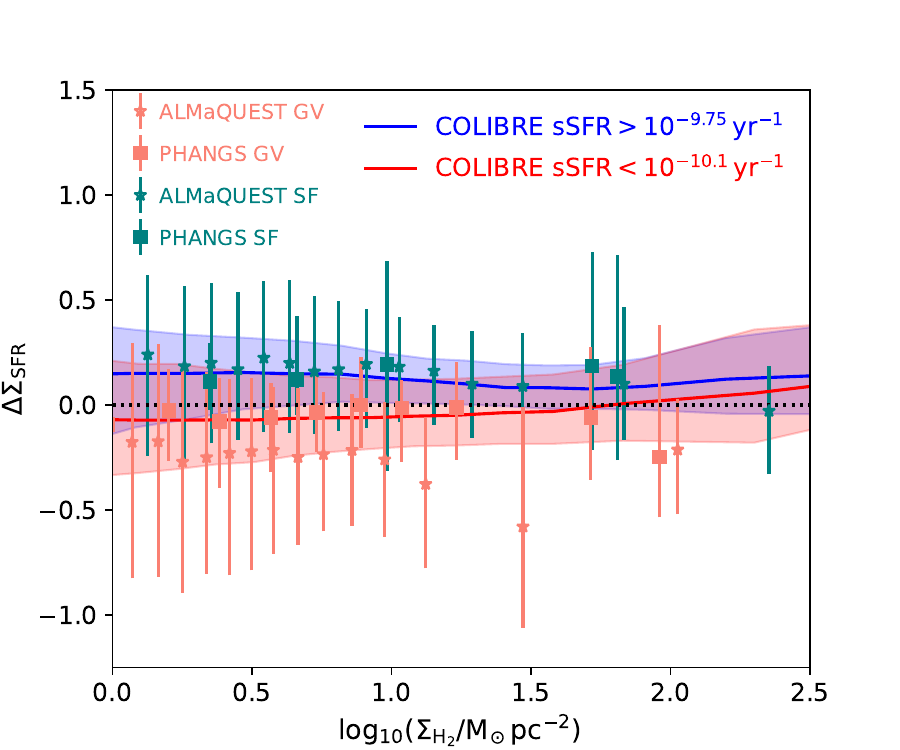}
\caption{As the top panel of Fig.~\ref{KS_ResolvedDepsALMAQUEST}, but for star-forming and green-valley galaxies separately, in \colibre\ (lines) and observations (symbols), as labelled. This is shown for galaxies at $z=0$. Symbols and lines show medians, while error bars and shaded regions show the $16^{\rm th}-84^{\rm th}$ percentile ranges. To select the two samples in \colibre, we apply thresholds in sSFR, as labelled, which give median sSFR similar to the median sSFR in ALMaQUEST and PHANGS for the two samples. 
 The agreement between \colibre\ and the observations is very good, particularly when comparing with PHANGS.}
\label{KS_GlobalDepsALMAQUEST}
\end{center}
\end{figure}

In this section, we present a series of comparisons between the predicted dependence of the scatter in the KS relation on galaxy properties, starting with the atomic gas and then moving to the molecular gas. Because most observational constraints are dedicated to the molecular KS relation, this section is devoted primarily to that.

\citet{Shi11} showed that there is a strong dependence of the scatter in the atomic gas KS relation on the stellar surface density, suggesting that it could be fundamentally driven by the dynamical equilibrium pressure. We compare the predicted relation between the local \ion{H}{i} depletion time ($\equiv \Sigma_{\rm HI}/\Sigma_{\rm SFR}$) and the local stellar surface density with observations in Fig.~\ref{KS_TauHIWang}. We show observations from the FEASTS \citep{Wang24} and PHANGS \citep{Sun22} surveys. FEASTS comprises  $17$ spiral galaxies with stellar masses  $10^{9.7}-10^{11}\,\rm M_{\odot}$, while PHANGS contains $70$ nearby galaxies with stellar masses between $10^{9}-10^{11}\,\rm M_{\odot}$. In both observational sets, the relation was computed using radial annuli with size $\approx 1\,\rm kpc$, similar to our method. 
The agreement between \colibre\ and the observations is excellent, even when we did not attempt to match the exact distribution of galaxy properties of the two surveys and limited to sampling a similar stellar mass range. This relation is very well converged with resolution, as shown in Fig.~\ref{KS_Resolution3}, so the agreement extends to the other \colibre\ simulations. 

As shown in Fig.~\ref{Delta_correlations}, $\Sigma_{\star}$ is one of the most significant drivers of the scatter in the \ion{H}{i} KS relation in \colibre, and the same appears to be the case in the observations. 
It is important to note that FEASTS appears to be slightly elevated compared to PHANGS, while \colibre\ lies in between. This most likely comes from the diffuse \ion{H}{i} FEASTS is sensitive to (due to the data coming from the single dish Five-hundred-meter Aperture Spherical Telescope), but that PHANGS misses (as it primarily employs the interferometric Very Large Array). According to \citet{Wang24}, diffuse \ion{H}{i} can account for up to $40$\% of the \ion{H}{i} mass, which directly translates into increasing $\tau_{\rm HI}$ in a direct proportion. In \colibre, the local dynamical equilibrium pressure does not enter the star formation model, implying that the $\tau_{\rm HI}-\Sigma_{\star}$ correlation does not necessarily directly emerge from that parameter. 

The scatter of the molecular gas KS relation in galaxies in the local Universe has been well studied with the ALMaQUEST \citep{Lin19}, EDGE-CALIFA \citep{Wong24} and PHANGS \citep{Sun22} surveys. Here, we compare our predictions in detail with ALMaQUEST and PHANGS. The ALMaQUEST sample consists of $46$ galaxies with stellar masses $10^{9.7}-10^{11.5}\,\rm M_{\odot}$, with $25$ of those being on the main sequence and $21$ being below the main sequence (referred to as ``Green Valley'' galaxies). Via a combination of MaNGA and ALMA CO(2-1) observations, they characterise the SFR, molecular gas, stellar surface density and gas metallicity of individual spaxels in galaxies, with spaxels having sizes of $0.3$~kpc. The PHANGS sample with H$_2$ profile measurements consists of $61$ galaxies (stellar mass range is as indicated above), and here we use the data provided in annuli of $1$~kpc width from \citet{Sun22}. Using the same sSFR criterion as \citet{Lin19}, the PHANGS catalogue consists of $19$ main sequence and $45$ Green Valley galaxies, thus PHANGS samples lower sSFR galaxies than ALMaQUEST on average. Another important difference between PHANGS and ALMaQUEST is that the SFRs of the regions in PHANGS are characterised using a combination of H$\alpha$, FUV, NUV and MIR, while in ALMaQUEST they come only from H$\alpha$. 
Although PHANGS extends to lower stellar masses compared with ALMaQUEST, most galaxies ($54/61$) have masses $\gtrsim 10^{9.8}\,\rm M_{\odot}$.  
Thus, to compare with observations, we select galaxies in \colibre\ at $z=0$ that have stellar masses $>10^{9.7}\,\rm M_{\odot}$ and $\rm SFR > 0$. We choose a slightly lower stellar mass limit than the minimum mass of the ALMaQUEST sample to allow for potential systematic uncertainties in the derivation of stellar masses of the order of those inferred at $z=0$ \citep{Bellstedt24}.
To perform the comparison with observations, we define $\rm \Delta\, \Sigma_{\rm SFR} \equiv log_{10}(\Sigma_{\rm SFR}) - \langle log_{10}(\Sigma_{\rm SFR}(\Sigma_{\rm H_2})\rangle$, where $\Sigma_{\rm SFR}$ corresponds to the value of an annulus (or spaxel in the case of ALMaQUEST) and $ \rm \langle log_{10}(\Sigma_{\rm SFR}(\Sigma_{\rm H_2})\rangle$ corresponds to the median of all annuli (or spaxels) within a bin of $\Sigma_{\rm H_2}$. We calculate this quantity for both simulation and observations, which helps us to focus on how the scatter of the relation correlates with properties of interest. 

We start by comparing the resolved dependence of the scatter of the molecular KS relation on the sSFR in Fig.~\ref{KS_ResolvedDepsALMAQUEST}. 
Here, we take all $z=0$ galaxies in \colibre\ with stellar masses  $>10^{9.7}\,\rm M_{\odot}$, and simply bin annuli by their local sSFR $\equiv \Sigma_{\rm SFR}/\Sigma_{\star}$. \colibre\ predicts a very strong dependence of the scatter in the molecular KS relation on the local sSFR that agrees remarkably well with the observed dependence in ALMaQUEST at $\rm sSFR > 10^{-10}\rm \, yr^{-1}$, which better samples the high $\rm s\Sigma_{\rm SFR}$ region. At lower sSFR values, we see that ALMaQUEST displays a lower normalisation of the KS relation than predicted by \colibre, while there is excellent agreement with PHANGS. This may be related to ALMaQUEST using H$\alpha$ to derive SFRs, which is known to systematically underestimate SFRs compared to UV/IR tracers at low SFRs \citep{Lee09}. PHANGS on the other hand, by using a combination of tracers, circumvents this limitation. 

We remind the reader that beyond the local sSFR, the scatter of the molecular KS relation is correlated with the local gas metallicity (Fig.~\ref{Delta_correlations}). 
We test this prediction with ALMaQUEST only, as the data for the metallicity profiles for PHANGS were not available (see \citealt{Sun22} for details). 
Fig.~\ref{KS_ResolvedDepsALMAQUEST2} shows $\Delta\,\rm \Sigma_{SFR}$ as a function of $\Sigma_{\rm H_2}$ for samples selected by local gas metallicity. 
To build Fig.~\ref{KS_ResolvedDepsALMAQUEST2},  we removed all spaxels in the ALMaQUEST sample that have too low signal-to-noise ratio ($\rm S/N<3$) to measure a gas metallicity, which removes $17$\% of the spaxels. The removed spaxels tend to be those with long $\rm H_2$ depletion times. Thus, the resulting sample tends to be biased towards short depletion times and high $\Sigma_{\rm H_2}$ (see the bottom panel of Fig.~\ref{KS_ResolvedDepsALMAQUEST2}) compared to what is seen in the bottom panel of Fig.~\ref{KS_ResolvedDepsALMAQUEST}.
Despite these differences, we see a similar trend in \colibre\ and ALMaQUEST: $\Delta\,\rm \Sigma_{\rm SFR}$ decreases as the gas metallicity increases. This is particularly clear in the $0.25\lesssim \rm log_{10}(\Sigma_{\rm H_2}/M_{\odot}\,pc^{-2})\lesssim 0.75$ range that is best sampled by the observations. 

We stress that from the ALMaQUEST sample alone, it is not clear whether gas metallicity is more weakly or strongly correlated with $\tau_{\rm H_2}$ than other properties are.
This could in part be due to the large fraction of spaxels we need to remove from the sample due to the low signal-to-noise ratio, but also to the limited stellar mass range of the sample, which is skewed towards massive galaxies. The mass-metallicity relation is steep at $M_{\star}\lesssim 10^{10}\,\rm M_{\odot}$ but flattens at higher stellar masses (e.g. \citealt{Tremonti04}). Hence, the overall metallicity range sampled by ALMaQUEST is rather small, with $90$\% of the spaxels having $\rm 8.57<12+\rm log_{10}(O/H)<8.8$. Future surveys sampling a larger dynamic range in gas metallicity would be needed to test the relative importance of different properties in setting the scatter of the molecular KS relation.

For the comparison above, we used all spaxels in ALMaQUEST and all annuli in  PHANGS regardless of whether galaxies were classified as star-forming or green valley, and only investigated the dependence of the scatter in the molecular KS relation on resolved properties. \citet{Lin22} argue that global galaxy properties also impact the resolved KS relation, and find that green-valley galaxies in ALMaQUEST have a lower global molecular-to-star formation conversion efficiency than star-forming galaxies. In Fig.~\ref{KS_GlobalDepsALMAQUEST}, we compare with those results by splitting our \colibre\ sample into star-forming and green valley galaxies, in such a way that the median sSFR of each sample resembles that of ALMaQUEST and PHANGS. We find that a selection $\rm sSFR>10^{-9.75}\,\rm yr^{-1}$ in \colibre\ provides a median $\rm sSFR\approx 10^{-9.64}\,\rm yr^{-1}$ while the median for the star-forming samples in ALMaQUEST and PHANGS is $\rm sSFR\approx 10^{-9.65}\,\rm yr^{-1}$ and $\approx 10^{-9.7}\,\rm yr^{-1}$, respectively. We select green-valley galaxies in \colibre\ to have $\rm sSFR<10^{-10.1}\,\rm yr^{-1}$, which produces a sample with a median $\rm sSFR\approx 10^{-10.65}\, yr^{-1}$, while the green-valley ALMaQUEST and PHANGS samples have medians of $\rm sSFR\approx 10^{-10.57}\, yr^{-1}$ and $\approx 10^{-10.45}\, \rm yr^{-1}$, respectively. We remind the reader that in \colibre, annuli are selected to have $\rm SFR > 0$. We find excellent agreement between the molecular KS relation in \colibre\ with both sets of observations for the SF galaxies, while for green-valley galaxies, the agreement is better with PHANGS. 
The latter resembles the findings of Fig.~\ref{KS_ResolvedDepsALMAQUEST}, where the H$\alpha$-derived SFRs at low sSFR in ALMaQUEST appear to underestimate the SFR values compared to the multi-wavelength approach of PHANGS.


The comparisons presented in  Figs.~\ref{KS_CompObs},~\ref{KS_ResolvedDepsALMAQUEST},~\ref{KS_ResolvedDepsALMAQUEST2}\, and \ref{KS_GlobalDepsALMAQUEST} show that \colibre\ predicts a resolved KS relation at $z=0$ in different gas phases in excellent agreement with the observations. This is not only the case for the average relation, but also for the dependence of the KS relation's scatter on local and global galaxy properties. This is a remarkable success, as the KS relation was not used to calibrate the free parameters in \colibre.

\subsection{KS relation: galaxy-to-galaxy variations}

\begin{figure}
\begin{center}
\includegraphics[trim=0mm 15mm 10mm 30mm, clip,width=0.49\textwidth]{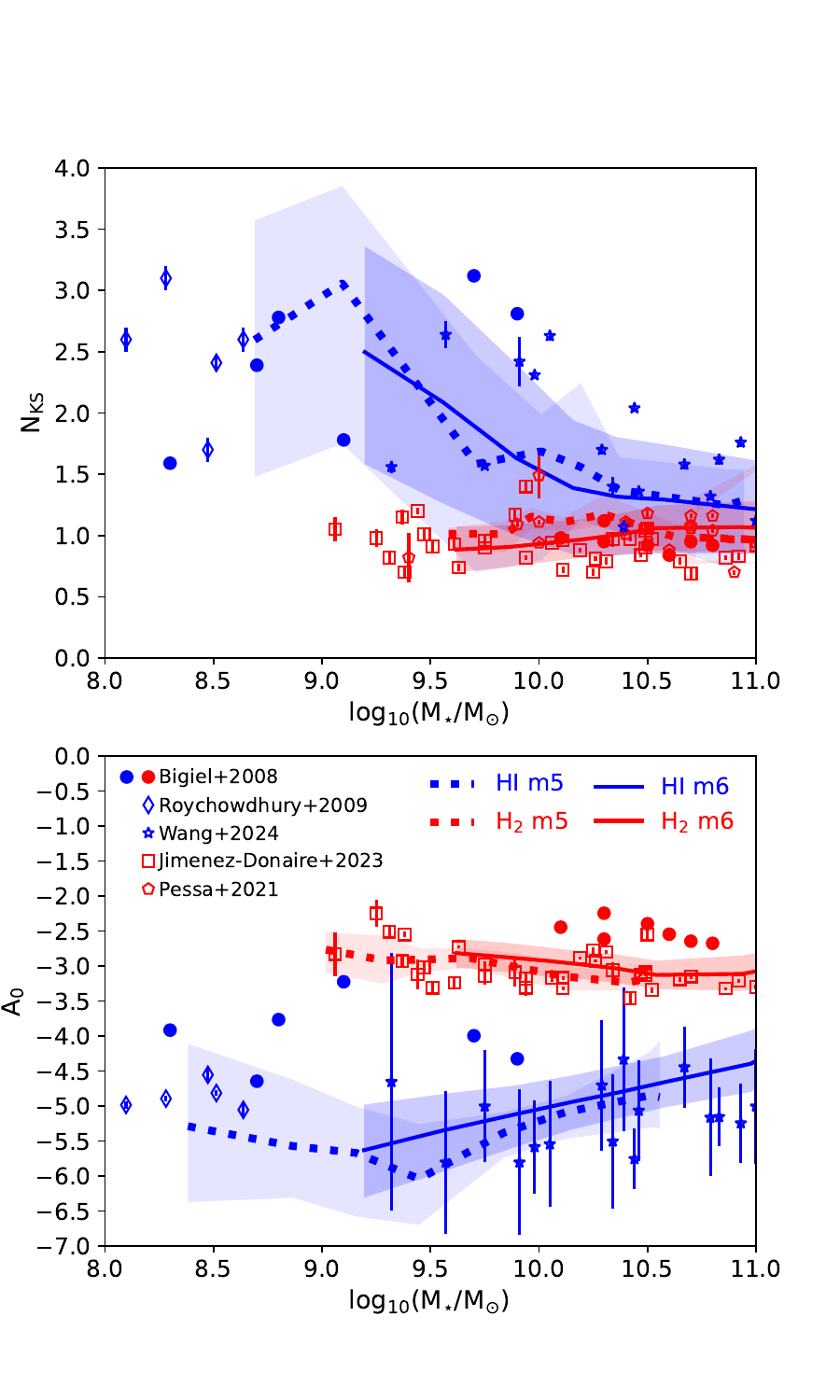}
\caption{The slope (top) and zero-point (bottom) of the KS relation fits for individual $z=0$ galaxies as a function of stellar mass in \colibre. We show results for the L200m6 (solid lines) and L025m5 (dotted lines) simulations, as labelled in the bottom panel. The equation being fitted is Eq.~\ref{KSfit}. We show the slope and zero-point of the fits for the case of fitting only \ion{H}{i} (blue) and H$_2$ (red). Lines with shaded regions show the median and the $16^{\rm th}-84^{\rm th}$ percentile ranges, respectively. Observational estimates from \citet{Bigiel08, Roychowdhury09,Pessa21,Jimenez-Donaire23,Wang24} are shown as symbols, as labelled in the bottom panel, using the same colour-code of \colibre.} 
\label{KS_IndividualTracks}
\end{center}
\end{figure}

We have explored the median KS relation and how the scatter correlates with resolved and global galaxy properties. In this section, we explore how much variation there is in the KS relation from galaxy to galaxy at $z=0$. To quantify this, we fit the relationship between $\Sigma_{\rm SFR}$ and $\Sigma_{\rm gas}$, with gas here being HI, H$_2$ or the sum of the two, using the function:

\begin{equation}
    \Sigma_{\rm SFR} = A_{\rm 0}\,\Sigma^{N_{\rm KS}}_{\rm gas}\label{KSfit}
\end{equation}

\noindent In this equation, $\Sigma_{\rm SFR}$ and $\Sigma_{\rm gas}$ take the units $\rm M_{\odot}\,yr^{-1}\,kpc^{-2}$ and $\rm M_{\odot}\,pc^{-2}$, respectively. The fits for \colibre\ are performed using only gas with a surface density $\ge 0.1\,\rm M_{\odot}\,pc^{-2}$. This is done to avoid the flattening of the KS relation at lower surface densities due to resolution limitations (see Fig.~\ref{KS_Resolution}) and to focus on regions that are directly comparable with observations. We do not see large variations in the individual fits as long as they are performed with a minimum surface density in the range  $0.1-1\,\rm M_{\odot}\,pc^{-2}$. 
In addition, we only fit galaxies with a KS relation sampled with $\ge 5$ rings. This means that we do not fit some of the low SFR galaxies in \colibre. The example galaxy shown at the bottom-right panel of Fig.~\ref{IndividualGalsexample2} has a well-defined fit for its \ion{H}{i} KS relation, but not for the H$_2$ KS relation. We find that the requirement above removes $37$\% of the galaxies with $M_{\star}\ge 10^9\,\rm M_{\odot}$ and $\rm SFR > 0$ at $z=0$. The removed galaxies tend to have lower stellar masses and sSFRs (with medians $\approx 10^{9.48}\,\rm M_{\odot}$ and $\approx 10^{-10.17}\,\rm yr^{-1}$, respectively) than the ones that are kept for the analysis (with medians $\approx 10^{9.73}\,\rm M_{\odot}$ and $\approx 10^{-10.02}\,\rm yr^{-1}$, respectively).
We tested different minimum numbers of bins between $3-7$ and found our results to be insensitive to that change.

Fig.~\ref{KS_IndividualTracks} shows the correlations of $N_{\rm KS}$ and $A_{0}$ with stellar mass. For \ion{H}{i} (blue), we find that the median slope of the entire population is $N_{\rm KS}\approx 2.15$, but the value of individual galaxies strongly anti-correlates with the stellar mass of the galaxy. Low-mass galaxies tend to have higher values of $N_{\rm KS}$, which is also seen for fits to the total gas (not shown here). The scatter around the median $N_{\rm KS}$ tends to increase with decreasing stellar mass for fits based on \ion{H}{i} or total gas. For HI, the zero point of the KS fits is positively correlated with stellar mass. For the molecular gas, we see that the slope and the zero point are close to constant, with values $\approx 1$ and $\approx -2.8$, respectively (in the units used in Eq.~\ref{KSfit}). We also find virtually no difference between the different \colibre\ resolutions.

The number of fits to the observed KS relation of individual galaxies is limited to a few dozens, even in the local Universe. We show in Fig.~\ref{KS_IndividualTracks} observational results from different surveys, which agree on a close to constant KS molecular gas relation. We highlight that the observations of \citet{Jimenez-Donaire23} correspond to Virgo cluster galaxies, with some of them being gas-deficient and clearly affected by their environment. Nevertheless, they follow a very similar relation. It is notable that the observations suggest a similar scatter in the slope and zero point of the molecular KS relation, as predicted by \colibre. The overall trends seen, including a potential weak dependence on stellar mass, also agree well with \colibre's predictions. 

\begin{figure*}
\begin{center}
\includegraphics[trim=25mm 0mm 35mm 12mm, clip,width=0.99\textwidth]{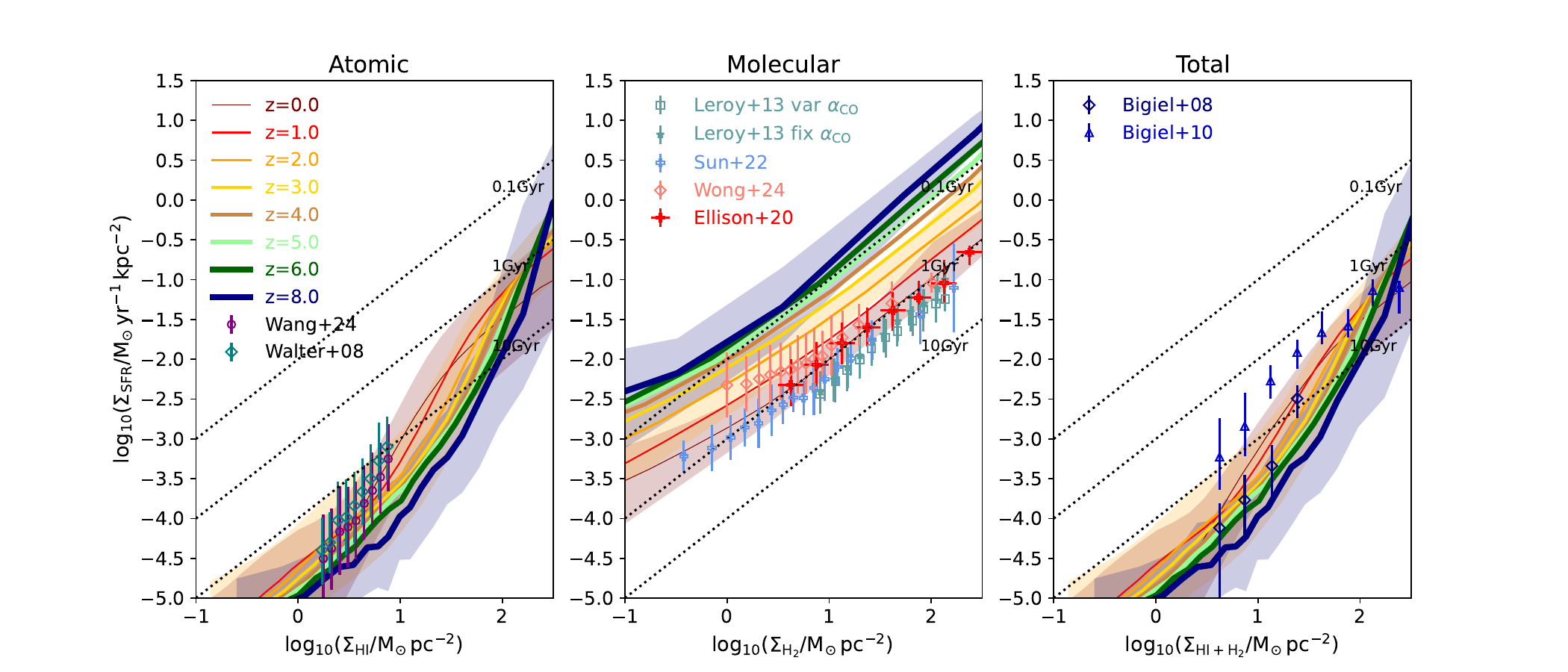}
\caption{As Fig.~\ref{KS_CompObs} but showing the evolution of the median KS relation for galaxies with $M_{\star}\ge 10^9\,\rm M_{\odot}$ and $\rm SFR>0$ from $z=8$ to $z=0$. We show the $16^{\rm th}-84^{\rm th}$ percentiles only for $z=0$, $2$ and $8$ for clarity. 
Redshifts are as labelled in the left panel. The observational data shown in all panels corresponds to $z\approx 0$.}
\label{KS_Zevo}
\end{center}
\end{figure*}
For HI, we again see very good agreement between the m5 and m6 resolutions. We note that the difference in $A_0$ for the \ion{H}{i} KS relation between the L200m6 and L025m5 runs at around $M_{\star}\approx 10^{9.5}\,\rm M_{\odot}$ is driven by the different box sizes, as the differences become negligible when we compare the L025m5 simulation with the L025m6 simulation instead of L200m6 (not shown here).
%
It is informative to show the L025m5 run along with the L200m6 run 
as the observational fits to the \ion{H}{i} KS relation are biased towards dwarf galaxies, which are \ion{H}{i} dominated, and better resolved in the m5 run. To compare \colibre\ with observations, we compiled data from \citet{Bigiel08,Roychowdhury09,Wang24}. \citet{Roychowdhury09} did not report stellar masses for their sample of dwarf galaxies, and hence we computed our own stellar masses from IRAC 3.6$\mu m$ photometry. We collected the IRAC 3.6$\mu m$ photometry from the NASA/IPAC Extragalactic Database for the \citet{Roychowdhury09} sample, and adopted a mass-to-light ratio corresponding to a \citet{Chabrier03} IMF following \citet{Leroy08}. 
Although the scatter is large, the observations are consistent with a dependence of the slope and zero point of the \ion{H}{i} KS relation on stellar mass that is similar to what \colibre\ predicts. Because the predicted scatter is so large, much larger observational samples would be required to validate or refute our predictions. 

\section{Redshift evolution of the Kennicutt-Schmidt relation}\label{sec:zevo}

\begin{figure}
\begin{center}
\includegraphics[trim=4mm 4mm 3mm 1mm, clip,width=0.47\textwidth]{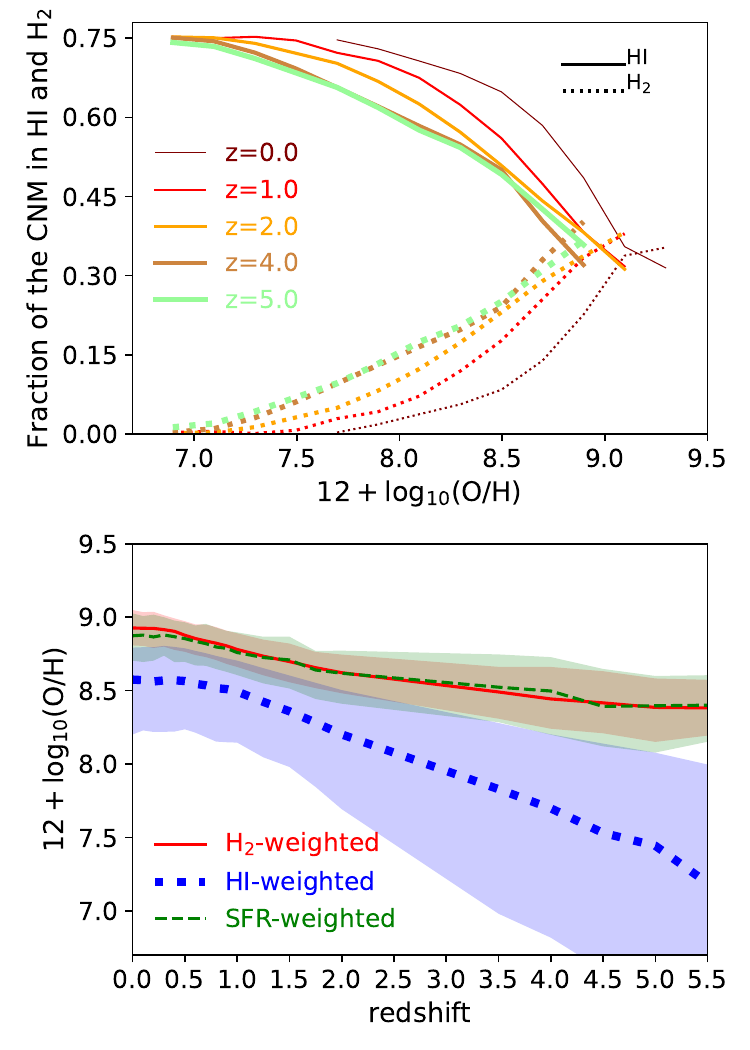}
\caption{
{\it Top panel:} The fraction of the CNM in \ion{H}{i} (solid lines) and H$_2$ (dotted lines) as a function of gas metallicity, from $z=0$ to $z=5$ ({see Eq.~\ref{CNMdef} for a definition of the CNM)}. We apply the same stellar and SFR selection of Fig.~\ref{FracCNMz0}. Note that we only include a subset of the redshifts shown in Fig.~\ref{KS_Zevo} to avoid overlapping lines. 
{\it Bottom panel:} The redshift evolution of the median gas metallicity (solid lines), and the $25^{\rm th}-75^{\rm th}$ percentile range (shaded regions), of galaxies with $M_{\star}\ge 10^9\,\rm M_{\odot}$, when weighted by the H$_2$ mass (solid line), \ion{H}{i} mass (dotted line) or SFR (dashed line).} 
\label{Fracts}
\end{center}
\end{figure}

\subsection{The evolving KS relation from $z=0$ to $z=8$}

Fig.~\ref{KS_Zevo} shows the evolution of the KS relation in HI, H$_2$ and total neutral hydrogen, from $z=8$ to $z=0$. \colibre\ predicts a monotonic evolution of the H$_2$ KS relation with time, so that the normalisation of the relation consistently decreases with time, maintaining a power-law index $N_{\rm KS}\approx 1$ at all redshifts (at least at $\rm log_{10}(\Sigma_{\rm H_2}/M_{\odot}\,pc^{-2}) \gtrsim 1$). The zero-point of the H$_2$ KS relation evolves very strongly with redshift, decreasing by a factor of $
\approx 20$ from $z=8$ to $z=0$. 

The case of \ion{H}{i} and total neutral hydrogen is more complex: at low gas surface densities $\lesssim 100\,\rm M_{\odot}\,pc^{-2}$, the normalisation of the KS relation decreases mildly with increasing redshift from $z=0$ to $z=8$ (by $\approx 0.8$~dex for \ion{H}{i} and $0.4$~dex for \ion{H}{i}+H$_2$), while also becoming slightly steeper. This evolution reverses at a gas surface density $\approx 100\,\rm M_{\odot}\,pc^{-2}$, resembling the evolution of the H$_2$ KS at higher surface densities. 

Interestingly, we do not see a strong evolution in the scatter of the H$_2$ KS relation, which is inconsistent with the findings reported in other simulations, such as {\sc NewHorizon}. \citet{Kraljic24} find that the scatter in the H$_2$ KS relation in {\sc NewHorizon} decreases by $\approx 0.25$~dex from $z=4$ to $z=0.25$. There are several differences between our work and theirs; for example, they include all galaxies with stellar masses $\ge 10^7\,\rm M_{\odot}$, while here we adopt a mass threshold that is $100$ times higher. Another important difference is that {\sc NewHorizon} does not directly model the formation of H$_2$, and hence it is assumed that thresholds in temperature and density are able to exclusively select H$_2$ gas. We demonstrated in Figs.~\ref{PhaseSpacez0}~and~\ref{FracCNMz0} that this is not the case and, in fact, as the metallicity decreases, more of the cold gas can remain atomic. These differences make it hard to pinpoint the driver of the discrepancies but they certainly serve as a motivation to constrain these relations observationally.
{On the other hand, \citet{Shen2026} presented an analysis of the KS relation for total neutral gas at $z>3$ for Thesan-zoom simulations measured at $1\,\rm kpc$-scales. They found a steep relationship at high surface densities, $\Sigma_{\rm HI+H_2}\gtrsim 10^{1.2}\,\rm M_{\odot}\,pc^{-2}$, which agrees well with ours. 
At lower surface densities, their simulation predicts much higher $\Sigma_{\rm SFR}$ compared to \colibre, but this is a regime where Thesan-zoom is not well sampled given the small number of galaxies and the overall smaller number of low density regions with $\rm SFR>0$.
The steep neutral KS relation is interpreted by \citet{Shen2026} as the result of 
turbulent energy balance in the ISM maintained by stellar feedback. The importance of turbulence in regulating star formation is also predicted by \colibre, as reported by \citet{Schaye25}.}

The strong evolution of the zero point of the H$_2$ KS relation in \colibre\ is in part due to the evolution of the  average gas metallicity of galaxies with $M_{\star}\ge 10^9\,\rm M_{\odot}$. Fig.~\ref{Fracts} shows the fraction of CNM in \ion{H}{i} and H$_2$ as a function of the gas metallicity from $z=0$ to $z=5$ (similar to what Fig.~\ref{FracCNMz0} shows), and the evolution of the gas metallicity when weighted by different properties. First, we see that, at all redshifts, there is a clear trend of an increasing amount of the CNM being \ion{H}{i} as the metallicity decreases. However, this relation is redshift dependent, with a decreasing fraction of the CNM in \ion{H}{i} with increasing redshift at fixed gas metallicity. The corresponding simultaneously decreasing metallicity with redshift, leads to an overall effect of less H$_2$ and more \ion{H}{i} being associated with a given amount of SFR with increasing redshift. This gives the impression that the molecular gas is more efficient at forming stars at high redshift, but this is primarily due to $\rm H_2$ becoming a poorer tracer of star formation at high redshift, at least on kpc scales. We remind the reader that the star formation model in \colibre\ assumes a universal star formation efficiency per free fall time of $1$\% and a universal gravitational instability criterion to select star forming gas (Eq.~\ref{alphacrit}), and hence there is no evolution built into the model. Thus, the strong evolution in the KS relation predicted by \colibre, particularly for H$_2$, is driven by the processes that govern the formation of H$_2$ and determine whether the gas is unstable and what the density distribution of the unstable gas is.  

Fig.~\ref{Fracts} shows that the gas metallicity weighted by the \ion{H}{i} mass is always lower than the values weighted by SFR or H$_2$, with the differences becoming larger at higher redshifts. This is not surprising, as \ion{H}{i} traces lower density gas than $\rm H_2$ and SFR, which is typically located in the outskirts of galaxies, with most galaxies in \colibre\ having negative metallicity gradients (see Serrano Rodriguez et al. in preparation).

\subsection{Comparison with observations of the resolved and global KS relation}
\begin{figure}
\begin{center}
\includegraphics[trim=1mm 14mm 15mm 26mm, clip,width=0.499\textwidth]{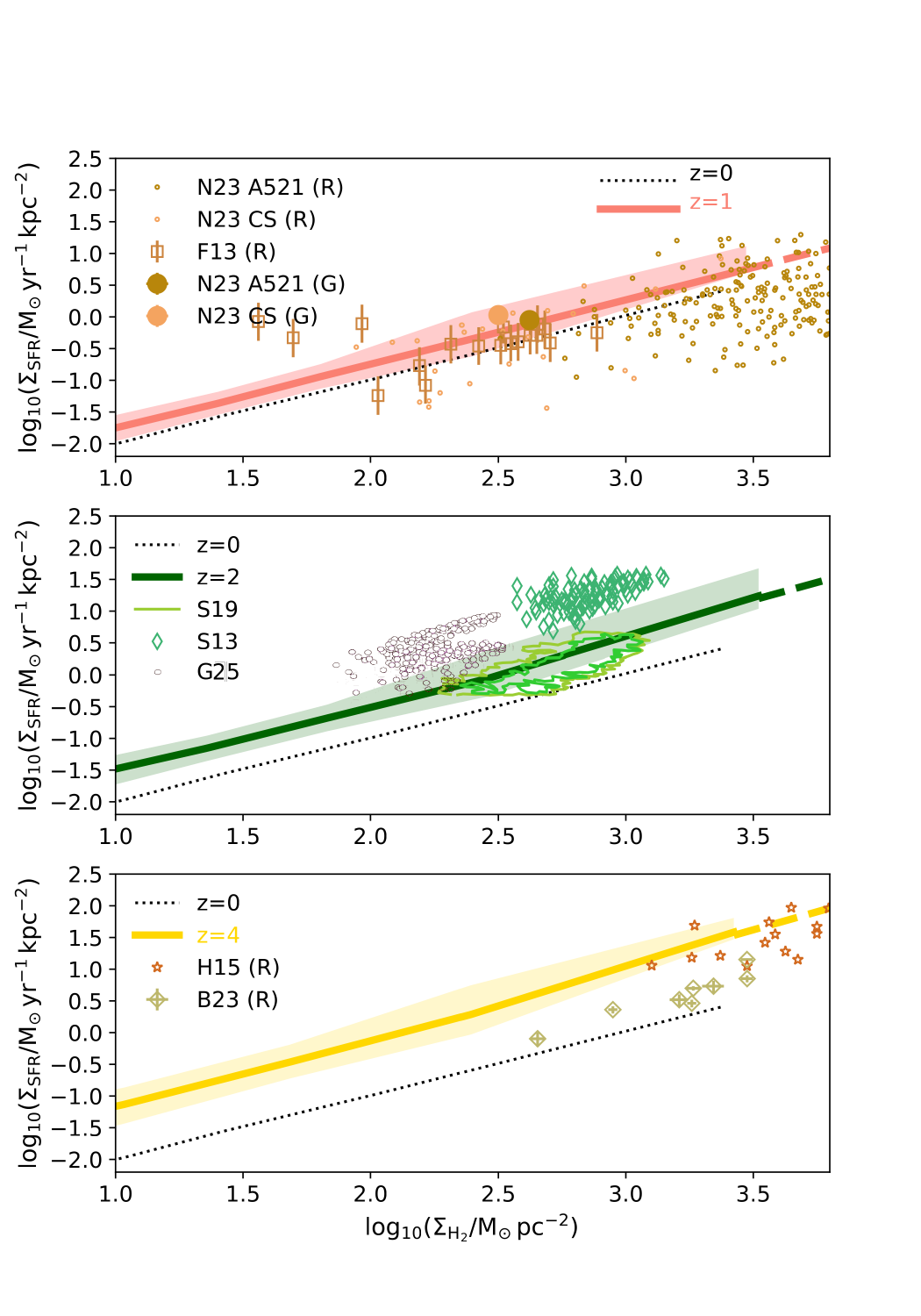}
\caption{The predicted KS relation at $z=1$ (top), $z=2$ (middle) and $z=4$ (bottom) {for galaxies with $M_{\star}\ge 10^9\,\rm M_{\odot}$}. We compare with a compilation of observations of the (resolved) KS relation in individual galaxies. At $z=1$, we show the resolved (R) and global (G) measurements of A521 and the Cosmic Snake from \citet{Nagy23} (N+23). For the resolved data in N+23, we use the data provided in $800$~pc apertures. We also show the resolved data from $4$ massive star-forming galaxies at $z\approx 1.2$ from \citet{Freundlich13}. At $z=2$ we compare with {four} galaxies; two from \citet{Sharon13} (S13; a sub-millimetre galaxy at $z=2.6$) and \citet{Sharon19} (S19; a main sequence galaxy at $z=2.26$); {and a pair of two hyper-luminous IR galaxies from \citet{Gomez25} (G25)}. For S13, we show the individual $\approx 2$~kpc pixels; for S19 we show the contours tracing one and three pixels, where each pixel is of $\approx 2$~kpc size (see \citealt{Sharon19} for details); {while for G25 we show individual $\approx 2$~kpc pixels for each galaxy.}    
At $z=4$, we show the results from $3$ massive, star-forming galaxies from the ALPINE survey \citep{Bethermin23} (B23) at $z\approx 4.5$, and one submillitre galaxy at $z\approx 4$ \citet{Hodge15} (H15). For \colibre, we show as dashed lines the extrapolation at high surface densities of the linear fit to the median relation (solid lines) measured at $1\le \rm log_{10}(\Sigma_{\rm H_2}/\rm M_{\odot}\,\rm pc^{-2})\le 3$.} 
\label{KS_Zevo_comp_obs}
\end{center}
\end{figure}

\begin{figure}
\begin{center}
\includegraphics[trim=4mm 4mm 0mm 3mm, clip,width=0.499\textwidth]{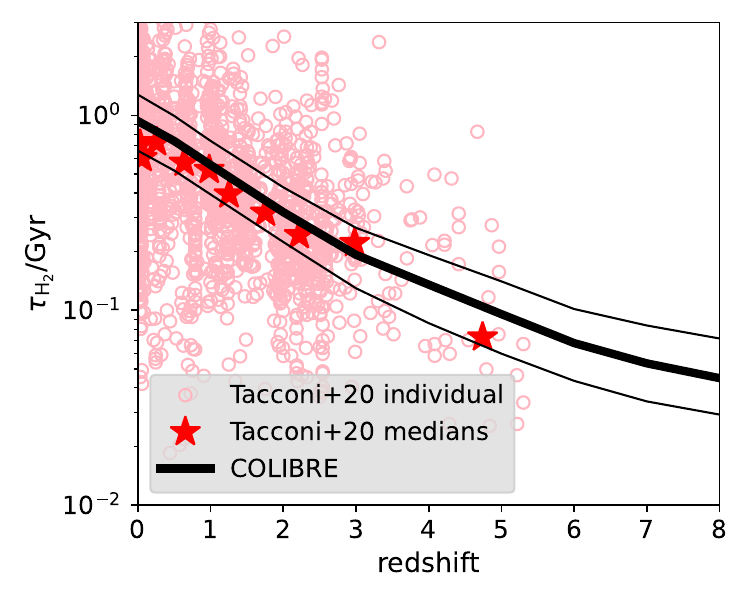}
\caption{The evolution of the H$_2$ depletion time compared with the compilation of observations presented in \citet{Tacconi20}. Small circles correspond to individual galaxies, while the stars correspond to the medians computed from the observations. The thick and thin solid lines correspond to the median and $16^{\rm th}-84^{\rm th}$ percentile ranges of all galaxies in \colibre\ with $M_{\star}\ge 10^9\,\rm M_{\odot}$ and $\rm SFR>0$. To compute $\tau_{\rm H_2}$, we only considered $1$~kpc annuli with $\Sigma_{\rm H_2} > 1\,\rm M_{\odot}\,pc^{-2}$, which correspond to the gas density regime where the L200m6 run is well converged (see Fig.~\ref{KS_Resolution}) and comfortably includes the regimes from which observational results are computed.} 
\label{KS_Zevo_comp_obs2}
\end{center}
\end{figure}
In Fig.~\ref{KS_Zevo_comp_obs}, we present comparisons with observations of the KS relation at $z>0$. There are only very limited samples of galaxies at intermediate and high redshifts for which one can measure a resolved KS relation, and they correspond primarily to strongly lensed galaxies (e.g. \citealt{Hodge15,Sharon13,Sharon19,Nagy23}). Note that at $z=1$ the observations correspond to $6$ galaxies ($4$ coming from \citealt{Freundlich13} and $2$ from \citealt{Nagy23}). At $z=2$, there are observations for {four} galaxies, one is a submilimetre highly star-forming galaxy \citep{Sharon13}; {two are a pair of hyper-luminous IR galaxies \citep{Gomez25} separated by $\approx 85$~kpc, one of which is mildly magnified}; and a main-sequence galaxy \citep{Sharon19}. At $z=4$ the observations correspond to $5$ galaxies ($4$ from \citealt{Bethermin23} and $1$ from \citealt{Hodge15}). In all cases, a constant (Milky-Way or starburst galaxy) CO-to-H$_2$ conversion factor was used. The conclusions we can draw from such a small sample are limited, but can serve as a reference to compare \colibre\ predictions with. 
We applied corrections to the SFR to convert them from a Salpeter or Kroupa IMF to a Chabrier IMF when necessary, and remove the helium contributions from the molecular gas if quantities reported include it. 

At $z=1$, we find that the global measurements of \citet{Nagy23} agree well with the predicted relation for \colibre, but the individual regions (which sample a $800$~pc scale) display a very large scatter, larger than the typical variations in \colibre. The resolved relations measured for the Cosmic snake and the sample of \citet{Freundlich13} agree well with \colibre, but the individual regions in A521 tend to scatter down towards longer H$_2$ depletion times. At $z=2$, we see that \colibre's median relation agrees better with the main sequence galaxy in \citet{Sharon19}, while the {highly starbursting} galaxies of \citet{Sharon13,Gomez25} lie (slightly) above the predicted relation. This shows that even within the small observational samples, there are hints that the normalisation of the KS relation varies with a galaxy's sSFR, in agreement with our predictions. 
At $z=4$, we find that the observations sample $\Sigma_{\rm H_2}$ to values higher than what is typically reached by \colibre\ at m6 resolution, but the extrapolation of the predicted relation agrees well with what \citet{Hodge15} report and is potentially above what \citet{Bethermin23} measure. Larger samples are required to start to   systematically probe the evolution of the resolved KS relation. This will become possible in the near future, for example with the extended Northern Extended Millimetre Array (NOEMA).

Unlike spatially resolved observations, integrated values for individual galaxies have been obtained by the thousands \citep{Tacconi20}. From these observations, it has been well established that the H$_2$ depletion time decreases with increasing redshift. This is shown for \colibre\ and a compilation of observations in Fig.~\ref{KS_Zevo_comp_obs2}. 
The observations come from a range of H$_2$ tracers: CO, sub-millimetre or millimetre emission, or FIR SED. Even though they are all subject to different systematic effects, \citet{Tacconi20} argue that they produce consistent H$_2$ scaling relations. 
Each circle represents an individual observed galaxy, while the stars show the median for the observations. The latter are computed with a minimum of $10$ galaxies. We show with lines the predictions of \colibre, which are in remarkable agreement with the observations. {However, we remind the reader that for \colibre\ we show individual $1$~kpc rings rather than global measurements for galaxies, so the comparison is not one-to-one. Another important caveat is that we show regions from all galaxies with $M_{\star}\ge 10^9\,\rm M_{\odot}$ and $\rm SFR>0$, while the observations are biased towards more massive galaxies with higher SFRs. Below we explore the effect of sSFR.}

\begin{figure}
\begin{center}
\includegraphics[trim=4mm 4.5mm 0mm 3mm, clip,width=0.499\textwidth]{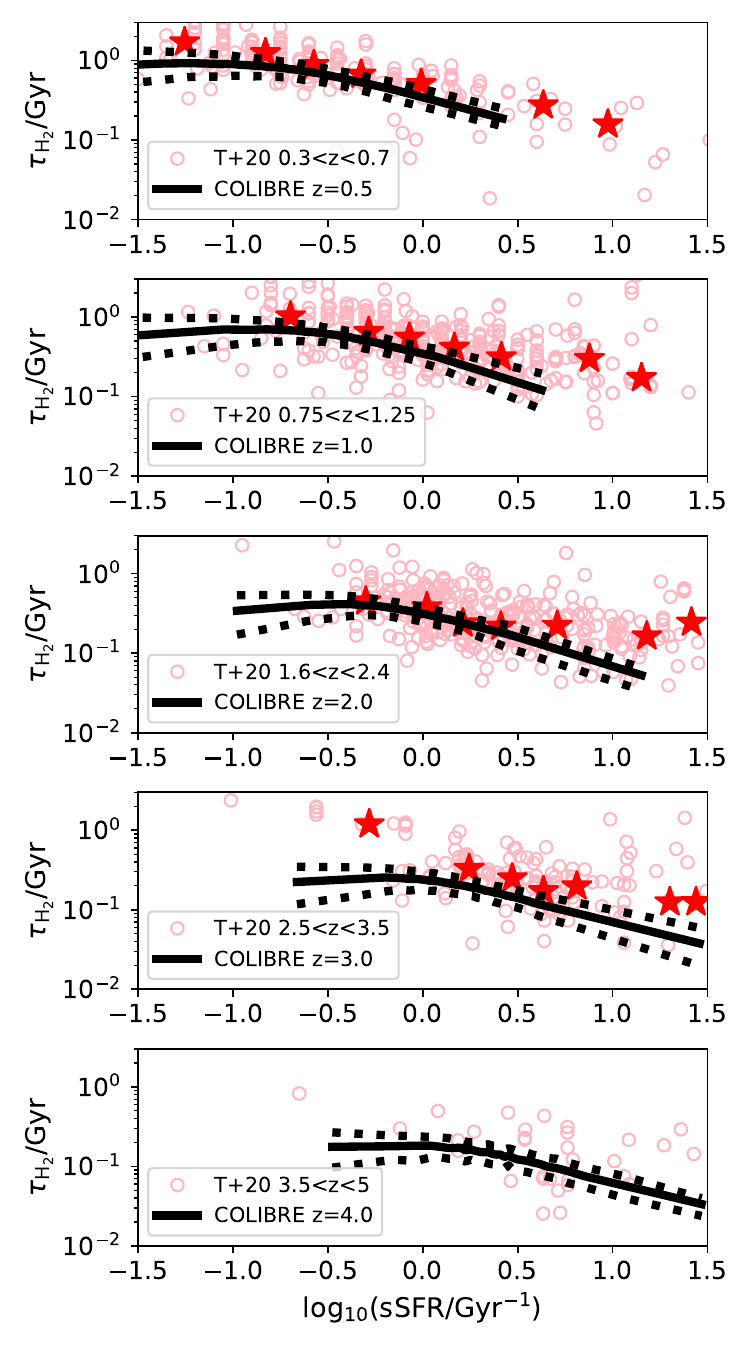}
\caption{The H$_2$ depletion time as a function of sSFR {for galaxies with $M_{\star}\ge 10^9\,\rm M_{\odot}$,} at different redshift bins from $z\approx 0.5$ to $z\approx 5$,  compared with a compilation of observations presented in \citet{Tacconi20}. The solid and dashed lines in each panel show the median and $16^{\rm th}-84^{\rm th}$ percentile ranges, respectively, for galaxies in \colibre.
Symbols are as in Fig.~\ref{KS_Zevo_comp_obs2}, and their corresponding redshift range are labelled in each panel. We do not show medians for the observations for the bottom panel due to the small number of galaxies.} 
\label{KS_Zevo_comp_obs3}
\end{center}
\end{figure}

The scatter in the observations is large and driven by physical parameters, such as the sSFR of galaxies, which we show in Fig.~\ref{KS_GlobalDeps} can systematically shift the H$_2$ depletion time. We demonstrate this at different redshifts in Fig.~\ref{KS_Zevo_comp_obs3}, where we plot the resolved H$_2$ depletion time as a function of a galaxy's sSFR in $5$ redshift bins. At all redshifts, there is a strong dependence of the local $\tau_{\rm H_2}$ on the global sSFR, with the zero point of the relation evolving strongly with redshift in \colibre. We also compute medians for the observations in redshift bins where there is enough data to have at least $10$ points per bin. 
The observations agree strikingly well with our predictions, noting though that the observational samples are not stellar-mass complete, which our simulated sample is by construction. We also note that at the highest sSFR, the observations prefer slightly longer H$_2$ depletion times than what \colibre\ predicts at kpc scales. In the future we will compare global measurements to perform a fairer comparison with these global observations. 


\begin{figure*}
\begin{center}
\includegraphics[trim=0mm 4.5mm 4mm 2mm, clip,width=0.49\textwidth]{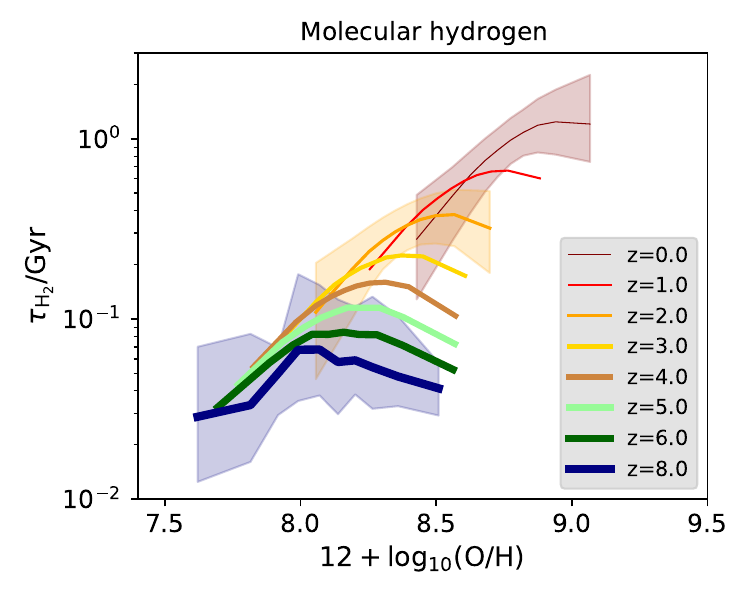}
\includegraphics[trim=0mm 4.5mm 4mm 2mm, clip,width=0.49\textwidth]{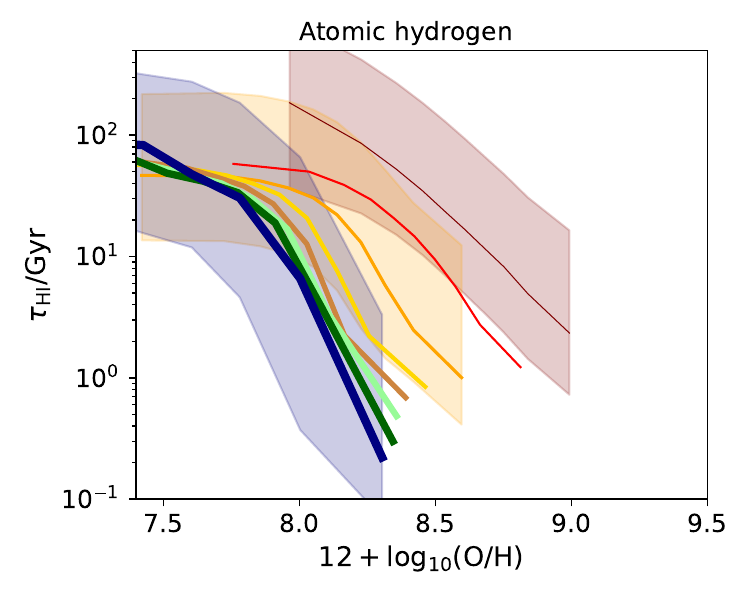}
\includegraphics[trim=0mm 4.5mm 4mm 2mm, clip,width=0.49\textwidth]{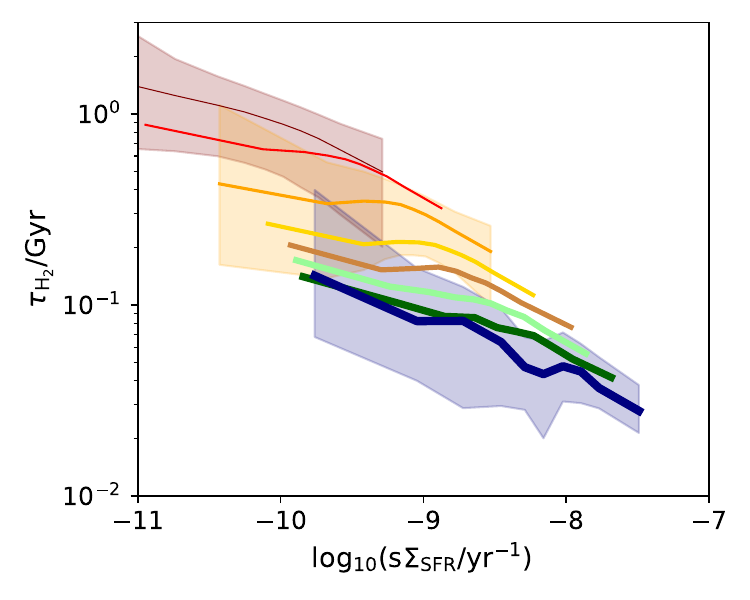}
\includegraphics[trim=0mm 4.5mm 4mm 2mm, clip,width=0.49\textwidth]{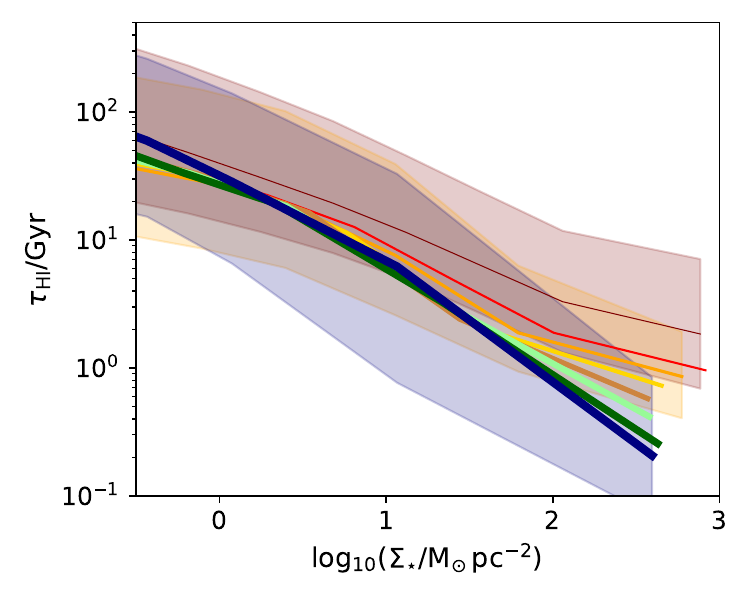}
\caption{Evolution of some of the  strongest resolved relations found in \colibre, between local galaxy properties and the kpc-scale H$_2$ (left) and \ion{H}{i} (right) depletion timescales, {for galaxies with $M_{\star}\ge 10^9\,\rm M_{\odot}$}. For H$_2$, we show the evolution of the 
 relation between the resolved $\tau_{\rm H_2}$ and the local gas metallicity (top-left panel) and sSFR surface density (bottom-left panel), between $z=0$ and $8$. Solid lines with shaded regions represent the medians with $16^{\rm th}-84^{\rm th}$ percentile ranges (only shown for $z=0$, $2$ and $8$), respectively, with the colours indicating different redshifts, as labelled in the top-left panel. For HI, we show the relation with local gas metallicity (top-right panel) and the stellar surface density (bottom-right panel).}
\label{tauH2_vs_z}
\end{center}
\end{figure*}
\subsection{The evolving dependence of the scatter in the KS relation with local physical properties}

Finally, in Fig.~\ref{tauH2_vs_z} we explore how some of the strongest correlations between $\tau_{\rm H_2}$ and $\tau_{\rm HI}$ with other local properties presented in Fig.~\ref{Delta_correlations} for $z=0$ evolve with time. For $\tau_{\rm H_2}$, we find that the correlation with local gas metallicity remains strong with increasing redshift, with the amplitude of the correlation being close to redshift invariant. However, the correlation displays a saturation at the highest metallicities at any one redshift, with a reversal of the correlation at the high-metallicity end. We confirmed that the $\tau_{\rm H_2}$-gas metallicity correlation does not change when we increase the resolution, nor when we adopt instead the hybrid AGN feedback model (using the L025m5 and L050m6 Hybrid runs listed in Table~\ref{TableSimus}). When using the L025m5 run, we find that the relation holds and the saturation manifests itself at the same $\tau_{\rm H_2}$ values as in the L200m6 run. Interestingly, this saturation happens at $\approx 0.1/H(z)$, where $H
(z)$ is Hubble's constant at redshift $z$. The latter corresponds to the free-fall time of a sphere of density $\sim 200\, \rho_{\rm crit}$, where $\rho_{\rm crit}$ is the critical density of the universe. Realistically, star formation cannot proceed any slower than that timescale, as $200\, \rho_{\rm crit}$ corresponds to the minimum gas density for gravitational collapse in a $\rm CDM$ universe.
The redshift invariance of the $\tau_{\rm H_2}$-gas metallicity correlation implies that the physical mechanism behind it remains the same throughout cosmic time: as gas metallicity decreases, more of the star-forming gas is in the form of HI, and hence the amount of H$_2$ associated with a fixed SFR decreases. This was shown in Fig.~\ref{Fracts}. 

The bottom-left panel of Fig.~\ref{tauH2_vs_z} shows that the correlation between $\tau_{\rm H_2}$ and $s\Sigma_{\rm SFR}$ remains strong even up to $z=8$, but with the normalisation of the relation decreasing strongly with redshift. This mirrors the strong evolution we see in the relation between $\tau_{\rm H_2}$ and the global sSFR of galaxies in Fig.~\ref{KS_Zevo_comp_obs3}. Ignoring the saturation, the close to redshift invariance of the $\tau_{\rm H_2}$-metallicity relation suggests a more fundamental physical connection than that between $\tau_{\rm H_2}$ and $s\Sigma_{\rm SFR}$, even if statistically the latter may be stronger in some regimes (as measured by the Spearman correlation coefficient).

In the top-right panel of Fig.~\ref{tauH2_vs_z}, we show that 
the correlation between $\tau_{\rm HI}$ and metallicity evolves very strongly with redshift, with $\tau_{\rm HI}$ decreasing by two orders of magnitude from $z=0$ to $z=8$ at fixed local gas metallicity. The bottom-right panel of Fig.~\ref{tauH2_vs_z} shows that the relation between $\tau_{\rm HI}$ and $\Sigma_{\star}$ is already in place at $z=8$ in \colibre, and evolves more mildly with redshift than the relation with local gas metallicity, with $\tau_{\rm HI}$ decreasing by a factor of $\approx 3$ from $z=0$ to $z=8$ at fixed $\Sigma_{\star}$. 
For \ion{H}{i} there is usually larger scatter in all the relations we explored compared with H$_2$, which is natural, as a significant fraction ($\gtrsim 50$\%) of the \ion{H}{i} is not in the CNM, but instead in the WNM, which is less directly related to star formation.

Overall, our results show that the kpc-scale KS relation is already in place by $z=8$ in \colibre. This is true for both the \ion{H}{i} and H$_2$ KS relations. This contrasts with the predictions presented in \citet{Kraljic24} for {\sc NewHorizon}, in which they found that the KS relation is well established only at $z\lesssim 2-3$. However, as we mentioned above, there are important differences between our sample and theirs in terms of the stellar mass of galaxies covered, while also having several differences in the physical models. In the future, we want to investigate the impact on the KS relation of switching on and off various processes in \colibre, including, but not limited to, early stellar feedback, the non-equilibrium chemistry and the chemical diffusion model.

\section{Discussion and Conclusions}\label{sec:conclusions}

We used the \colibre\ cosmological hydrodynamical simulations of galaxy formation to investigate the HI, H$_2$ and total neutral hydrogen Kennicutt-Schmidt (KS) relations and the scatter of these relations {for galaxies with $M_{\star}\ge 10^9\,\rm M_{\odot}$}, over $0\le z\le 8$. \colibre\ features on-the-fly non-equilibrium chemistry for H and He, with metals also included, in equilibrium but with non-equilibrium electron abundances,  detailed radiative cooling down to $\approx 10$~K, and live dust that impacts gas cooling and the formation of H$_2$. This detailed modelling enables direct predictions for the HI, H$_2$ and total neutral H KS relations. 

We measure the KS relation using circular rings of fixed width in galaxies oriented face-on. We find, however, that modifying the method to measure the KS relation has only minimal effects on our results (Appendix~\ref{convmeth}). Our main findings are summarised below:

\begin{itemize}
    \item \colibre\ accurately reproduces the observed HI, H$_2$, and total gas KS relations at $z = 0$ (Fig.~\ref{KS_CompObs}), including their scatter, without having been explicitly tuned to these data. \colibre's star formation model assumes an underlying constant gas-to-star formation efficiency per free fall time of $1$\%, with the gas required to be gravitationally unstable at the resolution limit to form stars. This model, coupled with the sophisticated ISM model in \colibre, naturally captures differences in the KS relation amplitude and scatter between \ion{H}{i} and H$_2$, with \ion{H}{i} showing longer depletion times and larger scatter than H$_2$, consistent with the HI's association with more diffuse gas phases (Figs.~\ref{PhaseSpacez0},  \ref{densities_CNM}, and \ref{FracCNMz0}). The kpc-average KS relations are very well converged in \colibre, even when the numerical resolution is varied by $2$ orders of magnitude. 
    \item \colibre\ predicts the scatter in the KS relation to  correlate strongly with local properties such as the stellar surface density, gas metallicity, and dust surface density (Figs.~\ref{KS_ResolvedDeps} and \ref{Delta_correlations}). At kpc scales, we find that lower metallicity/dust content regions form stars efficiently before significant H$_2$ formation, leading to apparently shorter H$_2$ and longer \ion{H}{i} depletion times. The strongest predictors of H$_2$ depletion time are the local sSFR and gas metallicity, while \ion{H}{i} depletion time correlates most strongly with the stellar surface density and dust content (Fig.~\ref{Delta_correlations}).
    \item \colibre\ agrees  remarkably well with spatially resolved observations at $z=0$. In particular, we compared with ALMaQUEST, PHANGS and \citet{Wang24} and found broad agreement in both the KS relation itself and its dependence on resolved properties such as sSFR and metallicity (Figs.~\ref{KS_TauHIWang}, \ref{KS_ResolvedDepsALMAQUEST} and \ref{KS_ResolvedDepsALMAQUEST2}). \colibre\ also reproduces the lower KS amplitude in green-valley galaxies relative to main-sequence systems, matching ALMaQUEST and PHANGS (Fig.~\ref{KS_GlobalDepsALMAQUEST}). We also explore KS fits performed on a galaxy-by-galaxy basis and find that \colibre\ predicts variations in the best fit parameters that resemble what observations indicate for \ion{H}{i} and H$_2$ (Fig.~\ref{KSfit}), including a dependence of the best fit parameters on stellar mass for HI. 
    \item While the global stellar mass plays a minor role in the normalisation of the KS relation (Fig.~\ref{KS_GlobalDeps}), the galaxy-wide sSFR has a stronger influence: high-sSFR galaxies have both larger cold gas reservoirs and higher star-formation efficiencies. We find that the enhanced efficiency is not purely due to gas content but also linked to higher conversion rates of gas into stars.
    \item The kpc-scale HI, H$_2$ and \ion{H}{i}+H$_2$ KS relations are already in place by $z=8$ in \colibre. At even higher redshifts, there are too few galaxies that meet our selection ($M_{\star}>10^9\,\rm M_{\odot}$ and $\rm SFR > 0$) to clearly establish the presence of a KS relation. The H$_2$ depletion time decreases by a factor of $\approx 20$ from $z = 0$ to $z = 8$ (Fig.~\ref{KS_Zevo}), largely driven by the lower local gas metallicities at higher redshift, leading to more HI-dominated cold neutral medium from which stars form (Fig.~\ref{Fracts}). The monotonic evolution of the H$_2$ KS relation contrasts with that of the HI, which shows a more complex evolution, with a transition at around $\Sigma_{\rm HI}\approx 100 \,\rm M_{\odot}\,pc^{-2}$. At lower gas surface densities, the amplitude of the \ion{H}{i} KS relation decreases mildly with redshift, while at higher densities, the trend reverses.
    \item The overall lower gas metallicities at high redshifts predicted by \colibre\ lead to higher star formation efficiencies per unit H$_2$ in early galaxies, which is  consistent with recent high-redshift constraints (Figs.~\ref{KS_Zevo_comp_obs}, \ref{KS_Zevo_comp_obs2} and \ref{KS_Zevo_comp_obs3}). The lower gas metallicity is associated with  a cold neutral medium that is \ion{H}{i} dominated, which once averaged on  kpc scales, looks like having an efficient H$_2$ to SFR conversion. 
    \colibre\ predicts a strong correlation between $\tau_{\rm H_2}$ and sSFR, which agrees very well with observational constraints coming from integrated galaxy properties. 
\end{itemize}

Our results demonstrate that the KS relation at kiloparsec scales for the \ion{H}{i} and H$_2$ gas is an emergent outcome of the ISM physics (cooling, dust grains, metal enrichment, stellar feedback and non-equilibrium chemistry) and the underlying universal star formation efficiency included in the \colibre\ model. This is the case not only for the median KS relation, but also for the dependence of the scatter on local and global galaxy properties, and its very strong redshift evolution. The agreement with spatially resolved observations of the KS relations is a significant success of the \colibre\ model because none of the model's free parameters were adjusted to match these observations.

In the future we want to use variations of the \colibre\ physics, e.g. assuming chemical equilibrium, turning off the live dust model, varying the parameters of the star formation model and the stellar feedback, to isolate the physical mechanisms setting the amplitude and scatter of the KS relation at different cosmic epochs. This will also allow us to understand  whether the KS relation can be used to study the interplay between galaxy-scale feedback effects and star formation.

One crucial limitation is the small observational samples of galaxies for which a resolved KS relation has been measured, and the ad-hoc galaxy selections. This complicates the testing of our predictions. Upcoming observations with ALMA of hundreds of local Universe galaxies\footnote{\url{https://www.kilogas.space/}} will allow the H$_2$ KS relation to be measured at kiloparsec scales for stellar-mass selected samples, and a robust characterisation of the scatter and its dependence on other local and global properties. 
Similarly, at higher redshifts, the combination of JWST and NOEMA/ALMA will allow increasing the sample of galaxies for which a resolved KS relation can be measured to dozens. 

\section*{Acknowledgements}

We thank Miroslava Dessauges, Anita Zanella, and Li-Hwai Lin for sharing their observational data in a format that facilitated the comparison with observations. We also acknowledge the PHANGS team for making the data presented in \citet{Sun22} publicly available. CL also thanks the CONDOR and KILOGAS teams for useful discussions. 

We acknowledge the use of SOAP \citep{McGibbon25} in our analysis, and thank their developers for the effort put in the code and documentation. 

This work used the DiRAC@Durham
facility managed by the Institute for Computational Cosmology on
behalf of the STFC DiRAC HPC Facility (\url{www.dirac.ac.uk}).
The equipment was funded by BEIS capital funding via STFC capital grants ST/K00042X/1, ST/P002293/1, ST/R002371/1 and
ST/S002502/1, Durham University and STFC operations grant ST/R000832/1. DiRAC is part of the National e-Infrastructure. 
This work was also supported by resources provided by The Pawsey Supercomputing Centre with funding from the 
Australian Government and the Government of Western Australia.
ABL acknowledges support by the Italian Ministry for Universities (MUR) program “Dipartimenti Di Eccellenza 2023-2027” within the Centro Bicocca di Cosmologia Quantitativa (BiCoQ), and support by UNIMIB’s Fondo Di Ateneo Quota Competitiva (project 2024-ATEQC-0050). 
CSF acknowledges support from the European Research Council (ERC) Advanced Grant
 DMIDAS (GA 786910). FH acknowledges funding from the Netherlands Organization for Scientific Research (NWO) through research
programme Athena 184.034.002.
JT acknowledges support of a STFC Early Stage Research and Development grant (ST/X004651/1).

\section*{Data Availability}

The data supporting the plots within this article are available on reasonable request to the corresponding author. The \colibre\  simulation data will eventually be made publicly available, although we note that the data volume (several petabytes) may prevent us from simply placing the raw data on a server. In the meantime, people interested in using the simulations are encouraged to contact the corresponding author. A public version of the {\sc Swift} code \citep{Schaller24} is available at \url{http://www.swiftsim.com}. The \colibre\ 
modules implemented in {\sc Swift} will be made publicly available after the public release of the simulation data.



\bibliographystyle{mn2e_trunc8}
\bibliography{SharkQ}




\appendix

\section{Convergence of the KS relation}\label{convergence}

\subsection{Convergence with the resolution of the simulation}\label{convres}

\begin{figure*}
\begin{center}
\includegraphics[trim=20mm 10mm 30mm 10mm, clip,width=0.95\textwidth]{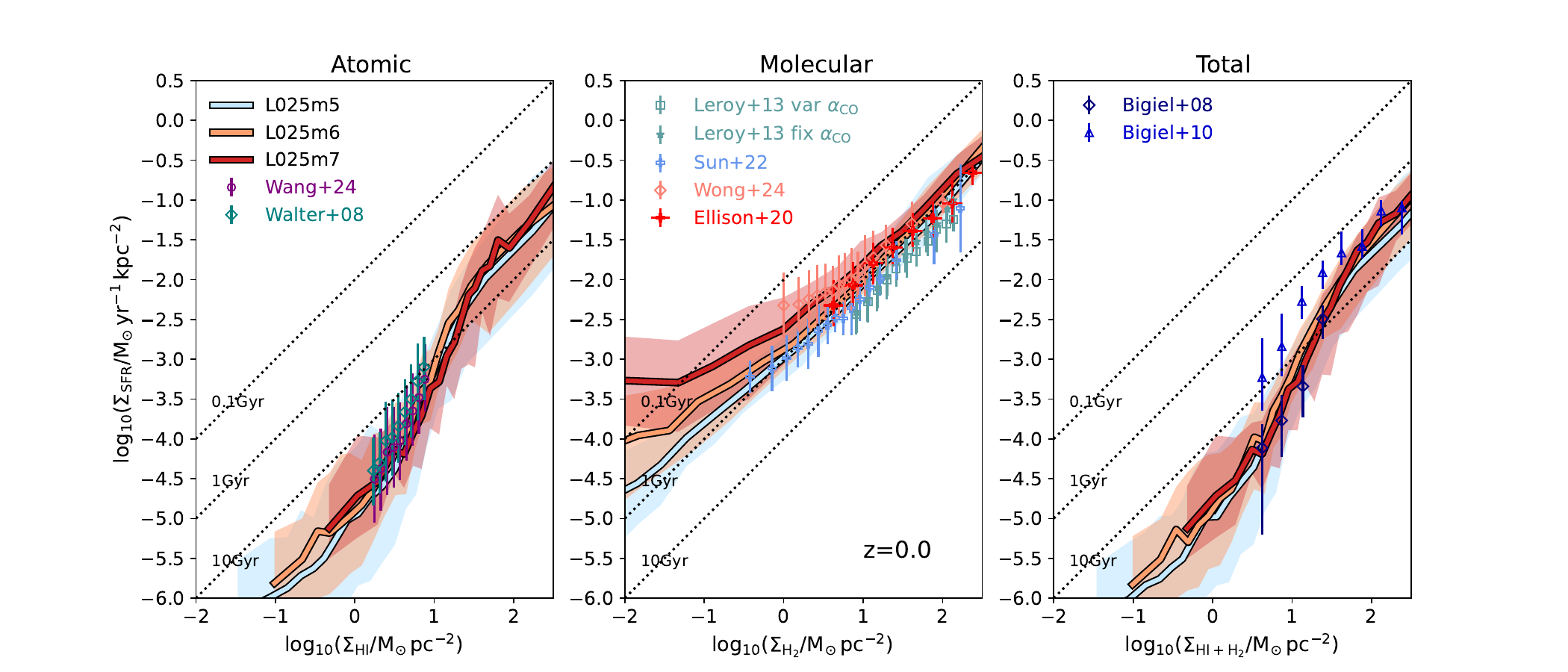}
\includegraphics[trim=20mm 0mm 30mm 17mm, clip,width=0.95\textwidth]{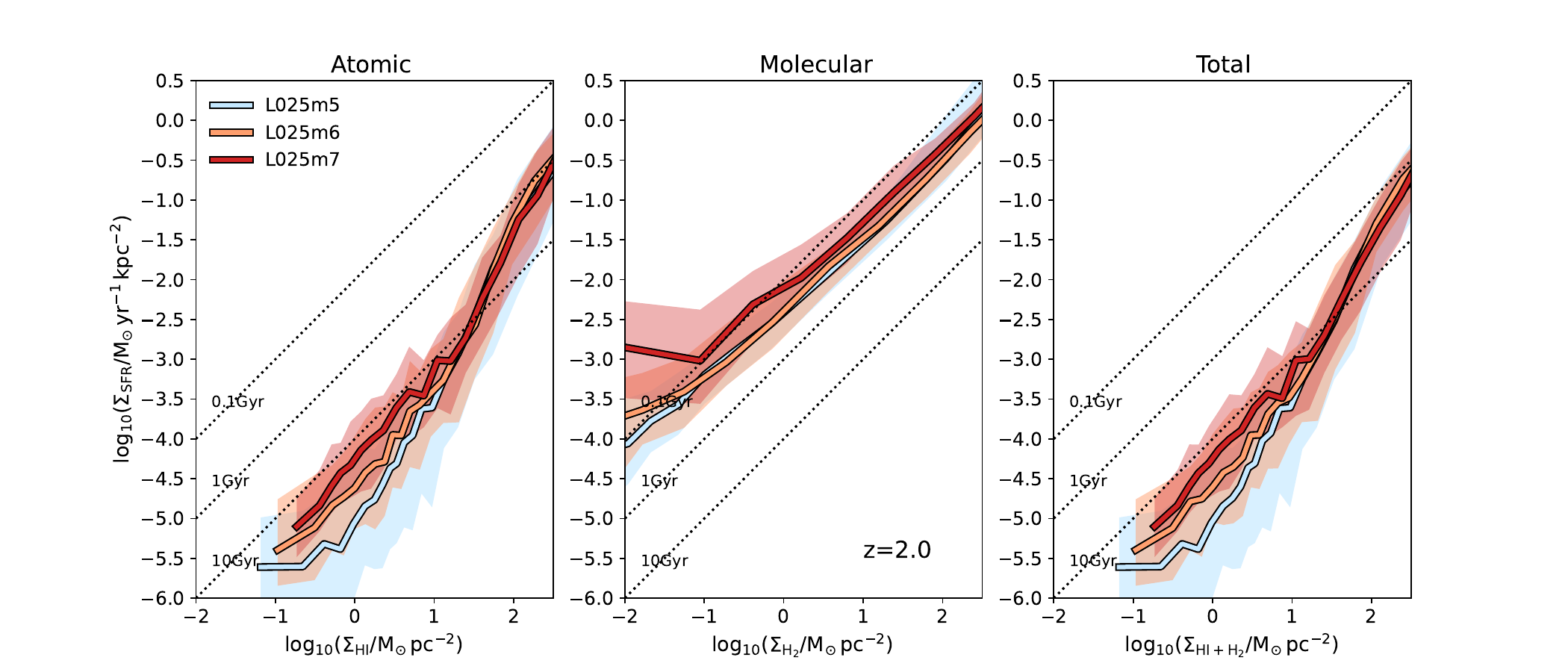}
\includegraphics[trim=20mm 0mm 30mm 17mm, clip,width=0.95\textwidth]{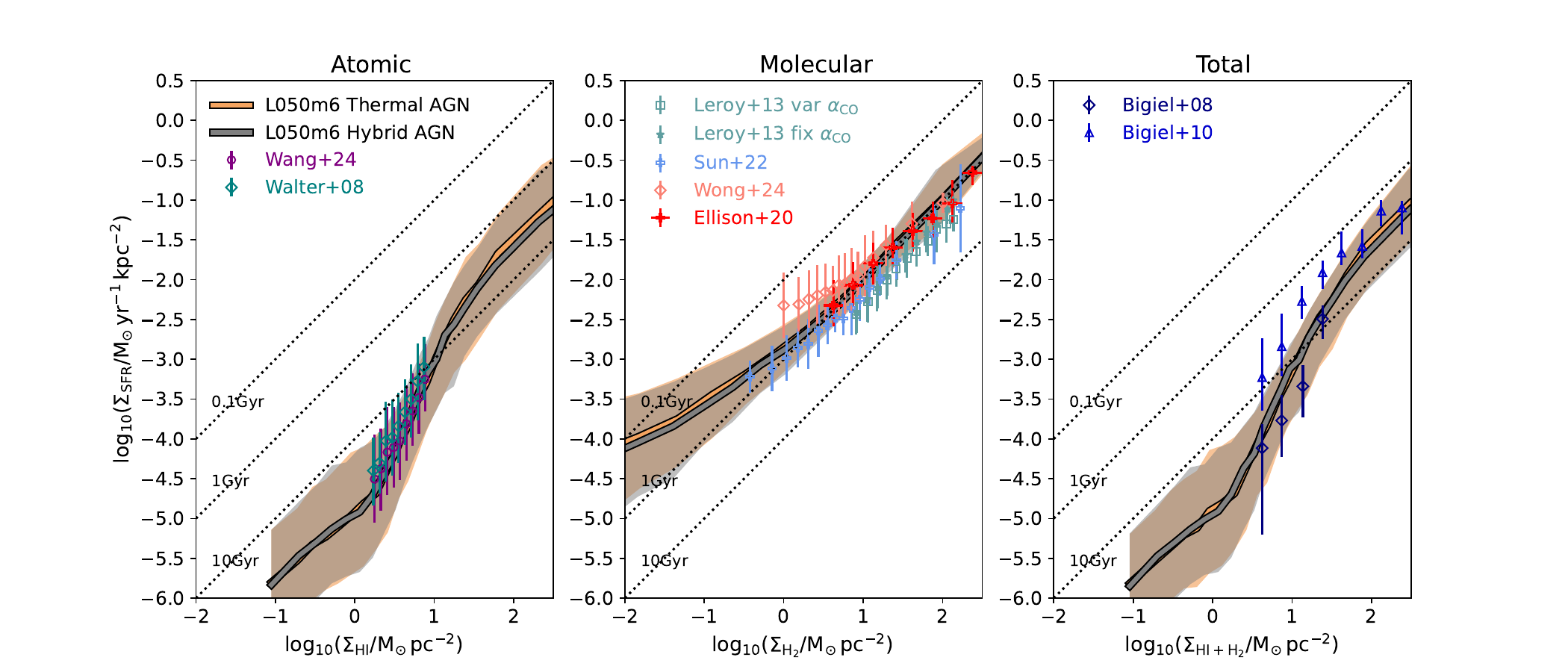}
\caption{As Fig.~\ref{KS_CompObs} but for the simulations L025m5, L025m6 and L025m7, as labelled, at $z=0$ (top) and $z=2$ (middle). The bottom panel shows the $z=0$ KS relation again, but for the runs L050m6 Thermal and Hybrid, as labelled. Table~\ref{TableSimus} summarises the runs used in this figure. 
Observations are only shown in the top and bottom panels.} 
\label{KS_Resolution}
\end{center}
\end{figure*}

\begin{figure*}
\begin{center}
\includegraphics[trim=10mm 0mm 5mm 5mm, clip,width=0.95\textwidth]{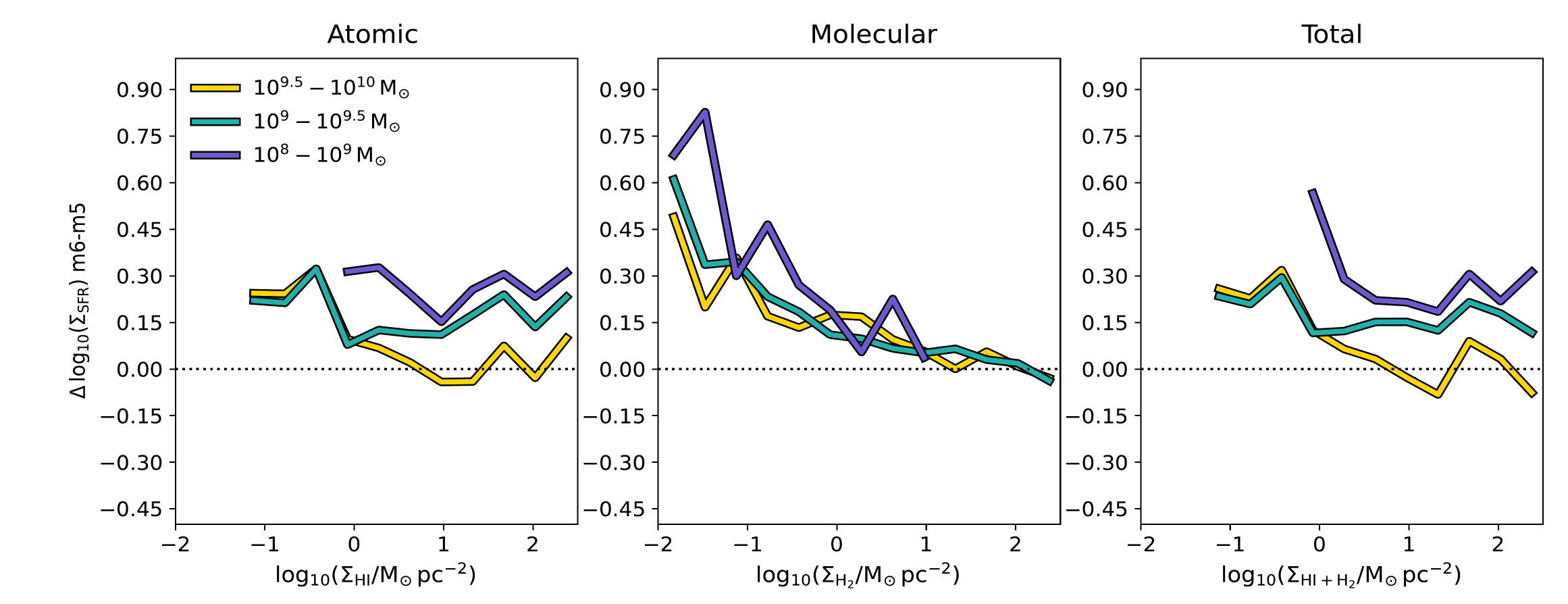}
\caption{The difference between $\rm log_{10}(\Sigma_{\rm SFR})$ in the L025m5 and L025m6 runs in bins of $\rm log_{10}(\Sigma_{\rm gas})$, for galaxies at $z=0$ selected in different ranges stellar mass, as labelled in the left panel. The differences become smaller as we move towards more massive galaxies. In the right panel, the lowest mass galaxies in the L025m6 display differences in $\rm log_{10}(\Sigma_{\rm SFR})$ of up to $0.6$~dex compared to L025m5.} 
\label{KS_Resolution2}
\end{center}
\end{figure*}

\begin{figure*}
\begin{center}
\includegraphics[trim=0mm 0mm 0mm 0mm, clip,width=0.45\textwidth]{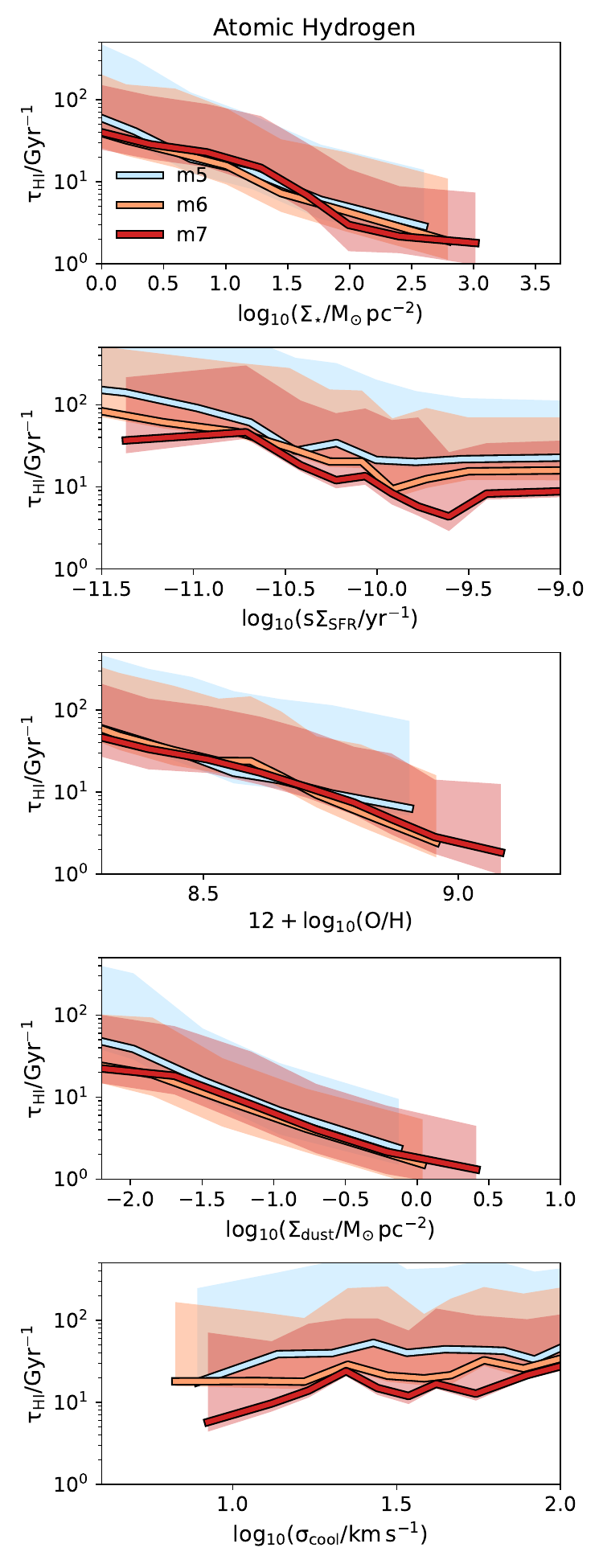}
\includegraphics[trim=0mm 0mm 0mm 0mm, clip,width=0.45\textwidth]{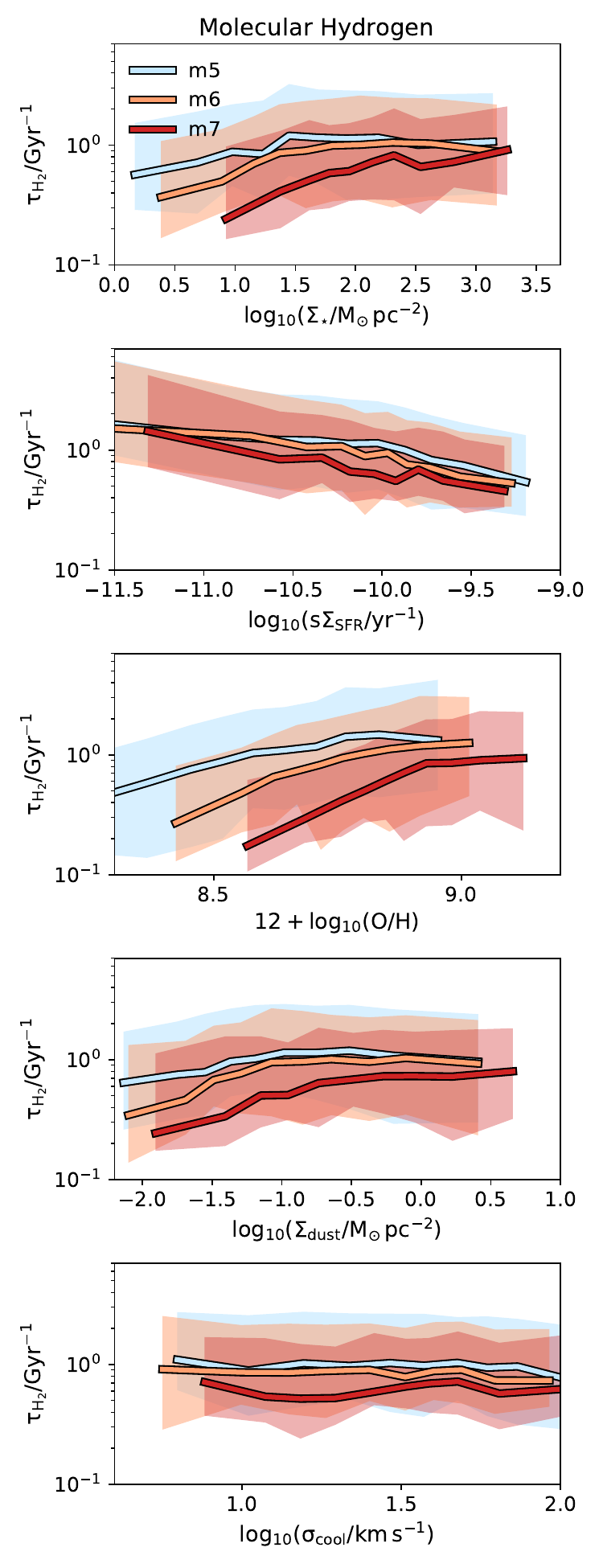}
\caption{The relation between the depletion times of \ion{H}{i} (left) and H$_2$ (right) and the 5 local properties analysed in Fig.~\ref{Delta_correlations}-from top to bottom stellar surface density, local sSFR, gas metallicity, dust surface density, and cool gas velocity dispersion. This is shown for the L025m5, L025m6 and L025m7 as labelled in the top panels. In the three runs we include galaxies with $M_{\star}>10^9\,\rm M_{\odot}$ and $\rm SFR>0$ at $z=0$.} 
\label{KS_Resolution3}
\end{center}
\end{figure*}

The top and middle panels of Fig.~\ref{KS_Resolution} show the KS relations for the atomic, molecular and total neutral hydrogen at $z=0$ and $2$ for the three L025 simulations listed in Table~\ref{TableSimus}.

For HI, we see that the convergence at $z=0$ is excellent, with differences between resolutions only apparent in the scatter of the relation. The resolution m5 produces the largest scatter (by $0.15$~dex), consistent with Fig.~\ref{KS_IndividualTracks}. At $z=2$, there are differences between the three resolutions at low surface densities of HI, $\rm log_{10}(\Sigma_{\rm HI}/M_{\odot}\,pc^{-2})\lesssim 1$. Below that surface density, m5 displays the steepest relationship (at $0 \lesssim \rm log_{10}(\Sigma_{\rm HI}/M_{\odot}\,pc^{-2})\lesssim 1$). In that regime, it follows the same slope as predicted for higher surface densities. 

For H$_2$, we see good convergence at high surface densities. At m5 resolution, the KS slope of $\approx 1$ is present at all densities $\gtrsim 0.1\,\rm M_{\odot}\, pc^{-2}$. For m6 resolution, we start to see a flattening of the KS relation and a deviation from m5 resolution below $\approx 1\,\rm M_{\odot}\, pc^{-2}$. For m7 resolution, this deviation occurs at higher densities, $\approx 10\,\rm M_{\odot}\, pc^{-2}$. At $z=2$, the deviations from the m5 resolution results move to lower H$_2$ surface densities, so that convergence is achieved already at $\gtrsim 0.1\,\rm M_{\odot}\, pc^{-2}$. \citet{Nobels24} used idealised galaxy disks to show that the molecular KS relation did not converge at low gas surface densities, which agrees with what we find in the fully cosmological set-up of \colibre. 

For the total \ion{H}{i}+H$_2$ KS relation, we see very similar behaviour to that obtained for the \ion{H}{i} KS relation.

For our fiducial m6 resolution, this means that our results are converged at the typical surface densities probed by observations, $\gtrsim 10\,\rm M_{\odot}\, pc^{-2}$ for H$_2$ and $\gtrsim 1 \rm M_{\odot}\, pc^{-2}$ for HI. 

Fig.~\ref{KS_Resolution2} shows the differences in the KS relation normalisation between m6 and m5 for galaxies at $z=0$ selected in bins of stellar mass. We find that galaxies at m6 resolution with stellar masses $10^8-10^9\,\rm M_{\odot}$ can show deviations of up to $0.6$~dex in the normalisation of the total KS relation (right panel) relative to the m5 run. We thus decide to impose a minimum stellar mass of $10^9\,\rm M_{\odot}$ for our analysis of the L200m6 run, but noting that there is no clear stellar mass threshold above which the KS relation is fully converged. This will depend on how star-forming/gas-rich a galaxy is, and   which gas surface densities we sample. Nevertheless, we consider the stellar mass threshold above to be a reasonable compromise between probing a large dynamic range while also avoiding galaxies that are more affected by resolution.  

The analysis above is for the predicted average KS relation. The convergence of the scaling relations between the molecular and atomic depletion times and the several local galaxy properties analysed in \S~\ref{resolveddeps} is not guaranteed and needs to be explicitly tested. Fig.~\ref{KS_Resolution3} shows the dependence of the H$_2$ and \ion{H}{i} gas depletion times and the $5$ properties shown in Fig.~\ref{Delta_correlations}. 
In the case of $\tau_{\rm HI}$, Fig.~\ref{KS_Resolution3} shows an overall good convergence between the three resolutions. This agrees with the convergence seen in the average \ion{H}{i} KS relation in Fig.~\ref{KS_Resolution}. The convergence is particularly good for the properties that are most strongly correlated with $\tau_{\rm HI}$, i.e. $\Sigma_{\star}$ and $\Sigma_{\rm dust}$. 

For $\tau_{\rm H_2}$ we see that the m5 and m6 resolutions produce very similar correlations with the dust surface density, local sSFR, cool gas velocity dispersion and stellar surface density. The m7 resolution clearly deviates from the relations obtained at higher resolutions. The gas metallicity (top-left panel) shows that the m5 and m6 resolution exhibit significant differences, particularly in low metallicity gas ($\rm 12 + log_{10}(O/H) \lesssim 8.7$). This may not be surprising given the different densities at which stars form in the m5 and m6 resolutions at low gas metallicity (Fig.~\ref{FracCNMz0}). \citet{Schaye25} shows that the stellar mass-gas metallicity relation predicted by \colibre\ at $z=0$ does not converge well between the runs at m5 and m6 resolution for low-mass, low-gas metallicity galaxies ($M_{\star}\lesssim 10^{9.3}\,\rm M_{\odot}$). 

\begin{table}
\begin{center}
  \caption{The Spearman correlation coefficient, $R$, for the correlation between $\rm log_{10}(\tau_{\rm H_2})$ or $\rm log_{10}(\tau_{\rm HI})$ and $5$ local properties listed in each row. This is shown for the \textcolor{cyan}{L025m5}, \textcolor{orange}{L025m6} and \textcolor{red}{L025m7} simulations for galaxies with $M_{\star}>10^9\,\rm M_{\odot}$ and $\rm SFR>0$ at $z=0$. Units used for each property are as  in Fig.~\ref{KS_Resolution3}.  Fig.~\ref{Delta_correlations} shows these values for the L200m6 run we use throughout the paper.}\label{tab:corrfactors}
\begin{tabular}{l|c|c}
\hline
           & $\rm log_{10}(\tau_{\rm HI})$ & $\rm log_{10}(\tau_{\rm H_2})$\\
          \hline
      $\rm log_{10}(\Sigma_{\star})$ & \textcolor{cyan}{-0.66}, \textcolor{orange}{-0.68}, \textcolor{red}{-0.74}& \textcolor{cyan}{0.32}, \textcolor{orange}{0.42}, \textcolor{red}{0.35}\\
       $\rm log_{10}(\rm s\Sigma_{\rm SFR})$ & \textcolor{cyan}{-0.35}, \textcolor{orange}{-0.34}, \textcolor{red}{-0.27}& \textcolor{cyan}{-0.52}, \textcolor{orange}{-0.56}, \textcolor{red}{-0.44}\\
      $12+\rm log_{10}(O/H)$  &  \textcolor{cyan}{-0.27}, \textcolor{orange}{-0.35}, \textcolor{red}{-0.38} & \textcolor{cyan}{0.24}, \textcolor{orange}{0.32}, \textcolor{red}{0.36}\\
      $\rm log_{10}(\Sigma_{\rm dust})$ & \textcolor{cyan}{-0.62}, \textcolor{orange}{-0.71}, \textcolor{red}{-0.77}& \textcolor{cyan}{0.26}, \textcolor{orange}{0.36}, \textcolor{red}{0.32}\\
      $\rm log_{10}(\sigma_{\rm cool})$ & \textcolor{cyan}{0.09}, \textcolor{orange}{0.14}, \textcolor{red}{0.14}& \textcolor{cyan}{-0.08}, \textcolor{orange}{-0.10}, \textcolor{red}{-0.03}\\
     \hline
\end{tabular}
\end{center}
\end{table}

Despite some differences found between resolutions, the strengths of the correlations shown in Fig.~\ref{Delta_correlations} converge quite well. We show this in Table~\ref{tab:corrfactors}. The strongest correlations in terms of Spearman's correlation coefficient $R$ are the same for the three resolutions. The general trend is that the m5 resolution tends to display a larger scatter in the correlations between $\tau_{\rm HI}$ or $\tau_{\rm H_2}$ and the five properties listed in  Table~\ref{tab:corrfactors}, leading to slightly lower values of $R$.

\subsection{Convergence with the adopted AGN feedback model}\label{convagn}

The bottom panels of Fig.~\ref{KS_Resolution} show the KS relations predicted by the L050m6 Thermal and Hybrid AGN feedback runs (listed in Table~\ref{TableSimus}). This is shown at $z=0$ and for the same stellar mass and SFR selection. The predicted relations are very similar including the medians and scatter.
The figure confirms that the adopted AGN feedback model has a negligible impact on the predicted KS relation in \colibre, and justifies our decision to carry out most of the analysis using runs with thermal AGN feedback only. 

\subsection{Convergence between the methods used to measure the KS relation}\label{convmeth}

Besides the method ``Annuli on face-on galaxy'', we also implemented other methods to measure maps of the SFR and gas surface densities with the purpose of studying how well the resulting KS converges between these different methods.

Below, we summarise the methods explored:
\begin{itemize}
    \item {\it Annuli on randomly-oriented galaxy:} This method starts by taking the simulation box XY projection, and then following the same procedure as described for the {\sc annuli-face} method described in \S~\ref{ResolvedMapsMethods}. We refer to this method as {\sc annuli-rand}.
    \item {\it Regular grid on face-on galaxy:} This method also starts from the galaxies oriented face-on, which is computed as described for the {\sc annuli-face} method. Then, all particles within a square of $50$~pkpc are selected. We then grid the space using regular square pixels of size $\rm \Delta r\times \Delta r$. We compute for each pixel the masses, SFRs and metallicities following the {\sc annuli-face} method but using only the particles that fall in the pixel. We refer to this method as {\sc grid-face}. Surface densities are then computed using the area of the pixel.
    \item {\it Regular grid on randomly-oriented galaxy:} This method follows the description provided for {\sc grid-face}, but the grid is constructed over the galaxy projected onto the simulation box XY's coordinates, as is done for {\sc annuli-rand}. We refer to this method as {\sc grid-rand}.
\end{itemize}

We remind the reader that our fiducial method bins galaxies in circular apertures of $\rm \Delta r = 1$~kpc after orienting them face-on, and with a  fiducial minimum number of particles per bin of $N_{\rm min}=10$. 

\begin{figure*}
\begin{center}
\includegraphics[trim=20mm 0mm 30mm 10mm, clip,width=0.93\textwidth]{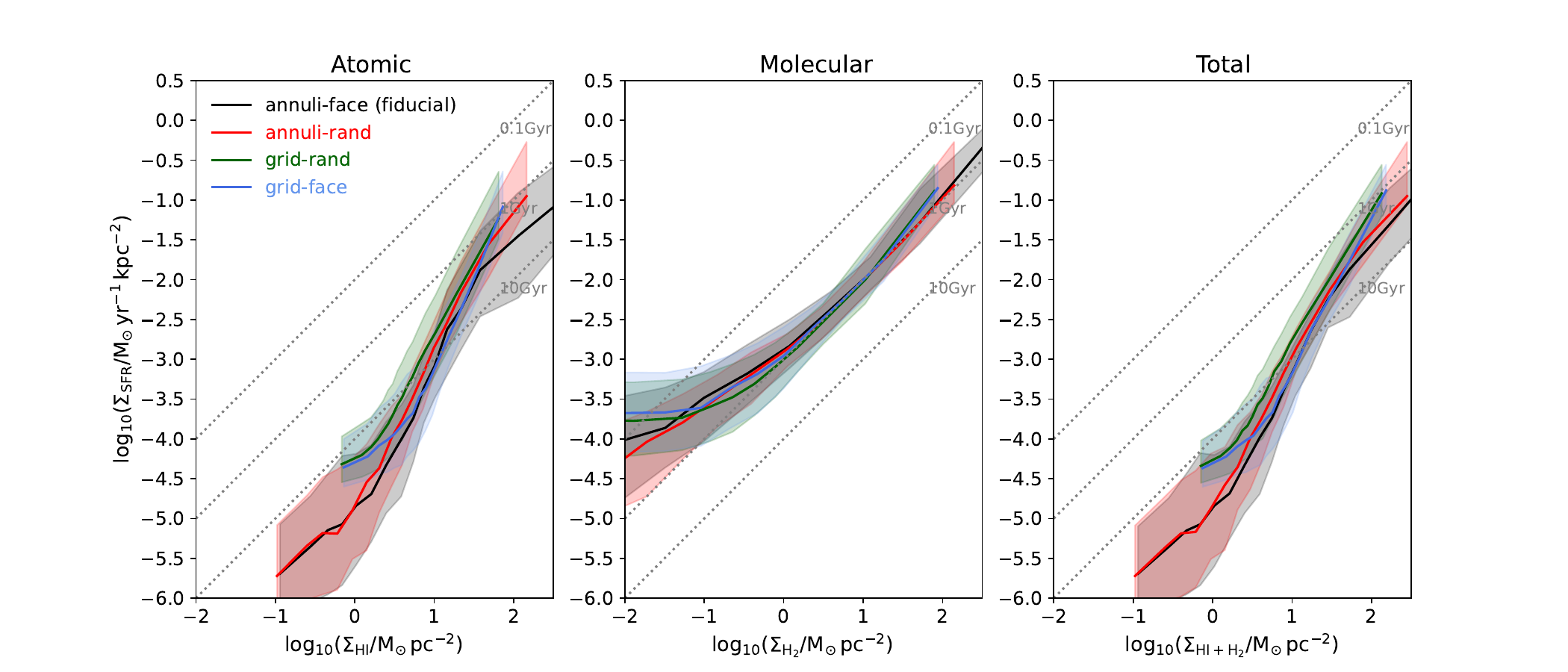}
\includegraphics[trim=20mm 0mm 30mm 17mm, clip,width=0.93\textwidth]{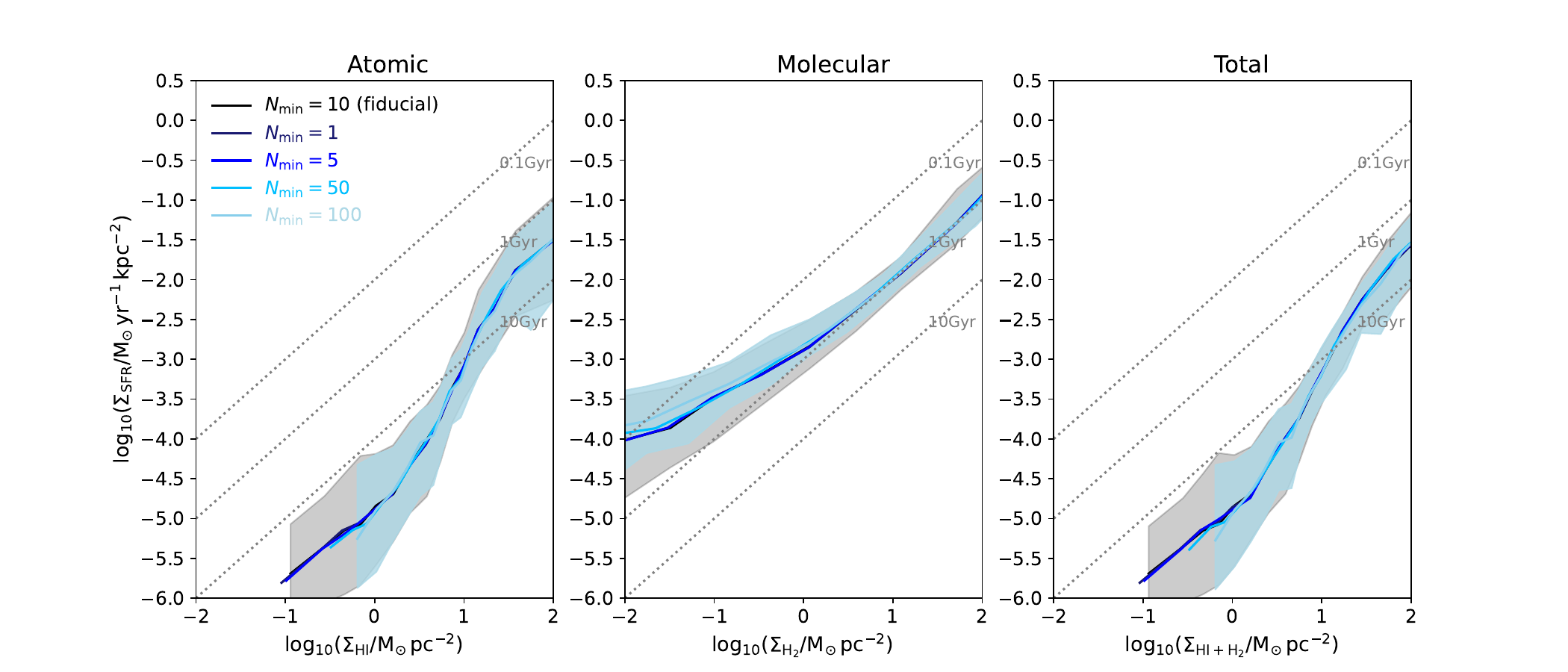}
\includegraphics[trim=20mm 0mm 30mm 17mm, clip,width=0.93\textwidth]{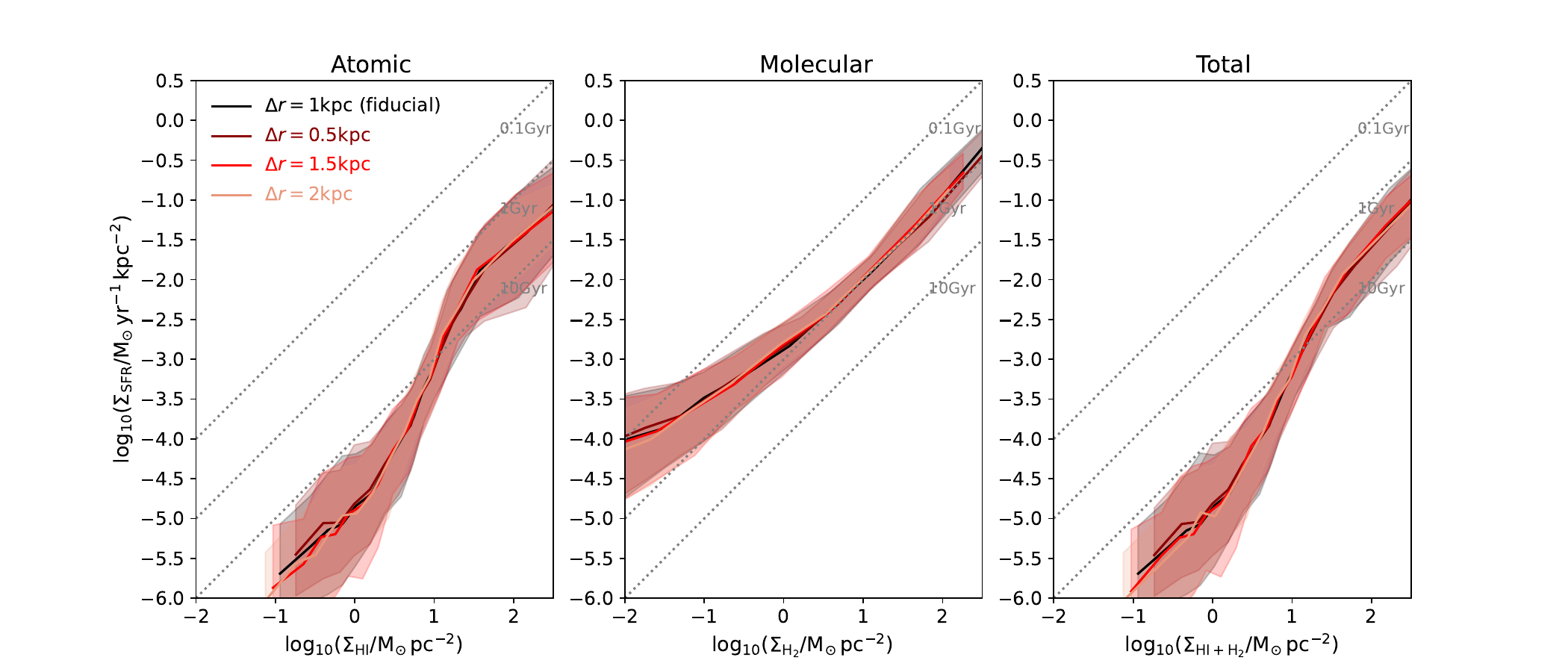}
\caption{The HI, H$_2$ and \ion{H}{i}+H$_2$ KS relations, as labelled at the top, at $z=0$ in the L025m6 run of Table~\ref{TableSimus}, measured using different methods (top); using our fiducial {\sc annuli-face} method but adopting different thresholds for the minimum number of gas particles in a annulus (middle); and adopting different widths for the annuli (bottom). In all panels lines show medians, while the shaded regions indicate the $16^{\rm }-84^{\rm th}$ percentile ranges. Not not overwhelm the figure, in the middle panel we only show the $16^{\rm }-84^{\rm th}$ percentile ranges for $N_{\rm min}=10$ and $=100$, while for the other panels we show medians and percentile ranges for all the samples. We find excellent convergence in the predicted KS relation against methods and the specific choices of each method.} 
\label{KS_Methods}
\end{center}
\end{figure*}

In the following, we compare different methods of measuring the KS relation at $z=0$ in the L025m6 run to determine whether our results are robust against the choices we make about how to measure the KS relation.

The top panels of Fig.~\ref{KS_Methods} shows a comparison of the $4$ different methods used in this paper to measure the KS relation. For H$_2$ (top-middle), we find excellent agreement between methods at $\rm log_{10}(\Sigma_{\rm H_2}/M_{\odot}\,pc^{-2}) \gtrsim 0.5-1.8$, which is the regime in which observations are available. At lower and higher surface densities, we start to see some differences in the median, but they are still well within the scatter of the correlation.

For \ion{H}{i} (top-left panel in Fig.~\ref{KS_Methods}), we see more systematic differences between methods,  with the ones adopting a random orientation leading to a slightly higher normalisation than those adopting a face-on orientation. This is naturally expected in situations where the \ion{H}{i} disk has a larger scaleheight than the disk traced by the instantaneous SFR. The latter is expected in \colibre\ given the hotter nature of \ion{H}{i} compared to the gas particles with $\rm SFR>0$ (see Fig.~\ref{PhaseSpacez0}). Regardless, the differences in normalisation are within $\approx 0.2$~dex, well below the intrinsic scatter of the \ion{H}{i} KS relation in \colibre.

For the total \ion{H}{i}+H$_2$ (top-right panel of Fig.~\ref{KS_Methods}), the systematic differences between methods are smaller than those seen for HI, but go in the same direction, with the methods using face-on orientation producing a slightly lower normalisation of the KS relation than those using random orientations.

The middle and lower panels of Fig.~\ref{KS_Methods} show the KS relations obtained for \colibre\ using the {\sc annuli-face} method, but varying $\Delta\,r$ (the width of the circular bins) and $N_{\rm min}$ (the minimum number of particles required per circular bin), as labelled. In both cases we see excellent convergence in the shape and scatter of the KS relation in all neutral hydrogen phases. The main difference we see is that increasing $N_{\rm min}$ leads to a poorer sampling of the low surface density regime, but without introducing a significant bias. 

Overall, Fig.~\ref{KS_Methods} demonstrate that the predicted KS relation in \colibre\ at the m6 resolution is robust against variants in the methods used to measure it. 

\section{Additional visualisation examples}

\begin{figure*}
\begin{center}
\includegraphics[trim=0mm 0mm 0mm 0mm, clip,width=0.33\textwidth]{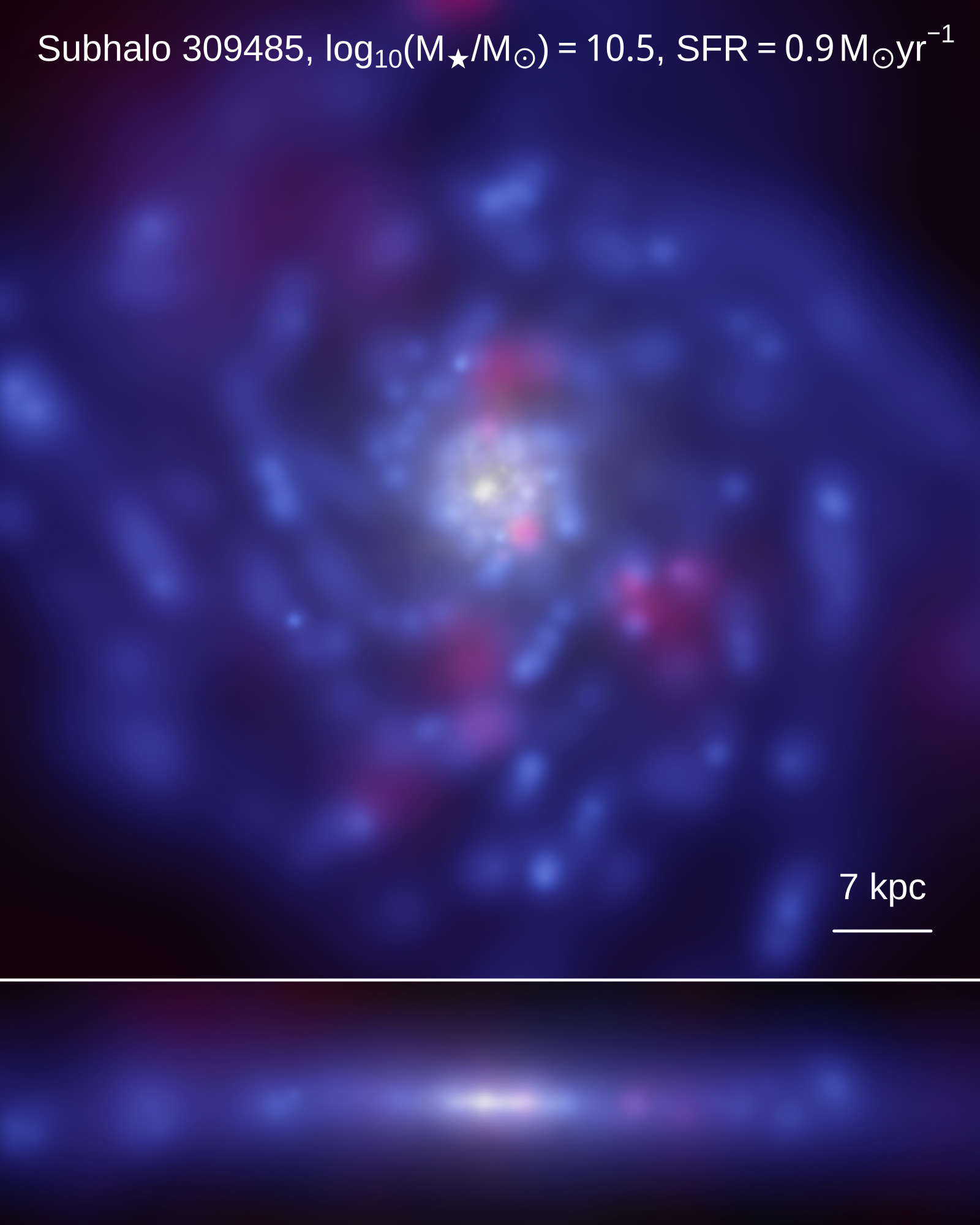}
\includegraphics[trim=0mm 0mm 0mm 0mm, clip,width=0.33\textwidth]{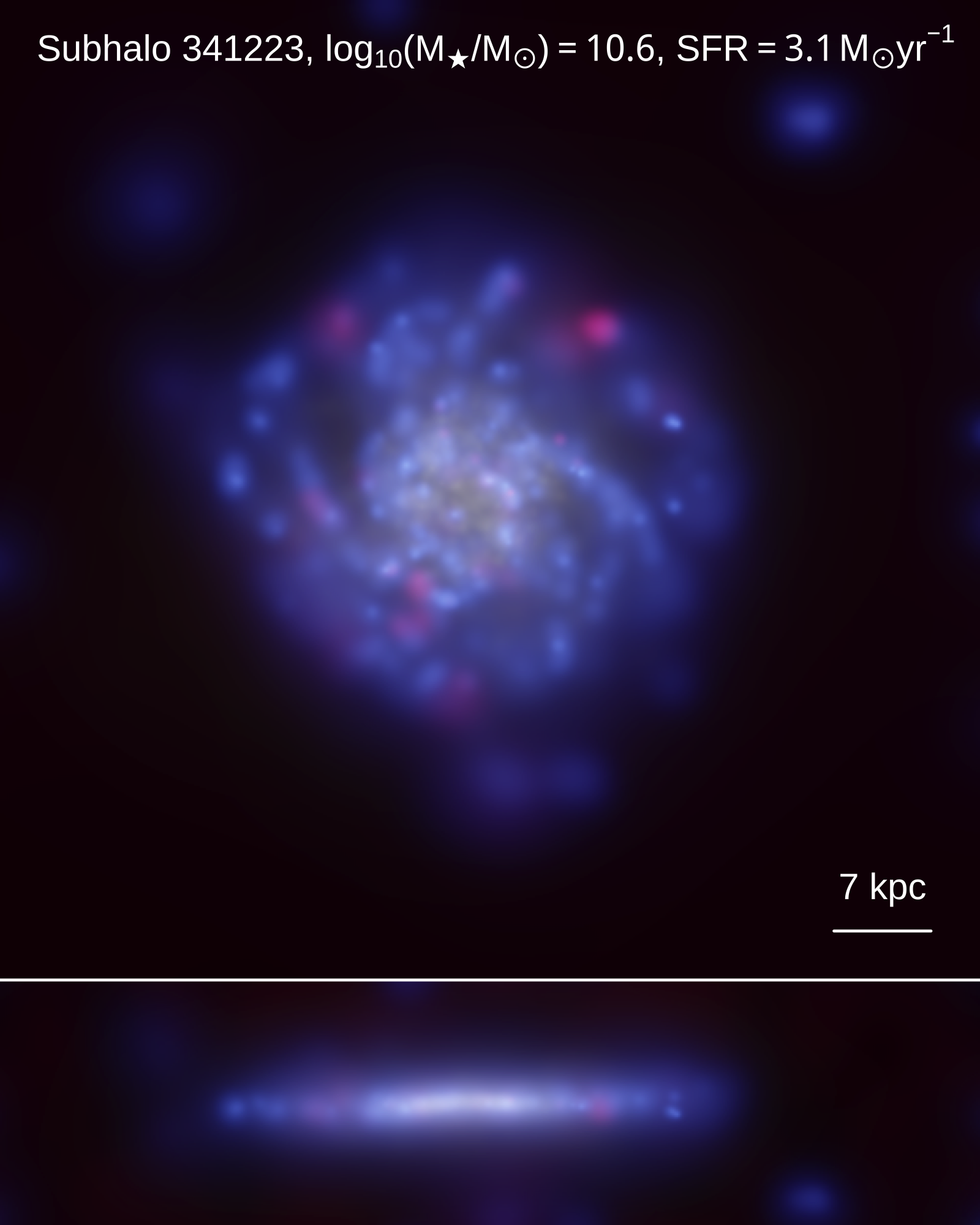}
\includegraphics[trim=0mm 0mm 0mm 0mm, clip,width=0.33\textwidth]{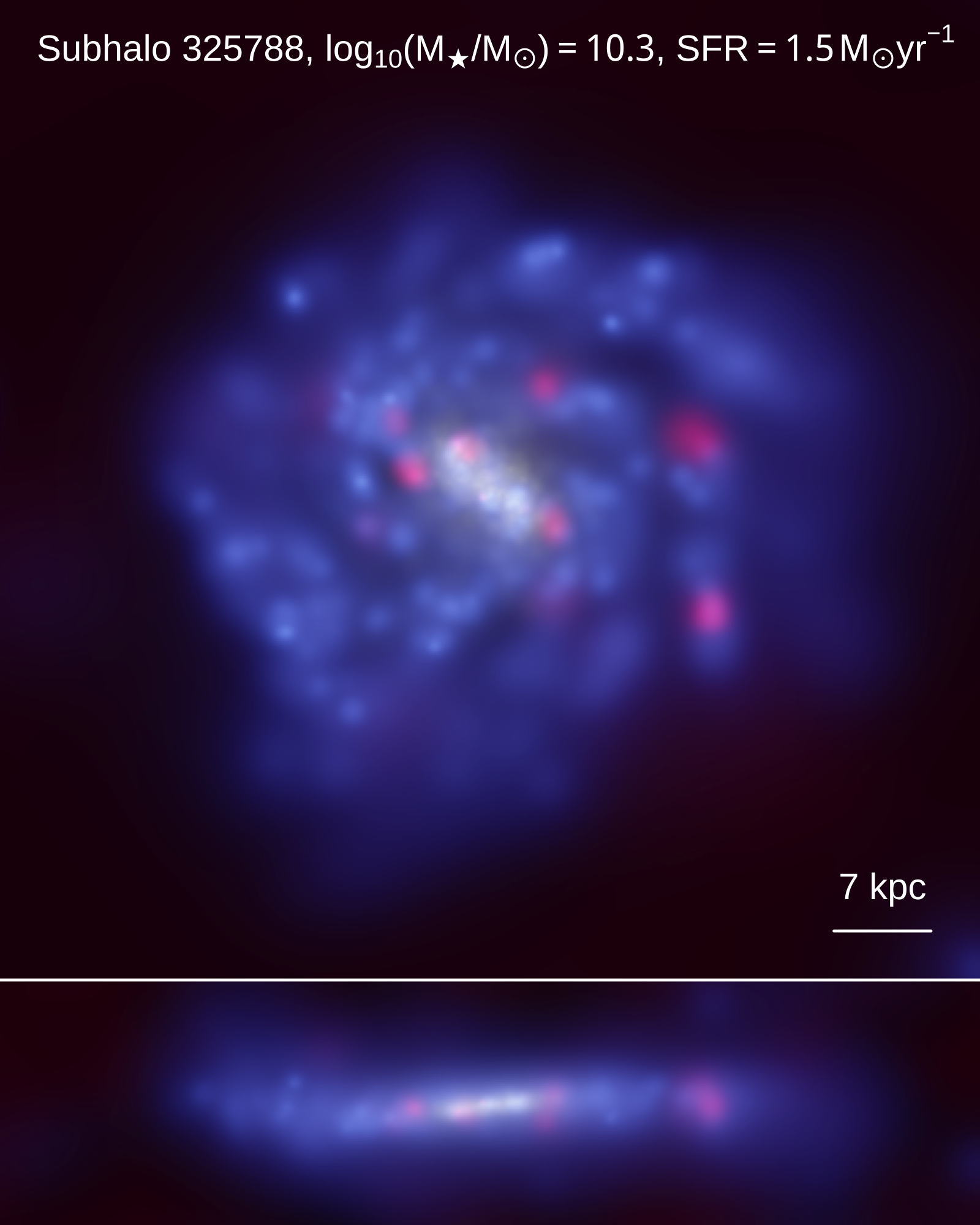}
\hspace*{-0.8cm}\includegraphics[trim=5mm 5mm 3mm 1mm, clip,width=0.37\textwidth]{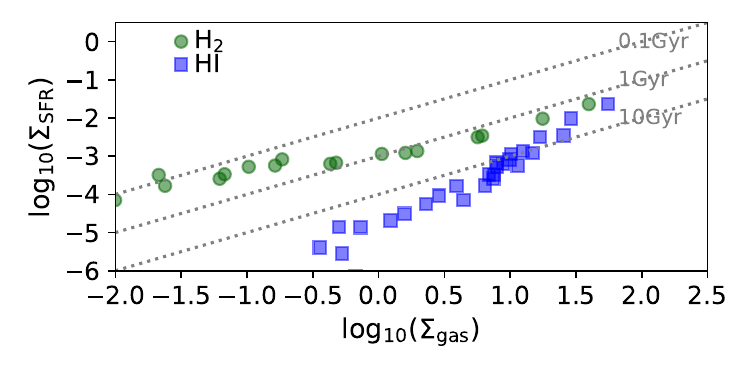}
\hspace*{-0.3cm}\includegraphics[trim=10mm 5mm 3mm 1mm, clip,width=0.35\textwidth]{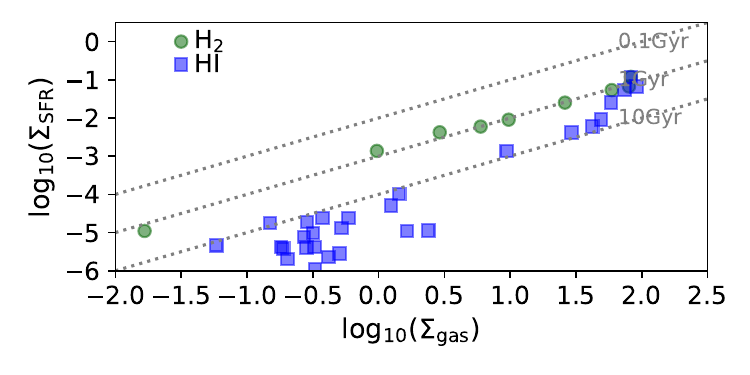}
\hspace*{-0.3cm}\includegraphics[trim=10mm 5mm 3mm 1mm, clip,width=0.35\textwidth]{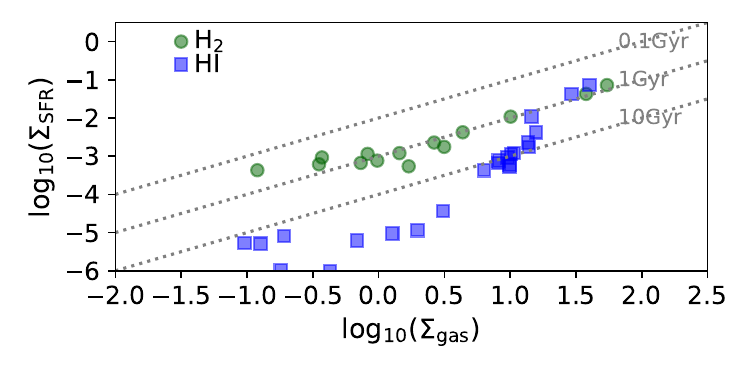}
\includegraphics[trim=0mm 0mm 0mm 0mm, clip,width=0.33\textwidth]{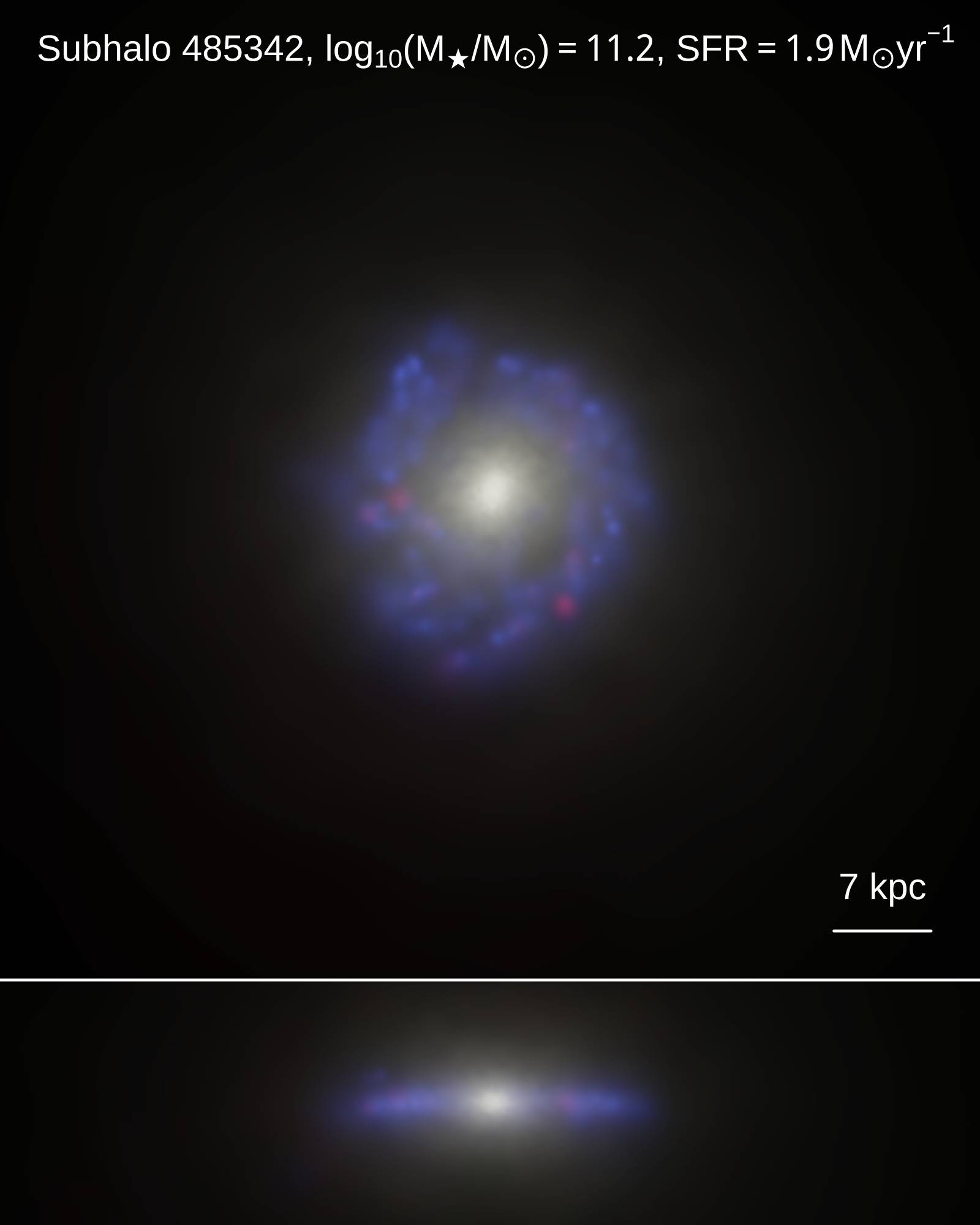}
\includegraphics[trim=0mm 0mm 0mm 0mm, clip,width=0.33\textwidth]{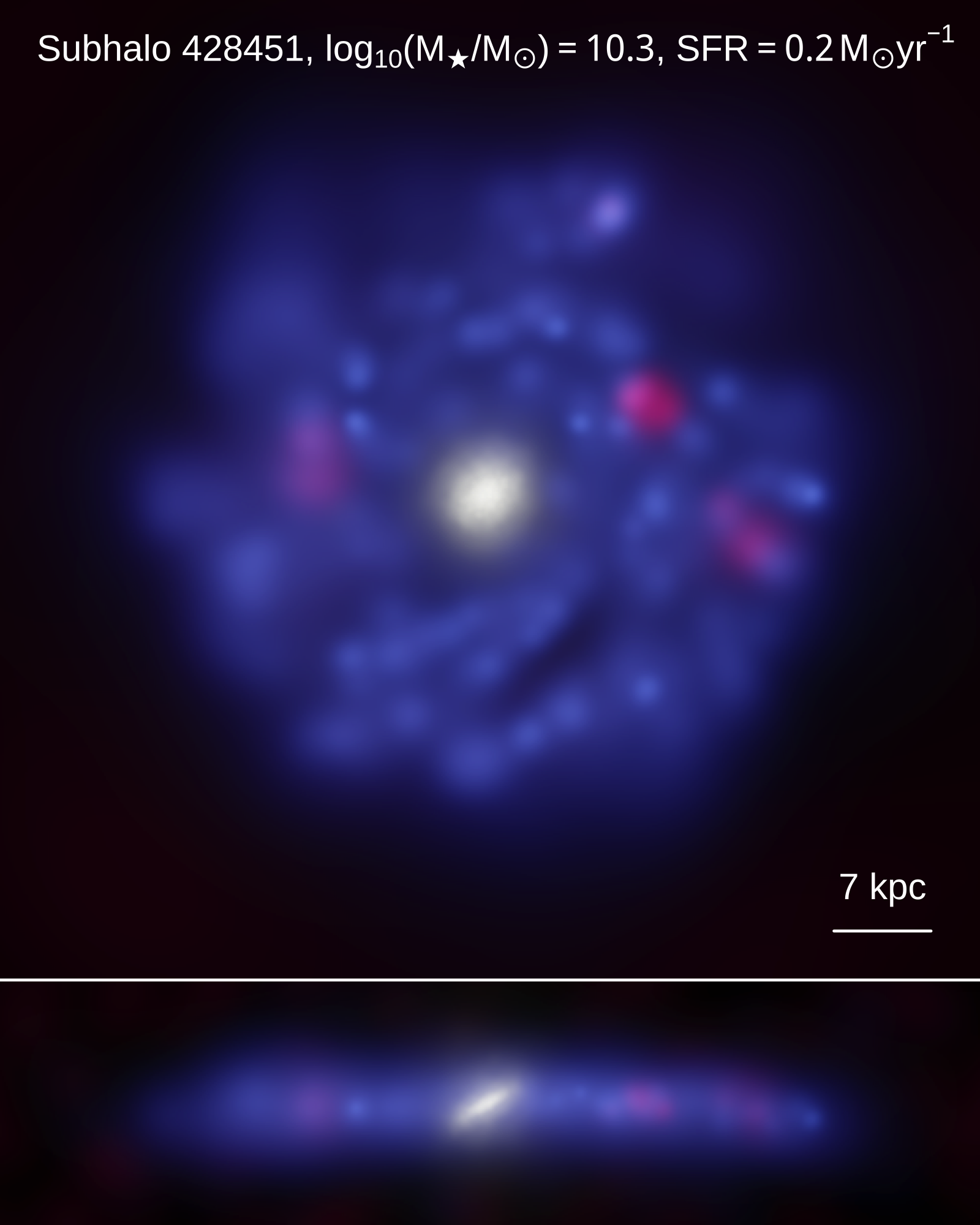}
\includegraphics[trim=0mm 0mm 0mm 0mm, clip,width=0.33\textwidth]{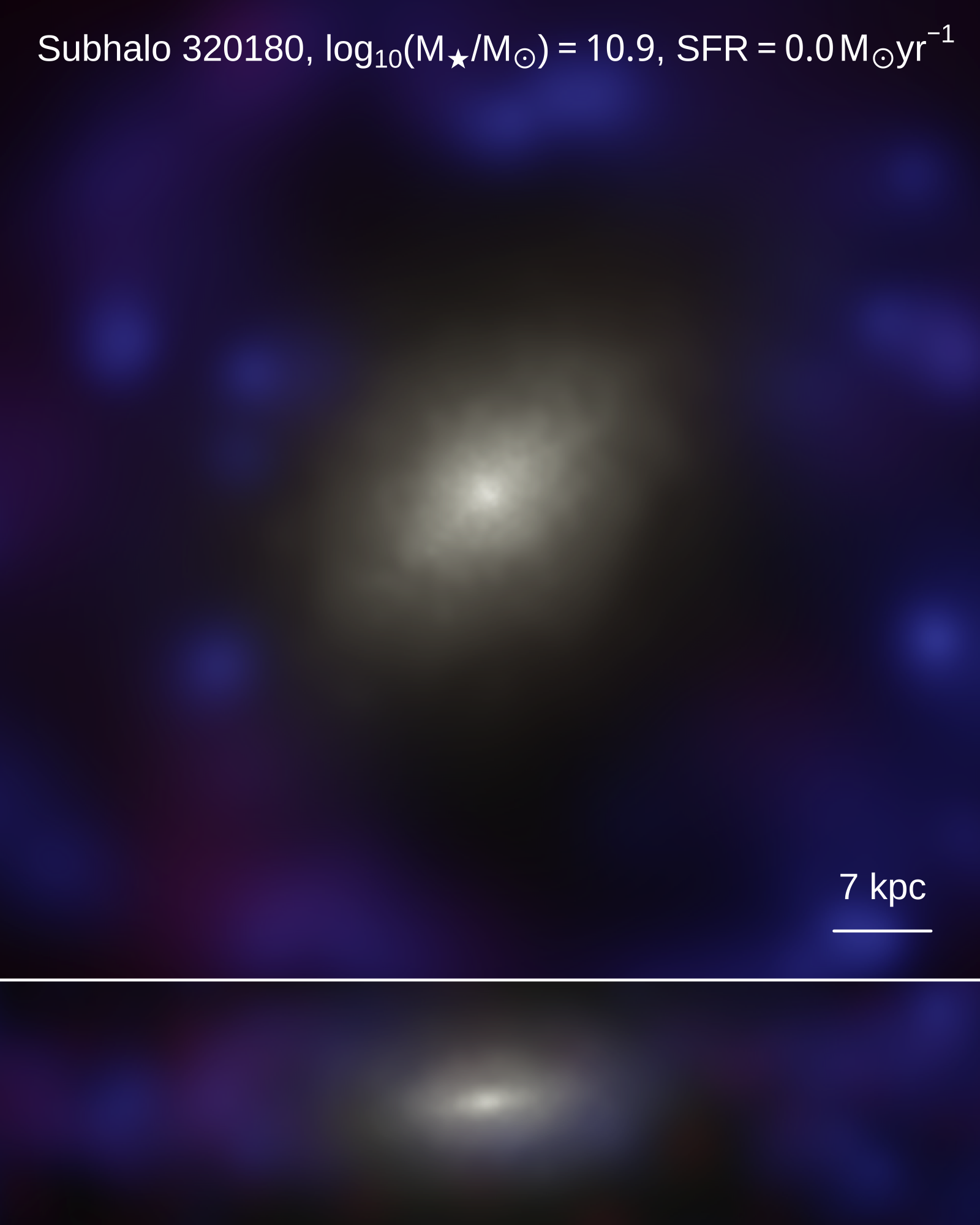}
\hspace*{-0.8cm}\includegraphics[trim=5mm 5mm 3mm 1mm, clip,width=0.37\textwidth]{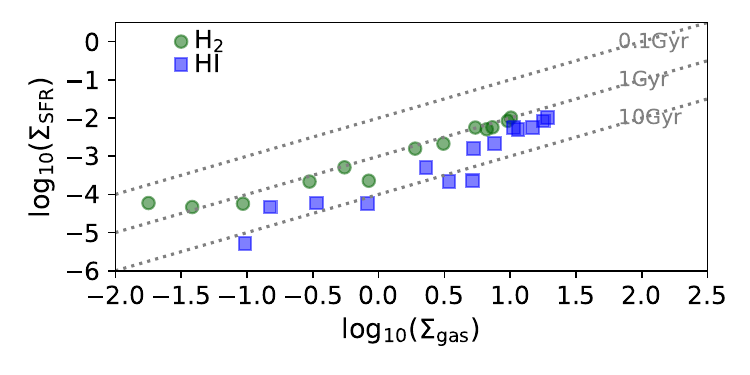}
\hspace*{-0.3cm}\includegraphics[trim=10mm 5mm 3mm 1mm, clip,width=0.35\textwidth]{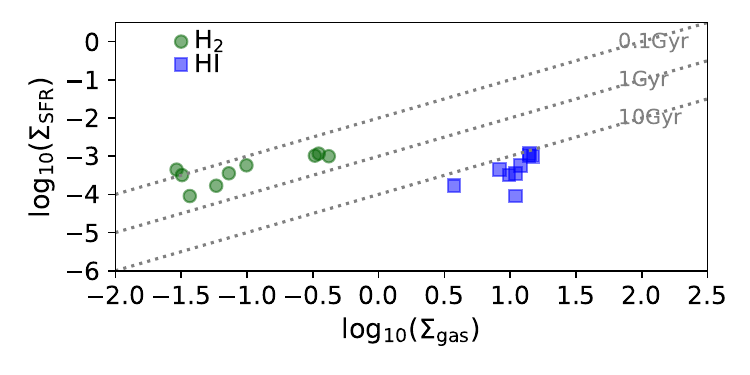}
\hspace*{-0.3cm}\includegraphics[trim=10mm 5mm 3mm 1mm, clip,width=0.35\textwidth]{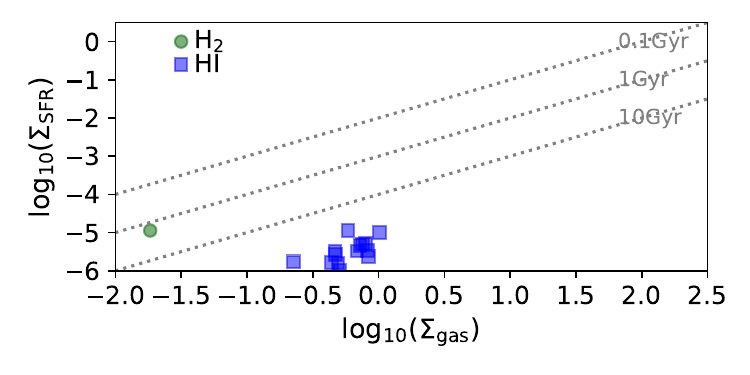}
\caption{As Fig.~\ref{IndividualGalsexample} but for another 6 galaxies at $z=0$ in the L025m6 run. The top row shows examples of star-forming galaxies, while the bottom row shows examples of galaxies below the main sequence but that still have some gas left to form stars.} 
\label{IndividualGalsexample2}
\end{center}
\end{figure*}

Fig.~\ref{IndividualGalsexample2} shows examples of star-forming galaxies (top) and passive galaxies (bottom) at $z=0$ in the L025m6 run. The stellar mass and its hierarchy (whether it is a central or a satellite galaxy) are shown in the top panels. In general and at this resolution, we see that star-forming galaxies sample H$_2$ surface densities in excess of $\rm log_{\rm 10}(\Sigma_{\rm H_2}/\rm M_{\odot}\,pc^{-2}) \gtrsim 1.5$, while galaxies below the main sequence rarely have H$_2$ gas that reaches such high densities.


\bsp	
\label{lastpage}
\end{document}